\documentclass[11pt]{article}
\usepackage{setspace,graphicx,amssymb,amsmath,latexsym,amsfonts,amscd,amsthm,multirow,ctable,mathdots,caption,array}
\usepackage{mathtools}
\usepackage{fancyhdr}
\usepackage{tabularx,cite,mathrsfs,color}
\usepackage{stmaryrd}
\usepackage{rotating}
\usepackage{authblk}
\usepackage{hyperref}
\usepackage{accents}
\usepackage{enumitem}
\usepackage{bbm}

\usepackage[margin=2cm]{caption}
\usepackage{geometry}

\usepackage{multirow}
\setlist[enumerate]{leftmargin=*}
\usepackage[font={small,it}]{caption}

\newcommand{\bea}{\begin{eqnarray}} 
\newcommand{\eea}{\end{eqnarray}} 
\newcommand{\bee}{\begin{eqnarray*}} 
\newcommand{\eee}{\end{eqnarray*}} 
\newcommand{\al}{\begin{align*}} 
\newcommand{\eal}{\end{align*}} 
\newcommand{\be}{\begin{equation}} 
\newcommand{\ee}{\end{equation}} 
\newcommand{\eq}[1]{(\ref{#1})} 
\newcommand{\bem}{\begin{pmatrix}} 
\newcommand{\eem}{\end{pmatrix}}

\def\b{\beta} 
\def\c{\gamma}

\def\inf{\infty}

\def\m{\mu}

\def\p{\pi}    
\def\pa{\partial}

\def\t{\tau} 
\def\th{\theta} 
\def\til{\tilde} 
\def\Tr{\text{Tr}}

\newcolumntype{R}{ >{$}r <{$}}
\newcolumntype{C}{ >{$}c <{$}}
\newcolumntype{L}{ >{$}l <{$}}
\newcolumntype{F}{>{\centering\arraybackslash}m{1.5cm}}

\def\ll{\ell}

\newcommand{\mc}[1]{\mathcal{#1}}

\newcommand{\comment}[1]{}

\newcommand{\RR}{{\mathbb R}}
\newcommand{\CC}{{\mathbb C}}
\newcommand{\ZZ}{{\mathbb Z}}
\newcommand{\QQ}{{\mathbb Q}}
\newcommand{\HH}{{\mathbb H}}

\newcommand{\tr}{\operatorname{{tr}}}

\newcommand{\ex}{\operatorname{e}} 

\newcommand{\SL}{\operatorname{\textsl{SL}}}      

\newcommand{\G}{\Gamma}	
\newcommand{\g}{\gamma}	

\newcommand{\Co}{\textsl{Co}}	

\newcommand{\cN}{{\mathcal N}}
\newcommand{\cA}{{\mathcal A}}

\newcommand{\gfac}[2]{\ensuremath(\tilde{\bf #1})^{#2}}

\newcommand{\pfac}[2]{\ensuremath({\bf #1})_+^{#2}}
\newcommand{\mfac}[2]{\ensuremath({\bf #1})_-^{#2}}

\newcommand{\Gfac}[2]{\ensuremath\lbrack\tilde{\bf #1}\rbrack^{#2}}
\newcommand{\Pfac}[2]{\ensuremath\lbrack{\bf #1}\rbrack_+^{#2}}
\newcommand{\Mfac}[2]{\ensuremath\lbrack{\bf #1}\rbrack_-^{#2}}

\newtheorem{thm}{Theorem}[section]

\theoremstyle{definition}
\newtheorem{defn}[thm]{Definition}

\theoremstyle{remark}

\numberwithin{equation}{section}

\makeatletter
\newsavebox\myboxA
\newsavebox\myboxB
\newlength\mylenA

\newcommand*\xunderline[2][0.75]{%
    \sbox{\myboxA}{$\m@th#2$}%
    \setbox\myboxB\null
    \ht\myboxB=\ht\myboxA%
    \dp\myboxB=\dp\myboxA%
    \wd\myboxB=#1\wd\myboxA
    \sbox\myboxB{$\m@th\underline{\copy\myboxB}$}
    \setlength\mylenA{\the\wd\myboxA}
    \addtolength\mylenA{-\the\wd\myboxB}%
    \ifdim\wd\myboxB<\wd\myboxA%
       \rlap{\hskip 0.5\mylenA\usebox\myboxB}{\usebox\myboxA}%
    \else
        \hskip -0.5\mylenA\rlap{\usebox\myboxA}{\hskip 0.5\mylenA\usebox\myboxB}%
    \fi}
\makeatother

\makeatletter
\newsavebox\myboxC
\newsavebox\myboxD
\newlength\mylenC

\newcommand*\xoverline[2][0.75]{%
    \sbox{\myboxC}{$\m@th#2$}%
    \setbox\myboxD\null
    \ht\myboxD=\ht\myboxC%
    \dp\myboxD=\dp\myboxC%
    \wd\myboxD=#1\wd\myboxC
    \sbox\myboxD{$\m@th\overline{\copy\myboxD}$}
    \setlength\mylenC{\the\wd\myboxC}
    \addtolength\mylenC{-\the\wd\myboxD}%
    \ifdim\wd\myboxD<\wd\myboxC%
       \rlap{\hskip 0.5\mylenC\usebox\myboxD}{\usebox\myboxC}%
    \else
        \hskip -0.5\mylenC\rlap{\usebox\myboxC}{\hskip 0.5\mylenC\usebox\myboxD}%
    \fi}
\makeatother

\makeatletter
\renewcommand*{\@fnsymbol}[1]{\ifcase#1\or$\Rbag$\or$\Lbag$\or$\varcurlyvee$\or$\varcurlywedge$\or$\bindnasrepma$\or$\binampersand$\or$\S$\or\else\@arabic{#1}\fi}
\makeatother

\title{
\vspace{-35pt}   
 \textsc
  {\huge{Properties of extremal CFTs with small central charge  }
        }
    }
\author[1]{{Francesca Ferrari}\thanks{f.ferrari@uva.nl}}
\author[2,3]{{Sarah M. Harrison}\thanks{sarharr@physics.mcgill.ca}}

\date{}

\affil[1]{\small{Institute of Physics, University of Amsterdam, Amsterdam, the Netherlands}}
\affil[2]{\small{Department of Mathematics and Statistics, McGill University, Montreal, QC, Canada}}
\affil[3]{\small{Department of Physics, McGill University, Montreal, QC, Canada}}
\begin{document}

\setstretch{1.2}
   \maketitle
\abstract{We analyze aspects of extant examples of 2d extremal chiral (super)conformal field theories with $c\leq 24$. These are theories whose only operators with dimension smaller or equal to $c/24$ are the vacuum and its (super)Virasoro descendents. The prototypical example is the monster CFT, whose famous genus zero property is intimately tied to the Rademacher summability of its twined partition functions, a property which also distinguishes the functions of Mathieu and umbral moonshine. However, there are now several additional known examples of extremal CFTs, all of which have at least $\mathcal N=1$ supersymmetry and global symmetry groups connected to sporadic simple groups. We investigate the extent to which such a property, which distinguishes the monster moonshine module from other $c=24$ chiral CFTs, holds for the other known extremal theories. We find that in most cases, the special Rademacher summability property present for monstrous and umbral moonshine does not hold for the other extremal CFTs, with the exception of the Conway module and two $c=12, ~\mc N=4$ superconformal theories with $M_{11}$ and $M_{22}$ symmetry. This suggests that the connection between extremal CFT, sporadic groups, and mock modular forms transcends strict Rademacher summability criteria.} 
\pagebreak
  \tableofcontents
  \section{Introduction}
  An extremal 2d (super)conformal field theory ((S)CFT) is a (S)CFT which has the minimal spectrum of primary operators consistent with both the (super)Virasoro algebra and modular invariance \cite{hoehn, Witten}. For the case of bosonic and $\mathcal N=1$ CFTs, Witten \cite{Witten} derived partition functions for putative extremal (S)CFTs assuming holomorphic factorization.
  Modular invariance and holomorphicity constrains the allowed values of the central charge to be $c=24k$ or $c=12k^*, ~ k,k^* \in \mathbb N,$ for bosonic and $\mathcal N=1$ CFTs, respectively. These CFTs, if they exist, were furthermore proposed to be holographically dual to pure (super)gravity in AdS$_3$.\footnote{There have been numerous works investigating this conjecture further (see, e.g., \cite{Gaiotto:2007xh,Gaberdiel:2007ve,Yin:2007gv,Yin:2007at,MalWit,Gaiotto:2008jt,Li:2008dq,Gaberdiel:2008pr,Gaberdiel:2010jf,Benjamin:2016aww,Bae:2016yna}); however as of now it has neither been proven nor disproven.} 
The authors of \cite{GGKMO} similarly derived elliptic genera for putative extremal SCFTs with $\mathcal N=2$ and $\mathcal N=4$ superconformal symmetry and conjectured theories with such elliptic genera, if they exist, would be dual to pure ($\mathcal N=2$ and $\mathcal N=4$) supergravity in AdS$_3$. Furthermore, they found that such theories can only exist for a finite set of small central charges due to constraints coming from the modular and elliptic properties of the elliptic genus. In particular, parameterizing the central charge for these theories as $c=6m, ~m\in \mathbb N$,\footnote{It would be interesting to consider a generalization to half-integer $m$ in the $\cN=2$ case.} such Jacobi forms exist only for $m\leq 13, ~m\neq 6,9,10,12$ and $m\leq 5$, in the $\mathcal N=2$ and $\mathcal N=4$ cases, respectively.\footnote{One interesting question is whether one can define a notion of ``near-extremal" CFT which extends to arbitrarily high central charge. 
See \cite{GGKMO,Benjamin:2016aww} for attempts in this direction.}

    Therefore one motivation for studying properties of extremal CFTs (ECFT) is to better understand three-dimensional quantum gravity.
Given the minimal mathematical input arising from physical reasoning via AdS/CFT, one surprise is that there are a number of known (chiral) CFTs with small central charge and extremal spectrum. Furthermore, they each have global symmetry groups related to sporadic finite simple groups. We summarize the existing extremal CFTs here and introduce notation we will use throughout the text.
 In the bosonic case, there is one known ECFT at $k=1$ \cite{Witten}, usually denoted as $\mc V^\natural$; this is the famous CFT with global symmetry group the monster group ($\mathbb M$), which was constructed by Frenkel, Lepowsky, and Meurman (FLM) in \cite{FLM}. In the case of $\mathcal N=1$ chiral CFTs, it was pointed out in \cite{Witten} that there are ECFTs with $k^*=1,2$ and symmetry related to the sporadic group $Co_0$ (``Conway zero") first built in \cite{FLM, Duncan} and \cite{DGH}, respectively. We refer to these as $\mc E^{\cN=1}_{k^*=1}:=\mc V^{s\natural}$ and $\mc E^{\cN=1}_{k^*=2}$. 
 
 Moreover, a number of additional extremal SCFTs with extended supersymmetry were constructed recently: $\mathcal N=2$ and $\mathcal N=4$ SCFTs with $m=2$ \cite{M5} ($\mc E^{\cN=2}_{m=2}(G)$ and $\mc E^{\cN=4}_{m=2}(G)$), SCFTs with $c=12$ and $SW(3/2,2)$ superconformal symmetry \cite{exceptional} ($\mc E^{\rm Spin(7)}(G)$), an $m=4$, $\mathcal N=2$ SCFT with $M_{23}$ symmetry \cite{extN2} ($\mc E^{\cN=2}_{m=4}$), and an $m=4$, $\mathcal N=4$ SCFT with $M_{11}$ symmetry \cite{extN4} ($\mc E^{\cN=4}_{m=4}$). Because there exist multiple extremal SCFTs with central charge 12 and extended superconformal symmetry, we distinguish them by specifying their global symmetry group $G$. 
 We will describe these theories in much greater detail in \S\ref{sec:c=12} and \S\ref{sec:c=24}.
 
 Finally, we would like to point out that the K3 non-linear sigma model is also an extremal $\mc N=4$ CFT with $m=1$ ($\mc E^{\rm K3}(G)$) according to the definition of \cite{GGKMO}. However, unlike the other known examples of ECFTs, it is not chiral. Interestingly, this theory also has a connection with sporadic groups, beginning with the connection between the character decomposition of its elliptic genus and the Mathieu group $M_{24}$ first observed in \cite{EOT}. Symmetry groups of K3 non-linear sigma models have since been classified \cite{Gaberdiel:2011fg} and are  in one-to-one correspondence with subgroups $G \subset Co_0$ such that $G$ preserves a 4-plane in the non-trivial 24-dimensional irreducible representation of $Co_0$, denoted as ${\bf 24}$. In Table \ref{tbl:ECFT} we present the list of known extremal CFTs, including their central charges, chiral algebras, and global symmetry groups.
 \begin{center}\begin{table}[htb]\begin{small}
  \begin{center}
  \def\arraystretch{1.2}
  \begin{tabular}{c|c|c|c}\toprule
 ECFT \,&$\quad c \quad$ & $\mathcal A$& Symmetry Group ($G$)\\\midrule
 $\mc V^\natural$ &  \multirow{4}{*}{24}  & Virasoro & $\mathbb M$\\
  $\mc E^{\mc N=1}_{k^*=2}$ &  &  $\mathcal N=1$&$Co_0$\\
    $\mc E^{\mc N=2}_{m=4}$ &  &  $\mathcal N=2$&$M_{23}$\\
 $\mc E^{\mc N=4}_{m=4}$  &  &  $\mathcal N=4$&$M_{11}$\\\midrule
  $\mc V^{s\natural}$& \multirow{4}{*}{12}  & $\cN=1$ & $Co_0$\\ 
 $\mc E^{\rm Spin(7)}$ & & $\mathcal{SW}(3/2,2)$&$\{G\subset Co_0| G \text{ fixes a 1-plane}\}$ \\
 $\mc E^{\mc N=2}_{m=2}$ &  &  $\mathcal N=2$&$\{G\subset Co_0| G \text{ fixes a 2-plane}\}$\\
 $\mc E^{\mc N=4}_{m=2}$  &  &  $\mathcal N=4$&$\{G\subset Co_0| G \text{ fixes a 3-plane}\}$\\\midrule
   $\mc E^{\rm K3}$& 6 & $\cN=4$ & $\{G\subset Co_0| G \text{ fixes a 4-plane}\}$\\
\bottomrule
  \end{tabular}\caption{\small Known extremal CFTs with central charge $c$, chiral algebra $\mc A$, and global symmetry group $G$. An $n$-plane corresponds to an $n$-dimensional subspace in the representation {\bf 24} of $Co_0$.}\label{tbl:ECFT} 
  \end{center}\end{small}
 \end{table}\end{center}
  Besides the potential connection to quantum gravity in AdS$_3$, another motivation for studying ECFTs stems from the appearance of sporadic groups as symmetry groups.  One of the most impressive mathematical results of the 20th century was the classification of finite simple groups. The result is that there are 18 infinite families of simple groups as well as the 26 so-called sporadic simple groups, which do not arise as part of any infinite family. Though they are known to exist, there is not yet a deep understanding of the role of sporadic groups in physics. Other places where these particular finite groups and their representation theory naturally arise are in connection to automorphism groups of error-correcting codes, unimodular lattices \cite{ConwaySloane} and as coefficients of automorphic forms (``moonshine"; e.g. \cite{CN,Duncan,EOT,UM,UMNL}, and \cite{Duncan:2014vfa} for a recent review). Furthermore, all known examples of extremal CFTs have some large finite automorphism group which is either a sporadic simple group or very closely related to one. So studying such theories may give us hints as to the underlying role of sporadic groups within physics.
  
For a 2d chiral conformal field theory with Hilbert space $\mc H$ and discrete symmetry group $G$, it is interesting to consider the so-called twined partition function, defined as
\be\label{eq:PFg}
\phi_g(\t):=\Tr_{\mc H} \,g q^{L_0-c/24}, \,\,\qquad \forall \,g\in G
\ee
where\footnote{This short-hand notation will be valid throughout the text.} $q=e(\t)=e^{2\pi i \t}$. This is a class function, as it only depends on the conjugacy class $[g]$ of the element $g$, and reduces to the usual partition function of the theory when $g$ is the identity element of the group. Furthermore, the functions $\phi_g$ are highly constrained as they must transform under the subgroup of the modular group $\G$ which preserves the corresponding $g$-twisted boundary condition on the torus, as we review in \S\ref{sec:orb}. 
Thus they naturally provide a link between the representation theory of $G$ and a distinguished set of modular forms.
  
  The best-studied ECFT is the FLM monster module, $\mc V^\natural$. As we will review in \S \ref{sec:monster}, it enjoys a number of striking properties, including the fact that its twining functions as defined in (\ref{eq:PFg}) (and known as ``McKay-Thompson series") furnish Hauptmoduln for genus zero groups. This is the famous ``genus zero" property of monstrous moonshine \cite{CN}, which was 
  shown in \cite{Duncan:2009sq} to be equivalent to a particular feature of their Rademacher sums: each of these functions can be expressed as a Rademacher sum with only a simple pole at the infinite cusp.
  That is to say, one can represent these functions as a sum over representatives of $\Gamma_\infty \backslash \Gamma$ about the pole $\left (q^{-1}\right )$, where $\Gamma_\infty$ is the subgroup of $\Gamma$ that fixes the $i\infty$-cusp. A similar property is crucial in the formulation of umbral moonshine \cite{ChengDunc,UM,UMNL}, where again the polar structure at the infinite cusp is sufficient to recover almost all the functions. Thus a natural question is: can the twining functions of the other examples of ECFTs be expressed as Rademacher sums at the infinite cusp? 
  
  This question is particularly compelling given the proposed connection between the Rademacher sum and the path integral of quantum gravity in AdS$_3$, beginning with         \cite{Farey}. Via the AdS$_3$/CFT$_2$ correspondence, one associates the partition function of the 2d CFT on a torus with the Euclidean quantum gravity path integral in three dimensions with asymptotically AdS boundary conditions. The bulk path integral is evaluated on a solid torus whose boundary is the torus of the 2d CFT; its semi-classical saddle points correspond to representatives of equivalence classes of contractible cycles of the solid torus and are thus labeled by elements of the coset $\Gamma_\infty \backslash \Gamma$  for $\Gamma=SL_2(\mathbb Z)$ and $\Gamma_\infty$ the subgroup which stabilizes the contractible cycle. The sum over saddle points precisely appears in the Rademacher expansion of the CFT partition function, as noted above in the case of the monster CFT, suggesting a physical interpretation of this expression via holography. An explicit connection between the monster CFT and a family of 3d chiral gravities \cite{Li:2008dq} was proposed in \cite{Manschot, Duncan:2009sq}. One caveat to a holographic interpretation of Rademacher sums appearing in monstrous moonshine, however, is that the AdS radius in three dimensions is proportional to the central charge of the CFT. Thus only for very large $c$ does one have reason to trust the semi-classical bulk path integral, which is decidedly not the case for the monster CFT, which has $c=24$. Nevertheless, it is striking that such an interpretation seems to remain valid in this context.
      
  In this work we propose to investigate the extent to which the other known cases of extremal CFTs have similar Rademacher summability properties. We study the $\cN=1$ ECFT with Conway symmetry, and a number of ECFTs with extended superconformal algebras. In the former case, as proven in \cite{DM-C}, all of the McKay-Thompson series of the theory can be formulated as Rademacher sums at the infinite cusp. In the latter case, we consider graded representations of $G$-modules arising from these theories which are encapsulated in vector-valued mock modular forms whose pole structures have not been studied in detail as of yet. This generalization of the usual twined partition functions defined in (\ref{eq:PFg}) is motivated by the decomposition of the partition function into characters of the relevant superconformal algebra. In particular, we answer the question: is it possible to reconstruct the twining functions of these ECFTs implementing a Rademacher sum at the infinite cusp?
 
Our results, summarized in Table \ref{tbl:ECFTresults} in section \S \ref{sec:disc}, are as follows. We find that the $\mc N=1$ ECFT with Conway symmetry satisfies very similar properties to that of the monster module: all of its twined partition functions can be written as Rademacher sums at the infinite cusp for a subgroup $\G_g$ of $SL_2(\mathbb R)$. This arises from the fact observed in \cite{DM-C} that these functions are all normalized Hauptmoduln for genus zero groups. On the other hand, when we consider the known extremal theories with extended supersymmetry at central charge 12 and 24, we find that very few of them satisfy such a Rademacher summability property. With the exception of certain $c=12$, $\mc N=4$ ECFTs with symmetry groups $M_{22}$ and $M_{11}$, all other ECFTs we investigate have at least one conjugacy class whose corresponding graded character cannot be written as a Rademacher sum at the infinite cusp. These results suggest that the connection between sporadic symmetry groups, mock modular forms, and 2d CFTs does not hinge on the strict Rademacher summability properties at the infinite cusp present in most cases of moonshine.
 
The outline of the rest of the paper is as follows. In \S \ref{sec:bkgd}, we discuss aspects of Rademacher sums and holomorphic orbifold CFTs relevant for our subsequent discussion of ECFTs. In \S \ref{sec:monster}, we review the construction of the monster CFT, the genus zero property, and its connection with the Rademacher sum. In \S \ref{sec:c=12} and \ref{sec:c=24} we review the other known ECFTs, in central charge 12 and 24 respectively. We present our results in \S \ref{sec:results} on the Rademacher summability of the twined partition functions of these other ECFTs. Finally, we conclude with a summary and discussion of open questions in \S \ref{sec:disc}. A number of appendices contain additional details which complement the main text.

\section{Mathematical background}\label{sec:bkgd}
In this section we briefly describe some mathematical background relevant to the properties of extremal CFTs we will discuss. We start with an introduction to the Rademacher sum, and continue with a short review of (holomorphic) orbifolds. As described in the next section, the Rademacher sum is a powerful tool which allows to completely reconstruct a mock modular form once its modular transformations and $q$-polar terms at the different cusps of the modular group are known.

\subsection{Rademacher sum}\label{sec:rad}
Consider $\Gamma$ to be a subgroup of $SL_2(\mathbb{R})$ commensurable\footnote{The group $\G_1$ is said to be commensurable with $\G_2$ when the index of $\G_1\cup \G_2$ in $\G_1$ and $\G_2$ is finite.} 
with $SL_2(\mathbb{Z})$ and containing $-\mathbb{I}$. The action of a generic element $\gamma \in \Gamma$ on the upper half-plane is given by $\gamma\tau = \frac{a\tau+b}{c\tau+d}$, where $\gamma=\begin{psmallmatrix} a&b\\c&d\end{psmallmatrix},\,  \, \tau \in \mathbb{H}$. We denote by $h\in\mathbb{Z}_{>0}$ the width of $\Gamma$ at infinity, that is to say the minimal positive integer such that $T^h=\left(\begin{smallmatrix} 1 & h\\ 0&1 \end{smallmatrix}\right) \in \Gamma$. Furthermore, a cusp of $\Gamma$ is defined as a point in $\mathbb{Q} \cup \{i \infty\}$ fixed by an element of the modular group $\G$.\footnote{Throughout the paper, when we refer to a cusp at $\tau=\zeta$ we mean all cusps equivalent to $\zeta$ under the action of $\Gamma$.} The subgroup of $\Gamma$ fixing the infinite cusp is then generated by $\Gamma_{\infty}=\langle T^h, -\mathbb{I} \rangle$. 

Given a modular group $\G$, a {\it modular function} is a complex-valued function defined on the quotient space $\G\backslash \mathbb H$. A generalization of this concept is provided by {\it modular forms}. 
\begin{defn}
A vector-valued\footnote{Here a vector is represented by an underlined greek letter and the dot stands for matrix multiplication. In the rest of the text we do not use an explicit vector notation, to avoid cluttered notation, but the nature of the object will be clear from the context.} modular form of weight $w$ and multiplier system $\rho$ with respect to the modular group $\Gamma$ is a {map} $\underline{\varphi} : \mathbb{H}\rightarrow \mathbb{C}^{d}$ which obeys the functional equation
\be
\label{eqn:modular}
\underline{\varphi}(\g \t)=j_w (\gamma,\tau)\rho(\gamma).\,\underline{\varphi}(\tau), \qquad \, \forall \,\gamma=\begin{psmallmatrix}a&b\\ c&d \end{psmallmatrix} \in \Gamma,\quad \tau\in\mathbb{H}\,.
\ee
\end{defn}
\noindent
Here $j_w (\gamma,\tau)$ denotes the automorphic factor $(c\tau+d)^w$, which, together with the multiplier system $\rho : \Gamma \rightarrow SU(d)$, satisfies the consistency condition 
\[j_w (\alpha\beta,\tau)\rho(\alpha\beta)= j_w (\alpha,\beta\tau)j_w (\beta,\tau) \rho(\alpha) .\rho(\beta).\]
We write the entries of the unitary diagonal matrix $\rho(T^h)$ as $e(\mu_i)$, where $i= 1, ..., d$, $0\le \mu_i <1$. The Fourier expansion around the infinite cusp of a vector-valued modular form takes the form
\be
\underline{\varphi}(\t)=\begin{pmatrix} q^{(\mu_1-n_1)/h}(a_1+b_1 q+ ...)\\
q^{(\mu_2-n_2)/h}(a_2+b_2 q+ ...)\\
...\\
q^{(\mu_d-n_d)/h}(a_d+b_d q+ ...)
\end{pmatrix}
\ee
If $a_i=0$ for $i=1, ... d$, the modular form is bounded at infinity and it is called a cusp form. 

Lastly, we introduce one of the central objects in our subsequent discussion: {\it vector-valued mock modular forms}. Although mock modular forms were first considered by Ramanujan at the beginning of 19th century \cite{Raman}, it was not until the work of Zwegers \cite{Zwegers2008} that a complete mathematical framework was established. \begin{defn}
A vector-valued mock modular form of weight $w$ and multiplier system $\rho$ with respect to $\Gamma$ is a holomorphic  vector-valued {function} $\underline{\varphi}(\tau)$ with at most exponential growth at the infinite cusp and such that there exists a non-holomorphic function 
\be
\label{eqn:complet}
\widehat{\underline{\varphi}}(\tau)= \underline{\varphi}(\tau) + \underline{g}\,^*(\tau), 
\ee
called the completion of $\varphi$, which transforms as a modular form of weight $w$ and multiplier system $\rho$ with respect to $\Gamma$. 
\end{defn}

\noindent
The {completion} $\widehat{\underline{\varphi}}(\tau)$ is related to the mock modular form by the addition of the (non-holomorphic) Eichler integral of the so-called {shadow}, $\underline{g}(\tau)$, 
\be
\label{eq:Eichler}
\underline{g}\,^*(\tau):= \Bigl(\frac{i}{2\pi}\Bigr)^{w-1}\int_{-\bar{\tau}}^{\,i\infty}\,(z+\tau)^{-w}\,\overline{\underline{g}(-\bar{z})}\,\mathrm{d}z \, ,
\ee
where $\underline{g}(\tau)$ is a cusp form of weight $2-w$ and multiplier system conjugate to the one of $\underline{\varphi}(\tau)$. Even though more general definitions are allowed, see for instance \cite{DMZ}, we restrict to the case where $\underline{g}(\tau)$ is a cusp form, and in particular a unary theta series as defined in \eqref{eq:app-shadow}.
Clearly, a modular form is simply a special case of a mock modular form with vanishing shadow.

Once the polar $q$-terms at the different cusps and the modular properties (i.e. the data appearing in equations \eqref{eqn:modular} and/or \eqref{eqn:complet}) are known, a modular object can be completely reconstructed through the so-called {\it Rademacher sum}. 
The origin of the Rademacher sum can be traced back to the Poincar\'e sum 
\be 
\label{eqn:Poincsum}
{\mathcal{P}}^{(n)}_{\Gamma, w, \rho}(\tau):= \sum\limits_{\gamma \in \Gamma_{\infty}\backslash \Gamma} j_w (\gamma,\tau)^{-1}\rho(\gamma)^{-1} e({n} \, \gamma \tau).
\ee
This expression encodes the simple idea that the function $e({n}\tau)$ can be made invariant under $\Gamma$ by averaging over the images of the $\Gamma$-action. The sum is well-defined as a sum over elements of the right-coset $\Gamma_{\infty}\backslash \Gamma$ so long as the addends are invariant under the action of $\Gamma_{\infty}$. This holds {for $(n h-\mu) \in \mathbb{Z}$, where $h$ and $\mu$ are defined as above}.
Due to the absolute convergence of the sum in \eqref{eqn:Poincsum}, for particular weights and multiplier systems, the above expression can easily be shown to transform under the action of $\Gamma$ as a modular form of weight $w$ and multiplier system $\rho$. 

The analysis of Poincar\'e \cite{Poinc} was restricted to modular forms of even weight greater than two and trivial multiplier system with respect to the full modular group, $SL_2(\mathbb{Z})$.\footnote{Later Petersson \cite{Peters1,Peters2} generalized this discussion to different groups and multipliers.}
However, the absolute convergence of the series, which holds for $w>2$, is lost for smaller weights. Already at $w=2$ the sum requires a regularization procedure to be conditionally convergent. It was not until the studies of Rademacher  \cite{Rad,Radem,Radem2} and Rademacher, Zuckermann in \cite{RadZuc} that a compact formula, later defined as Rademacher sum, appeared for smaller weights. In particular, for $w=0$ Rademacher obtained a regularized expression which encodes the Fourier coefficients of the $J$-function 
\be
\label{eqn:Jfct}
J(\tau) +12= e(-\tau) +\text{lim}_{K\rightarrow \infty} \sum\limits_{\substack {{\gamma} \in \Gamma_{\infty}\backslash \Gamma^{*}_{K,K^2}}} \hspace{-1mm} \Bigl(e(-\gamma \tau)- e(-\gamma \infty) \Bigr)\, .
\ee
Here $J(\t)$ is the unique modular function with respect to $SL_2(\mathbb Z)$ with expansion 
  \be\nonumber
  q^{-1} + O(q) \qquad \text{as} \,\, \tau \to i \infty\,, 
  \ee
that is
  \be
  \label{eqn:Jexpans}
  J(\t) = q^{-1} + 196884 q + \ldots.
  \ee
The sum in \eqref{eqn:Jfct} is taken over representatives of the right coset of $\Gamma^{*}_{K,K^2}= \{\begin{psmallmatrix}a&b\\ c&d \end{psmallmatrix}\in \Gamma \, |\,  0< c <K, -K^2 <d <K^2\}$ by $\Gamma_{\infty}$.
Due to the conditional convergence of the series, the sum has to be taken in a particular order: specifically the addends are chosen with increasing $c$. 
This form was later generalized by Niebur in \cite{Nieb} for $w \le 0$
\be
 \label{eqn:rad}
\mathcal{R}^{(n)}_{\Gamma, w, \rho}(\tau):= \Delta+  \hspace{-1mm}\sum\limits_{\gamma \in \Gamma_{\infty}\backslash \Gamma_{K,K^2}}  \hspace{-1mm}\mathfrak{R}^n_w(\gamma, \tau)\,  j_w (\gamma,\tau)^{-1}\rho(\gamma)^{-1}e(n \, \gamma \tau)\, .
\ee
In contrast to \eqref{eqn:Jfct}, here the sum over coset representatives includes a term with vanishing $c$ and a constant $\Delta$, which vanishes for $\mu\neq0$ and it is otherwise defined in \eqref{eq:delta}. 
Lastly, the regularization factor is $$\mathfrak{R}^n_w(\gamma, \tau)= \frac{\bar{\gamma}(1-w, 2\pi i n (\gamma \tau - \gamma \infty))}{\Gamma(1-w)},$$ where $\bar{\gamma}$ denotes the lower incomplete gamma function. Specializing this compact formula to the case with $w=0$, $n=-1$, and trivial multiplier system we recover the Rademacher expression for the $J$-function {up to the constant $\Delta$. The addition of a constant in this case does not modify the modular properties of the function at hand; however for general weights it is a necessary ingredient to simplify the modular transformation of the object.\footnote{A detailed analysis on the role of the constant term is presented in \cite{Duncan:2009sq}.}}

Niebur proved that the Rademacher construction defined by the above regularization gives rise to a conditionally convergent series, that he referred to as {\it automorphic integral}. The latter is defined as a holomorphic map $\varphi:\mathbb{H}\rightarrow \mathbb{C}$ satisfying
\be
\varphi(\gamma \tau)=(c\tau+d)^w \rho(\gamma)\Bigl( \varphi(\tau) -p(w,\gamma^{-1},g)\Bigr)
\ee
where $$p(w,\gamma^{-1},g):= \frac{1}{\Gamma(1-w)}\int_{-\bar{\tau}}^{i\infty} (z+\tau)^{-w}\overline{g(-\bar{z})}dz\, ,$$ 
and $g$ is a cusp form of weight ${(2-w)}$ and conjugate multiplier system, $\bar{\rho}(\gamma)$. 
The regularization procedure was thus proven to lead to what is now known as a mock modular form. Consequently, if the space of cusp forms of dual weight is empty the automorphic integral reduces to a modular form. This happens, for instance, in the case of the $J$-function and more generally for all the McKay-Thompson series arising in monstrous moonshine.

In addition, Niebur showed that the Rademacher sum gives a basis for the vector space of automorphic integrals of negative weight $w$ and multiplier system $\rho$ for a generic modular group $\Gamma$. The technique was further developed in \cite{Knopp1,Knopp2,Fay} to quote just a few, and generalized to weight 1/2 mock modular forms in \cite{Pribit,ChengDunc,BO06,BO08,BO10}.

Although until now we focused on scalar-valued Rademacher sums, the main objects of the next sections are vector-valued Rademacher sums, recently constructed in \cite{KnoMas,Whalen,UM,ProofUM,weightUM}.
Following these results, the definition \eqref{eqn:rad} can readily be generalized to the vector-valued case
\be
 \label{eqn:radem}
\mathcal{R}^{(n_i)}_{\Gamma, w, \rho}(\tau)_j= \Delta_j+  \hspace{-1mm}\sum\limits_{\gamma \in \Gamma_{\infty}\backslash \Gamma_{K,K^2}} \hspace{-1mm} \mathfrak{R}^{n_i}_w(\gamma, \tau)  j_w (\gamma,\tau)^{-1}\rho(\gamma)_{ji}^{-1}e(n_i \gamma \tau) \,.
\ee
This corresponds to the contribution of the $n_i$-th pole at the infinite cusp\footnote{The definition of the Rademacher sum at different cusps of $\Gamma$ can be found in \cite{Duncan:2009sq,Whalen}.} to the $j$-th component of the Rademacher sum. If multiple polar terms are present in the Fourier expansion of the mock modular form then all polar contributions must be taken into account. 
This sum was proved to be convergent for negative weights in \cite{Whalen} and for $w=1/2$ and a particular multiplier system with respect to the modular groups $\G_0(N)$ in \cite{ProofUM,weightUM}. The latter results are the ones that we will mostly use in the following. 

Through the Lipschitz summation formula, the Fourier expansion of $\vec{\mathcal{R}}^{(n_i)}_{\Gamma, w, \rho}(\tau)$ can be recovered from \eqref{eqn:radem},
\be
\label{eqn:radvec}
\hspace{-4mm}\mathcal{R}^{(n_i)}_{\Gamma,w,\rho}(\tau)_j = \delta_{ij}q^{n_i}\, + \, 2 \Delta_j \,+ \hspace{-2mm} \sum\limits_{\substack{k_j>0 \\ hk_j \in \mathbb{Z}+\mu_j}} \hspace{-2mm}q^{k_j} \sum\limits_{c>0} \,S_{n_i,k_j}(c,\rho)_{ji} \, \frac{-2\pi i}{ch}\biggl( -\frac{k_j}{n_i}\biggr)^{\frac{w-1}{2}} \hspace{-1.8mm}J_{1-w}(\frac{4\pi i }{c}\sqrt{-k_j n_i})
\ee
where $J_{s}(x)$ is the $J$-Bessel function and 
\be\label{eq:delta}
\Delta_j = \begin{cases} 
-\frac{(2\pi i)^{2-w} \, (-n_i)^{1-w}}{2h \, \Gamma(2-w)} K_{n_i,0}(1-\frac{w}{2}) & \mu_j=0 \, ,\\
0 &\mu_j\neq 0 \, . \end{cases}
\ee 
The Kloosterman Selberg Zeta function and the Kloosterman sum are defined respectively by 
\begin{align}
&K_{n_i,0}(1-w/2)_{ji}= \sum\limits_{c>0}\, \frac{S_{n_i,0}(c,\rho)_{ji}}{c^{2(1-w/2)}} \, ,\\
&S_{n_i,k_j}(c,\rho)_{ji} = \hspace{-1mm}\sum_{\gamma \in \Gamma_{\infty}\backslash \Gamma / \Gamma_{\infty}}  \hspace{-1mm}e(n\gamma\infty -k_j \gamma^{-1} \infty)\rho(\gamma)^{-1}_{ji} \, .
\end{align}
Equation (\ref{eqn:radvec}) expresses once again the contribution of the $i$-th component, which has a pole of order $n_i$ at the infinite cusp, to the $j$-th component. 

Apart from furnishing an efficient method to reconstruct (mock) modular forms, the Rademacher sum prescription underlies the (re)formulation of monstrous moonshine in \cite{Duncan:2009sq} as well as umbral moonshine \cite{UM,UMNL}. 

\subsection{Holomorphic orbifolds}\label{sec:orb}
In this section we briefly review aspects of holomorphic orbifold CFTs which are relevant to chiral CFTs with a discrete symmetry group. Denote by $\phi(\tau)$ the partition function of a chiral CFT with Hilbert space $\mc H$ and central charge $c$,
\be
\phi(\tau)=\Tr_{\mathcal{H}} (q^{L_0 -c/24})\, ,
\ee
where $L_0$ represents the Virasoro generator. 
The above partition function corresponds to a path integral on a torus with complex structure parameter $\tau$ and periodic boundary conditions along the two cycles.
Given an automorphism group $G$ of the theory, it is possible to define twining functions 
\be
\phi_g(\tau)=\Tr_{\mathcal{H}} (g\, q^{L_0 -c/24}), \qquad \forall g\in G  
\ee
where the $g$-insertion stands for the representation of the element $g$ acting on the Hilbert space of the theory. 
Moreover, one can build the invariant subspace with respect to the action of $g$ by defining a projection operator, $\mathcal{P}$, whose action for an element of order $n$ is 
\be
\Tr_{\mathcal{H}} (\mathcal{P}\, q^{L_0 -c/24})= \frac{1}{n}\sum_{i=0}^{n-1}\Tr_{\mathcal{H}} (g^{i} \, q^{L_0 -c/24}),
\ee
This is the first step in the construction of an orbifold partition function. 

Additionally, one must include states arising from the $g$-twisted sectors, i.e.
\be
\phi_{e,g}(\tau)=\Tr_{\mathcal{H}_g} ( q^{L_0 -c/24}), \qquad \forall \, g\in G \,. 
\ee
The latter are defined as traces over twisted Hilbert spaces, $\mathcal{H}_g$, which consist of states defined modulo a $g$-transformation. Throughout we denote by $e$ the identity element of the group under consideration. 
Analogously, on the torus twisting and twining correspond to changing the boundary conditions along one of the cycles of the torus. Thus, we are led to define a twisted-twined function, whose boundary conditions along the two cycles are dictated by elements of the group $G$.  From a Hamiltonian approach, the twisted-twined function is defined as
\be
\phi_{g,h}(\tau,z)=\Tr_{\mathcal{H}_h} (g \, q^{L_0 -c/24}), \qquad g\in C_G(h), \,\, h \in G\, . 
\ee
Since the action on the spectrum is well defined so long as $g$ and $h$ commute, the twining element $g$ belongs to the centralizer of $h$ in $G$, $C_G(h)= \{g\in G| gh=hg\}\,$. 
In the case of chiral CFTs these functions are class functions up to a phase. In order to obtain a consistent orbifold one has to impose certain constraints which prevent anomalous phases from appearing under modular transformations which fix the boundary conditions. 

Different twisted-twined functions can be related to each other by modular transformations. In fact, $\phi_{g,h}$ satisfies the following functional equation 
\be
\phi_{g,h}(\g \t)= \rho_{g,h}\begin{psmallmatrix}a&b\\ c&d \end{psmallmatrix} \, \phi_{h^b g^d, h^a g^c}(\tau), \qquad \g=\begin{psmallmatrix}a&b\\ c&d \end{psmallmatrix} \in \G_{g,h}\,, 
\ee
defining a modular function with multiplier system $\rho$ with respect to the modular group $\G_{g,h}$ which fixes the pair $(g,h)$.

The complete $\langle h \rangle$-orbifold partition function therefore takes the form
\be
\phi_{orb}(\t)= \frac{1}{|C_G(h)|}\sum_{[h]}\sum_{g\in C_G(h)} \phi_{g,h}(\t),
\ee
where the first sum is over representatives of the conjugacy classes of $h$, and the second sum is over elements commuting with $h$. 

Examples of holomorphic orbifolds are the ones obtained from the monster CFT, coined by Norton as {\it Generalized moonshine}. Before considering their properties in the next section, we generalize the above concepts to superconformal field theories.  

A similar reasoning can be applied to SCFTs with a non-trivial current algebra. Instead of focusing on its partition function, we consider the elliptic genus (EG). The latter is defined for an $\mathcal{N}=2$ SCFT by
\be
\psi(\tau,z)=\Tr_{\mathcal{H}} ((-1)^F q^{L_0 -c/24}\bar{q}^{L_0 -c/24}y^{J_0})
\ee
where $z$ is the $U(1)$-chemical potential and $y=e(z)$.
Once again the modular properties of $\psi(\tau,z)$ and its twisted-twined companion $\psi_{g,h}(\tau,z)$ can be used to define the EG of the orbifolded theory, which this time depends on the two variables $\tau$ and $z$. 
Under a modular transformation $\g \in \G_{g,h}$, $\psi_{g,h}(\tau,z)$ transforms as a weight 0 index $m$ Jacobi form
\be
\psi_{g,h}(\g \t,\g z)\,=\, e\Bigl(\frac{mcz^2}{c\tau+d}\Bigr)\;  \psi_{h^b g^d, h^a g^c}(\tau,z)\,. 
\ee

\section{Properties of the monster CFT}\label{sec:monster}
In this section we review the construction of the monster CFT and discuss some of its defining properties, which include the genus zero property, the Rademacher summability of its twining functions and the connection between these properties and holomorphic orbifolds of the theory.

  Given a positive-definite even unimodular lattice $\Lambda$ of rank $24k$ one can construct a bosonic chiral conformal field theory with modular invariant partition function by compactifying the theory of $24k$ chiral bosons on the torus $\mathbb R^{24k}/\Lambda$. In the case of $k=1$, there are 24 such lattices: the Leech lattice, $\Lambda_L$, which has no roots, and the 23 Niemeier lattices, $\Lambda_N$, which can be uniquely specified by their root systems. These are a union of simply-laced root systems with the same Coxeter number and of total rank 24. We will call the 23 chiral bosonic CFTs on $\mathbb R^{24}/\Lambda_N$ the {\it Niemeier CFTs}, and the theory on $\mathbb R^{24}/\Lambda_L$ the {\it Leech CFT}. We label their associated modules as $\mathcal{V}^N$ and $\mathcal{V}^L$, respectively. 
  
  The partition function of each of these theories is simply given by 
  \be\label{eq:NiemZ}
  \mc Z^{\Lambda_N}(\t):=\Tr_{\mathcal{V}^N}\, q^{L_0-c/24}= {\Theta_{\Lambda_N}(\t)\over \eta^{24}(\t)} = J(\t)+24(h+1),
  \ee
  where $\Theta_{\Lambda_N}$ is the lattice theta function and $h$ is the Coxeter number associated to $\Lambda_N$. The constant comprises the contribution of the length-squared two vectors (roots) and the level-one bosonic states. In the case of the Leech lattice, we define $h=0$ for $\Lambda_L$ so that the partition function of the Leech CFT is simply
  \be
   \mc Z^{\Lambda_L}(\t):=\Tr_{\mathcal{V}^L}\,q^{L_0-c/24}= {\Theta_{\Lambda_L}(\t)\over \eta^{24}(\t)} = J(\t)+24.
  \ee
The monster CFT \cite{FLM} is constructed from a $\mathbb Z_2$ orbifold of the Leech CFT. The $\mathbb Z_2$ acts on the 24 coordinates as 
  \be\nonumber
  h: x_i \mapsto -x_i, ~~\forall i=1,\ldots, 24,\ee and the Hilbert space $\mathcal H$ of the Leech CFT splits into two Hilbert spaces $\mathcal H_\pm$ consisting of states which are either invariant or anti-invariant under the orbifold action:
  \be\label{eq:Horb}
  \mathcal H_\pm := \{\psi \in \mathcal H |h\psi = \pm \psi\}.
  \ee
  Furthermore, there is a twisted sector Hilbert space $\mathcal H^{tw}$ arising from the fixed points of the orbifold action; this is once again the direct sum of two Hilbert spaces comprised of twisted sector states which are invariant or anti-invariant under the orbifold action:
    \be\label{eq:Htwistorb}
  \mathcal H^{tw}_\pm := \{\psi^{tw} \in \mathcal H^{tw} |h\psi^{tw} = \pm \psi^{tw}\}.
  \ee
  The resulting Hilbert space of $\mathcal{V}^\natural$ is 
  \be
  \mc H_{\mathcal{V}^\natural}:= \mc H_+ \oplus \mc H^{tw}_+.
  \ee
  The partition function of the theory is given by 
  \be
  \mc Z_{\mathcal{V}^\natural}(\t)= \Tr_{\mathcal{V}^\natural} \,q^{L_0-c/24}= J(\t).
  \ee
Following \cite{Witten}, this is the partition function of a bosonic ECFT with smallest possible central charge. 

The action of the monster group on $\mathcal{V}^\natural$ allows one to define, for each conjugacy class $g \in \mathbb M$, the so-called ``McKay-Thompson" series $T_g(\tau)$, by
  \be\label{eq:MTseries}
  T_g(\tau) = \Tr_{\mathcal{V}^\natural} \, g q^{L_0-c/24}\,,
  \ee
which is a {modular function} for $\Gamma_g$, an Atkin-Lehner type subgroup of $SL_2(\mathbb R)$. See appendix \ref{app:modgrs} for the precise definition of $\Gamma_g$. 
Therefore, it follows that one can interpret $\mc V^{\natural}$ as an infinite-dimensional $\mathbb Z$-graded $\mathbb M$-module, $$\mc V^{\natural}= \bigoplus_{n=-1}^\infty \mc V^\natural_n$$ whose graded trace reproduces the McKay-Thompson series via $$T_g(\tau)= \sum_{n=-1}^\infty (\Tr_{\mc V^\natural_n}g) q^n$$ and where one gets $J(\tau)$ by taking $g=e$, the identity element of $\mathbb M$.
Moreover, each $T_g(\tau)$ is a Hauptmodul for $\Gamma_g$; i.e. it defines an isomorphism between the compactified fundamental domain\footnote{The compactified fundamental domain is constructing from the fundamental domain ${\mathcal{F}} = \mathbb{H}/\Gamma_g$ adding the cusps of $\G_g$.} and the Riemann sphere with finitely many points removed as
\be
\label{eqn:iso}
T_g :\overline{\mathcal{F}}\rightarrow \mathbb{C}\cup \{i \infty\},
\ee
such that any meromorphic $\Gamma_g$-invariant function can be expressed as a rational function of $T_g(\tau)$. Due to the isomorphism in \eqref{eqn:iso}, $\Gamma_g$ is called a {\it genus zero group}.
This distinguishing  feature of the McKay-Thompson series (genus zero property) was first conjecture by Conway and Norton in \cite{ConwayNort}, later confirmed via an explicit construction of the module \cite{FLM} and finally proved in \cite{Borch}.

The genus zero property reflects the pole structure of $T_g(\t)$ in the following way. For any element $g\in \mathbb M$, $T_g(\t)$ has a unique simple pole at the infinite cusp and is bounded at all the other cusps of $\G_g$, or in other words has exponential growth in $\overline{\HH} := \HH \cup \QQ \cup i\infty$ only at the images of $ i\infty$ under $\Gamma_g$. These extends the defining property of the $J$-function \eqref{eqn:Jexpans} to non-trivial conjugacy classes of the monster group. 

There are many more beautiful properties which distinguish the monster CFT from, say, the Leech and Niemeier CFTs or other bosonic chiral CFTs with central charge 24. We focus specifically on the following results, which elucidate the connection between the genus zero property of the McKay-Thompson series, their Rademacher summability, and the nature of $g$-orbifolds of $\mc V^\natural$.

Firstly, in \cite{Duncan:2009sq}, Duncan and Frenkel showed that the Hauptmodul property could be rephrased in terms of the Rademacher summability of $T_g$ around the infinite cusp as long as the modular group has width one at the infinite cusp.
\begin{thm}[Duncan-Frenkel] \label{thm:DunFren} 
For all $g\in G$, there is a (Atkin-Lehner type) $\Gamma_g < SL_2(\RR)$ and a multiplier system $\epsilon_g: \Gamma_g \to \CC^\times$ such that 
  \be
  T_g(\tau) = \mathcal{R}^{(-1)}_{\Gamma_g,0,\epsilon_g}(\tau) - 2\Delta(g)\,,
  \ee
  is the normalized Hauptmodul for $\Gamma_g$.
\end{thm} \noindent
In the above, the constant $\Delta(g)$ depends on the conjugacy class of $g$ and is given by the formula in equation (\ref{eq:delta}).
In addition, they proved that $\G_g$ has genus zero if and only if the Rademacher sum $\mathcal{R}^{(-1)}_{\Gamma_g,0,\epsilon_g}(\tau)$ is a function invariant under $\G_g$.  Therefore, for each $\G_g$ the associated Rademacher sum reduces to a modular function, specifically the Hauptmodul.\footnote{A Hauptmodul is said to be normalized when no constant term appears in its Fourier expansion.} As discussed in \S \ref{sec:rad}, this must be due to the absence of a cusp form for dual weight (weight two) and conjugate multiplier system. In fact, the space of cusp forms of weight two is isomorphic to the space of holomorphic differentials on $\bar{\mathcal{F}}$, which is empty when $\bar{\mathcal{F}}$ is a Riemann sphere. 

Secondly, besides constraining the Fourier expansion of $T_g(\t)$, the genus zero property was shown to correspond to a condition on the vacuum structure of sectors twisted by elements of the monster group, \cite{tuite1992}. 
This is a consequence of the modular properties of a twisted-twined function, which connect the expansion of a twining function at a particular cusp to the ground state energy of a twisted sector. 
That is to say, depending on the group $\G_g$, the $g$-twisted sector is related to the $g$-twined function by either an element of $\G_g$ or an element belonging to the normalizer group of $\Gamma_0 (N)$ in $SL_2(\RR)$, defined in appendix \ref{app:modgrs}. In \cite{tuite1992} it was proved that the former case corresponds to a twisted sector with one negative energy state, while in the latter case no negative energy state appears. 

Therefore, for all $g\in \mathbb M$ the $g$-twisted sector of the {$\langle g \rangle$-orbifold }theory is either completely determined by the untwisted sector and the cusp corresponding to the $g$-twisted sector is equivalent to the infinite cusp (under the action of $\G_g$) or the $g$-twisted sector spectrum has no negative energy states and the cusp corresponding to the $g$-twisted sector is inequivalent to the cusp at $\infty$. This condition together with a closure condition (c.f. appendix \ref{app:modgrs}) which relates the different cusps is sufficient to ensure that $T_g$ is a Hauptmodul for $\G_g$. 
Again this property is directly encoded in the Rademacher expression for $T_g$. 

Finally, another result by Tuite \cite{Tuite1995} relates the orbifold partition function for several conjugacy classes in $\mathbb M$ to the (conjectured) uniqueness of the module. 
\begin{thm}[Tuite]  Assuming the uniqueness of $\mathcal{V}^{\natural}$, then the genus zero property holds if and only if orbifolding $\mathcal{V}^{\natural}$ with respect to a monster element reproduces the monster module itself or the Leech theory.
\end{thm}
\noindent
 We would like to understand the extent to which (suitable generalizations of) the above mentioned properties hold in other cases of extremal CFTs. Specifically, one can ask if there is an analogue of Theorem 3.1 for the twining functions of other extremal CFTs. We investigate this question in \S \ref{sec:results} and comment on a possible extension of Theorem 3.2 in \S \ref{sec:disc}.

  \section{Central charge $12$}\label{sec:c=12}
  In this section we discuss a family of extremal superconformal field theories with central charge 12. Each of the theories discussed in this section arises from the same underlying chiral SCFT whose Neveu-Schwarz (NS) and Ramond (R) sectors are vertex operator algebras which, following \cite{DM-C}, we will refer to as $\mathcal{V}^{s\natural}$ and $\mathcal{V}^{s\natural}_{tw}$, respectively. As we will see, $\mathcal{V}^{s\natural}$ is in many senses the supersymmetric analogue of the monster CFT, $\mathcal{V}^{\natural}$. In \S \ref{sec:Vsnat} we review the construction of $\mathcal{V}^{s\natural}$. In \S \ref{sec:morec=12} we describe a number of extremal SCFTs which arise upon viewing $\mathcal{V}^{s\natural}$ as a module for a $c=12$ superconformal algebra (SCA) with extended supersymmetry.
  We discuss the symmetry groups of each of these theories in \S \ref{sec:symms12}, as well as 
 the mock modular forms whose coefficients encode the graded character of the corresponding $G$-module for each of these extremal CFTs. The material reviewed in this section primarily arises from the papers \cite{Duncan,DM-C,M5,spin7,exceptional}.

 \subsection{The Conway ECFT}\label{sec:Vsnat}
 Before defining $\mathcal{V}^{s\natural}$, we begin by {describing} a closely related theory, $\mathcal{V}^{sE_8}$, which we call the {\it super--$E_8$ CFT}. 
 The latter is the $\mathcal N=1$ SCFT obtained by compactifying eight chiral bosons on the {eight-dimensional} torus $\mathbb R^8/\Lambda_{E_8}$ with their eight chiral fermionic superpartners, where $\Lambda_{E_8}$ is the $E_8$ root lattice. The theory has {two sectors} corresponding to whether the fermions have 1/2-integer (NS) or integer (R) grading along the spatial direction. From this description the partition functions are easily determined to be
 \be\label{eq:E8NS}
 \mc Z^{sE_8}_{\rm NS}(\t)=\Tr_{\rm NS} \,q^{L_0-c/24}= {E_4(\t)\theta_3(\t,0)^4\over \eta^{12}(\t)}= {1\over \sqrt q} + 8 + 276 q^{1/2} + 2048 q + \ldots
 \ee
 in the NS sector, and
   \be\label{eq:E8R}
\mc Z^{sE_8}_{\rm R}(\t)=\Tr_{\rm R} \,q^{L_0-c/24}= {E_4(\t)\theta_2(\t,0)^4\over \eta^{12}(\t)}= 16 + 4096 q+ 98304 q^2 + \ldots
 \ee
 in the R sector. The functions (\ref{eq:E8NS}) and (\ref{eq:E8R}) are invariant under the modular groups $\Gamma_\theta$ and $\Gamma_0(2)$, respectively, and they are part of a vector-valued representation of $SL_2(\mathbb Z)$.\footnote{See appendix \ref{app:modgrs} for the definitions of the relevant modular groups.}
 
 The super--$E_8$ CFT is not extremal because of the eight fermions of dimension 1/2 in the NS sector. However, by taking a $\mathbb Z_2$ orbifold of the theory, one can remove these states and construct an extremal $\mathcal N=1$ theory, $\mc V^{s\natural}$. This is analogous to the $\mathbb Z_2$ orbifold which removes the 24 dimension one currents of the Leech CFT in the construction of $\mathcal{V}^\natural$.
In fact, $\mathcal{V}^{s\natural}$ has two distinct but equivalent constructions:
  \begin{description}
  \item[(A)] A $\mathbb Z_2$ orbifold of the theory on the eight-torus $\mathbb R^8/\Lambda_{E_8}$ which acts as $X_i \to -X_i$ on the eight chiral bosons and as $\psi_i \to -\psi_i$ on their eight fermionic superpartners.
  \item[(B)] A $\mathbb Z_2$ orbifold of 24 free chiral fermions, $\lambda_\alpha$, which acts as $\lambda_\alpha\to -\lambda_\alpha$.
  \end{description}
  Construction (A) was first discussed in \cite{FLM}. Construction (B) was first discussed in \cite{Duncan}, where the two constructions were shown to be equivalent as vertex operator superalgebras, and further in \cite{DM-C}, where it was shown that certain graded traces in this theory furnish normalized Hauptmoduln analogous to the McKay-Thompson series in monstrous moonshine. It is apparent that {(A)} is an $\mathcal N=1$ supersymmetric extension of the $E_8$ current algebra. Furthermore, {(B)} enjoys a hidden $\mathcal N=1$ superconformal symmetry as well. In particular, there are $2^{12}=4096$ dimension--${3\over 2}$ twist fields arising from zero modes of the $\lambda_i$ acting on the twisted sector ground state. In \cite{Duncan} it is shown that there exists a linear combination of these twist fields that satisfies the OPEs for a supercurrent of an $\mathcal N=1$ SCA with central charge 12. Moreover, the subgroup of Spin(24) which preserves this choice of supercurrent is the discrete subgroup $Co_0$ \cite{DM-C}.
  
The partition function of $\mathcal{V}^{s\natural}$ can be computed using either construction. {In the NS sector, the result is}
\bea\nonumber
\mc Z^{s\natural}_{\rm NS}(\t)=\Tr_{\rm NS}\, q^{L_0-c/24}&=&{1\over 2}\left ( {E_4(\t)\theta_3(\t,0)^4\over \eta^{12}(\t)}+16{\theta_4^4(\t,0)\over\theta_2^4(\t,0)}+16{\theta_2^4(\t,0)\over\theta_4^4(\t,0)}\right)\\\nonumber&=&{1\over 2}\sum_{i=2}^4{\theta_i^{12}(\t,0)\over \eta^{12}(\t)}\\\label{NSsnatural}&=&{1\over \sqrt q} + 276 q^{1/2} + 2048 q + 11202 q^{3/2}+ \ldots\\\nonumber
&=&K(\t)-24,
\eea
where the formula in the first line arises from construction (A) and that in the second from construction (B). In the last line we have introduced an expression in terms of $K(\t)$, a Hauptmodul for the modular subgroup $\Gamma_\theta$ (c.f. appendix \ref{app:modgrs}). The lack of constant term in this partition function indicates that $\mathcal{V}^{s\natural}$ furnishes an example of an extremal $\mathcal N=1$ SCFT with $k^*=1$, according to \cite{Witten}; i.e., there are no primary fields of dimension smaller than or equal to $c/24=1/2$. 
 For more details on the moonshine properties of $\mathcal{V}^{s\natural}$, see the papers \cite{Duncan,DM-C}.

 \subsection{More extremal theories} \label{sec:morec=12}
{Focusing on  construction (B)} of $\mathcal{V}^{s\natural}$, it is straightforward to construct a number of additional extremal SCFTs where the chiral algebra is an extension of the $\mathcal N=1$ superconformal algebra. In \cite{exceptional} theories with an $\mathcal{SW}(3/2,2)$ SCA are discussed, whereas in \cite{M5}, theories with $\mathcal N=2$ and $\mathcal N=4$ SCAs are discussed. In each case the approach is the same: given a choice of supercurrent $\mathcal W$ which generates an $\mathcal N=1$ SCA with $c=12$, one can pick an additional one, two, or three fermions to generate {a chiral} algebra which enhances the $\mathcal N=1$ SCA to an extended version. 
  That each of these theories furnishes an example of an ECFT is straightforward to see from the character decomposition of their (flavored) partition functions. At $c=12$, the extremal constraint forces the states of conformal dimension smaller than $1$ in the NS sector to be superconformal descendants of the identity. 
  We review this for each of these cases in turn. 
  \begin{enumerate}
  \item  If one chooses one of the 24 fermions, say $\lambda_1$, one can generate a chiral $c=1/2$ Ising model. This enhances the $\mathcal  N=1$ SCA to a $c=12$ $S\mc W(3/2,2)$, i.e. the SCA which arises on the worldsheet theory of a non-linear sigma model with target space a manifold of Spin(7) holonomy \cite{ShVafa}. See appendix \ref{app:spin7char} for a summary of the representation theory and characters of this algebra. It follows from the discussion in \cite{exceptional} and the appendix that the partition function of $\mathcal{V}^{s\natural}$ has the following decomposition into Spin(7) characters,
  \be\label{eq:spin7Z}
 \mc Z^{s\natural}_{\rm NS}(\t)=\tilde\chi_0^{\rm NS}(\t) +0\tilde\chi_{1\over 16}^{\rm NS}(\t) +23 \tilde\chi_{1\over 2}^{\rm NS}(\t) + \sum_{n=1}^\infty b_n \chi_{0,n}^{\rm NS}(\t) + \sum_{n=1}^\infty c_n \chi_{{1\over 16},n}^{\rm NS}(\t),
  \ee
  where the constraint of extremality is satisfied by the fact that the coefficient in front of $\chi_{1\over 16}^{\rm NS}(\t)$ is zero. We will denote this theory by $\mc E^{\rm Spin(7)}(G)$, where the group $G$ is the symmetry group of the theory, and depends on the choice of fermion $\lambda_1$.
  \item   If one chooses two of the 24 fermions, one can generate a $\widehat{u(1)}_2$ current algebra which, together with the $\mc N=1$ supercurrent, satisfies the OPEs of a $c=12$ $\mc N=2$ SCA \cite{M5}. The partition function of $\mathcal{V}^{s\natural}$ graded by this additional $U(1)$ is a weak Jacobi form for $SL_2(\mathbb Z)$ of weight zero and index two, which takes the form 
  \be \label{EGoftheModel}
\mc Z^{s\natural}_{\rm R}(\t,z)  = \Tr_{\rm R} (-1)^F q^{L_0-c/24}y^{J_0}=\frac{1}{2}\frac1{\eta^{12}(\t)} \sum_{i=2}^4 (-1)^{i+1} \th_i(\t,2z) \th_i^{11}(\t,0)\, ,
\ee
and admits the following decomposition into $c=12$, $\mathcal N=2$ characters
\begin{align} \notag
\mc Z^{s\natural}_{\rm R}(\t,z) & = 23 \,{\rm ch}^{\cN=2}_{\frac{3}{2};\frac{1}{2},0} + \,{\rm ch}^{\cN=2}_{\frac{3}{2};\frac{1}{2},2}+ \Bigl(770 \,({\rm ch}^{\cN=2}_{\frac{3}{2};\frac{3}{2},1}+{\rm ch}^{\cN=2}_{\frac{3}{2};\frac{3}{2},-1})\, + \\ \label{Nis2decomposition_of_Z}
 &\;\,+13915 \,({\rm ch}^{\cN=2}_{\frac{3}{2};\frac{5}{2},1}+{\rm ch}^{\cN=2}_{\frac{3}{2};\frac{5}{2},-1})+\dots \Bigr) + \Bigl(231 \,{\rm ch}^{\cN=2}_{\frac{3}{2};\frac{3}{2},2}+5796 \,{\rm ch}^{\cN=2}_{\frac{3}{2};\frac{5}{2},2}+\dots \Bigr).
\end{align}
From the discussion of the representation theory of the $\mathcal N=2$ superconformal algebra in appendix \ref{sec:N=2chars}, one sees from this character decomposition that the theory is an extremal $\mc N=2$ theory. We will denote this theory by $\mc E^{\mc N=2}_{m=2}(G)$ where $G$ is the global symmetry group of the theory and depends on the choice of $\mc N=2$ superconformal algebra.
\item Finally, by choosing three fermions one can generate an $\widehat{su(2)}_2$ current algebra which becomes part of a $c=12$ $\mathcal N=4$ SCA when combined with the $\mc N=1$ supercurrent \cite{M5}. The partition function of the theory with an additional grading by the Cartan of the $SU(2)$ coincides with the expression in \eqref{EGoftheModel}. Furthermore, it admits the following decomposition into $c=12$, $\mc N=4$ superconformal characters
  \begin{align} \notag
\mc Z^{s\natural}_{\rm R}(\t,z) & = 21\, {\rm ch}^{\cN=4}_{2;\frac{1}{2},0}+ {\rm ch}^{\cN=4}_{2;\frac{1}{2},1} + \big( 560\, {\rm ch}^{\cN=4}_{2;\frac{3}{2},\frac{1}{2}} +8470\, {\rm ch}^{\cN=4}_{2;\frac{5}{2},\frac{1}{2}}  +70576\,{\rm ch}^{\cN=4}_{2;\frac{7}{2},\frac{1}{2}} + \dots \big)
\\ \label{decom_N_is_4_1} &\;
+ \big( 210\, {\rm ch}^{\cN=4}_{2;\frac{3}{2},1} +4444\, {\rm ch}^{\cN=4}_{2;\frac{5}{2},{1}}  +42560\,{\rm ch}^{\cN=4}_{2;\frac{7}{2},{1}} + \dots \big).
\end{align}
The representation theory of the $\mathcal N=4$ SCA is reviewed in appendix \ref{sec:N=4chars}; with this information one can check that the above theory furnishes an extremal $\cN=4$ superconformal theory. We will denote this theory by $\mc E^{\mc N=4}_{m=2}(G)$, where again $G$ is the global symmetry group of the theory and depends on the choice of $\mc N=4$ superconformal algebra.
  \end{enumerate} 
 See table \ref{tbl:algebras} for a summary of the relation between $\mathcal{V}^{s\natural}$ and these different superconformal algebras.
    \begin{center}\begin{table}[htb]\begin{small}
  \begin{center}
  \def\arraystretch{1.2}
  \begin{tabular}{c|c|c|c}\toprule
 ESCFT \,& Fermions& Chiral algebra& $\mathcal A$\\\midrule
 $\mc E^{\rm Spin(7)}$ &1& Ising & $\mathcal{SW}(3/2,2)$\\
 $\mc E^{\mc N=2}_{m=2}$ & 2 & $\widehat{u(1)_2}$ & $\mathcal N=2$\\
 $\mc E^{\mc N=4}_{m=2}$  & 3 & $\widehat{su(2)_2}$ & $\mathcal N=4$\\
\bottomrule
  \end{tabular}\caption{\small Superconformal algebras (with central charge 12) generated by a subset of fermions using construction (B) of $\mathcal{V}^{s\natural}$.}\label{tbl:algebras}
  \end{center}\end{small}
 \end{table}\end{center}

    \subsection{Symmetry groups and twining functions}\label{sec:symms12}
In this section we consider the above mentioned ECFTs in more detail, beginning with an analysis of their global discrete symmetry groups. In order to do this, we restrict to construction (B), where the discrete symmetries are most transparent. 

Viewed as a theory with no supersymmetry, the continuous group Spin(24) has a natural action on the 24 fermions as signed permutations. In \cite{Duncan} it was shown that the choice of $\mathcal N=1$ supercurrent in $\mathcal{V}^{s\natural}$ breaks the Spin(24) symmetry of the 24 fermions to the discrete group $Co_0$, the group of automorphisms of the Leech lattice. 
Likewise, for each choice of superconformal algebra $\mc A$ introduced in the previous section, there is a distinct ECFT whose global symmetry group $G$ is the subgroup of $Co_0$ which preserves the choice of fermions used to construct $\mc A$.  There is moreover a geometrical interpretation of these symmetry groups: the distinct choices of superconformal algebra constructed from $n$ fermions are in one-to-one correspondence with subgroups $G <Co_0$ which preserve an $n$-dimensional subspace in the unique non-trivial irreducible 24-dimensional representation of $Co_0$ (${\bf 24}$) \cite{exceptional,M5}.
We refer to such a group as $n$-plane preserving if it preserves an $n$-dimensional subspace in the representation ${\bf 24}$. 
In the following table we have listed examples of $G$ which arise as subgroups of $Co_0$ preserving a choice of relevant superconformal algebra.
  \begin{center}
  \def\arraystretch{1.2}
  \begin{tabular}{c|c|c}\toprule
 ESCFT \,&  $\mathcal A$& Symmetry group ($G$)\\\midrule
 $\mathcal{V}^{s\natural}$& $\mathcal N=1$&$Co_0$\\
    $\mc E^{\rm Spin(7)}(G)$& $\mathcal{SW}(3/2,2)$&$M_{24}, Co_2, Co_3$\\
    $\mc E^{\mc N=2}_{m=2}(G)$ & $\mathcal N=2$&$M_{23}, M_{12}, McL, HS$\\
    $\mc E^{\mc N=4}_{m=2}(G)$  & $\mathcal N=4$&$M_{22}, M_{11}, U_4(3)$\\
\bottomrule
  \end{tabular}
  \end{center}
 For each of these theories one can construct character-valued twined partition functions for each conjugacy class $[g] \in G$. The twined functions are completely characterized by the action of $g$ on the 24 fermions, and thus by the eigenvalues of $g$ in the irreducible representation ${\bf 24}$. 
 
 Firstly, for every $[g] \in {G}$, where ${G}$ is either $\Co_0$ or a subgroup of $\Co_0$ preserving a vector in the ${\bf 24}$, the corresponding $g$-twined partition function in the NS sector is
 \be
 \label{eq:twine1}
  \mc Z^{s\natural}_{{\rm NS},g}(\t)=\Tr_{\rm NS} \,g q^{L_0-c/24} = {1\over 2}\sum_{i=1}^{4}\epsilon_i(g)\prod_{k=1}^{12} {\theta_i(\t,\rho_{g,k})\over \eta(\t)}   \ee
where the definition of the $\epsilon_i(g)$ can be found in \cite{DM-C}. Also, we have defined $e(\rho_{g,k})=\lambda_{g,k}$, where $k=1,\ldots,24$, $\rho_{g,k} \in [0,1/2]$ and $\lambda_{g,k}$ is an eigenvalue of $g$. The latter corresponds to one of the 24 roots of the rational polynomial 
   \be\label{eq:charpoly}
  \prod_{\ell|n}(t^\ell-1)^{k_\ell}\, ,
  \ee
  where $n=o(g)$ is the order of $g$, $\ell$'s are the positive divisors of $n$, and $k_\ell$'s are integers defined by the 24-dimensional irreducible representation of $g$. The data encoded in \eqref{eq:charpoly} can be succinctly written in terms of a formal product: the Frame shape of $[g]$,
  \be\label{eq:Frame}  
  \pi_g :=\prod_{\ell|n}\ell^{k_\ell}\,.
  \ee   
In \cite{DM-C} it was proved that, similar to the case of $\mathcal V^\natural$, $\mathcal{V}^{s\natural}$ furnishes a ${1\over 2}\mathbb Z$-graded $Co_0$-module, 
whose graded characters are encoded in the coefficients of the twined functions   $\mc Z^{s\natural}_{{\rm NS},g}(\t)$. 
Furthermore, for all $g \in Co_1$, the functions \eqref{eq:twine1} together with $\mc Z^{s\natural}_{{\rm R},g}(\t)$ and $\mc Z^{s\natural,-}_{{\rm NS},g}(\t),$\footnote{The upper index ``$-$'' stands for the insertion of $(-1)^F$ in the trace over the NS Hilbert space.} form a vector-valued representation of a modular group $\Gamma_g <SL_2(\mathbb R)$ with $\Gamma_0(o(g)) \subseteq \Gamma_g$.\footnote{For $g \in Co_0$ but $g\notin Co_1$, a slightly different set of functions forms a vector-valued representation of $\G_g$.}

For every $[g] \in {G}$ where ${G}$ is a subgroup of $\Co_0$ preserving (at least) a 2-plane in ${\bf 24}$, the corresponding $U(1)$-graded $g$-twined function in the R Hilbert space reads
\begin{align}\label{twining_Z_1}
 \mc Z^{s\natural}_{{\rm R},g}(\t,z) &= \Tr_{\rm R} (-1)^F q^{L_0-c/24}y^{J_0}\\\nonumber
&=\frac{1}{2}\frac{1}{\eta(\t)^{12}} \sum_{i=1}^4 (-1)^{i+1} \epsilon_{g,i}\, {\theta_i(\t,2z) }\prod_{k=2}^{12} \theta_i(\t,\rho_{g,k})\,.
\end{align}
Moreover, it was shown in \cite{exceptional,M5} that $\mathcal{V}^{s\natural}$, equipped with a choice of extended superconformal algebra $\mathcal A$ (either $\mathcal{SW}(3/2,2)$, $\cN=2$, or $\cN=4$) furnishes a $G$-module for the discrete group $G$ which preserves $\mc A$ and whose graded characters are encoded in the coefficients of a set of vector-valued mock modular forms whose corresponding shadows are (vector-valued) unary theta series. We summarize these results here and report the necessary definitions in Appendix \ref{app:funs} and \ref{app:B}.
    \begin{enumerate}\setcounter{enumi}{0}
    \item $\mc E^{\rm Spin(7)}(G)$: Let $\mc A$ be the choice of Spin(7) algebra, and $G$ the symmetry group preserving $\mc A$. $G$ is a subgroup of $Co_0$ which fixes a one-plane. From the discussion in appendix \ref{app:spin7char} on the representation theory of the Spin(7) algebra, it is apparent that one can rewrite the graded partition function of equation (\ref{eq:spin7Z}) as
      \be \label{eq:DecSpin7}
       \mc Z^{s\natural}_{\rm NS}(\t)=\mc P(\t)\left (24 \mu^{NS}(\tau)+h^{\rm Spin(7)}_{1}(\t)\Theta^{NS}_{1\over 16}(\t)+h^{\rm Spin(7)}_{7}(\t)\Theta^{NS}_{0}(\t)\right ), 
      \ee
      where $h^{\rm Spin(7)}$ is a weight 1/2 vector-valued mock modular form for $SL_2(\mathbb Z)$ with shadow given by 24$\underline{\tilde S}$,  multiplier system given by the inverse of $\underline{\tilde S}$. The definition of $\underline{\tilde S}$ is given in \eqref{eq:spin7shadow}.
      Moreover, the $g$-twined functions for all conjugacy classes $g\in G$ have a similar expansion given by
           \be\label{eq:twineSpin7}
       \mc Z^{s\natural}_{{\rm NS},g}(\t)=\mc P(\t)\left (\chi_g \mu^{NS}(\tau)+h^{\rm Spin(7)}_{g,1}(\t)\Theta^{NS}_{1\over 16}(\t)+h^{\rm Spin(7)}_{g,7}(\t)\Theta^{NS}_{0}(\t)\right ),
      \ee
       where $\chi_g= \Tr_{\bf 24} g$, and $h^{\rm Spin(7)}_g$ is a weight 1/2 vector-valued mock modular form for $\Gamma_0(n)$, $n=o(g)$ with shadow $\chi_g \underline{\tilde S}$ and, whenever $\chi_g \neq 0$, multiplier system given by the inverse multiplier system of $\underline{\tilde S}$ restricted to $\Gamma_0(n)$.\footnote{When $\chi_g=0$, the multiplier system is more complicated. This case is described in \cite{exceptional}.}
      \item $\mc E^{\mc N=2}_{m=2}(G)$: Now we let $\mc A$ be a choice of $\mc N=2$ superconformal algebra, and $G$ the two-plane preserving subgroup of $Co_0$ which preserves $\mc A$. In \cite{M5} it was shown that one can rewrite equation (\ref{Nis2decomposition_of_Z}) as
      \be
      \label{eq:DecN2}
  \mc Z^{\cN=2}_{m=2}(\t,z) = e\Bigl({3\over 4}\Bigr)(\Psi_{1,-{1\over 2}}(\tau,z))^{-1}  \biggl( 24 \,\tilde\m_{{3\over 2};0}(\t,z) +\sum_{j-{3\over 2}\in \ZZ/3\ZZ} h_j^{\cN=2}(\t) \th_{{3\over 2},j}(\t,z)   \biggr), 
  \ee
  where $h^{\cN=2}$ is a weight 1/2 vector-valued mock modular form. 
  Furthermore, $h^{\cN=2}$ has shadow given by $24S_{3/2}$ and inverse multiplier system to that of $S_{3/2}$, where $S_{3/2}$ is defined in \eqref{eq:app-shadow}. For all conjugacy classes $g\in G$, one can also write
  \be\label{eq:twineN2}
  \mc Z^{\cN=2}_{m=2,g}(\t,z) = e\Bigl({3\over 4}\Bigr)(\Psi_{1,-{1\over 2}}(\tau,z))^{-1}  \biggl( \chi_g \,\tilde\m_{{3\over 2};0}(\t,z) +\sum_{j-{3\over 2}\in \ZZ/3\ZZ} h_{g,j}^{\cN=2}(\t) \th_{{3\over 2},j}(\t,z)    \biggr),
  \ee
  where $h_g^{\cN=2}$ is a weight 1/2 vector-valued mock modular form for $\Gamma_0(n)$ with shadow $\chi_g S_{3/2}$ and multiplier given by the inverse multiplier of $S_{3/2}$ restricted to $\Gamma_0(n)$ whenever $\chi_g\neq 0$. When $\chi_g=0$,  $h_g^{\cN=2}$ is modular and has a more complicated multiplier system (c.f. \cite{M5}.)
   \item $\mc E^{\mc N=4}_{m=2}(G)$: Finally, let $\mc A$ be a choice of $\mc N=4$ superconformal algebra, and $G$ the three-plane preserving subgroup of $Co_0$ which preserves $\mc A$. It follows from \cite{M5} that equation (\ref{decom_N_is_4_1}) can be rewritten as
      \be
      \label{eq:DecN4}
  \mc Z^{\cN=4}_{m=2}(\t,z) =(\Psi_{1,1}(\tau,z))^{-1}   \biggl( 24 \,\m_{3;0}(\t,z) +\sum_{j\in \ZZ/6\ZZ} h_j^{\cN=4}(\t) \th_{3,j}(\t,z)   \biggr), 
  \ee
  where $h^{\cN=4}$ is a weight 1/2 vector-valued mock modular form and with shadow given by $24S_{3}$ and inverse multiplier system to that of $S_{3}$, \eqref{eq:app-shadow}. For all conjugacy classes $g\in G$, one can also write
  \be\label{eq:twineN4}
  \mc Z^{\cN=4}_{m=2,g}(\t,z) = (\Psi_{1,1}(\tau,z))^{-1}  \biggl(  \chi_g \,\m_{3;0}(\t,z) +\sum_{j\in \ZZ/6\ZZ} h_{g,j}^{\cN=4}(\t) \th_{3,j}(\t,z)    \biggr),
  \ee
  where $h_g^{\cN=4}$ is a weight 1/2 vector-valued mock modular form for $\Gamma_0(n)$ with shadow $\chi_g S_{3}$ and multiplier given by the inverse multiplier of $S_{3}$ restricted to $\Gamma_0(n)$ whenever $\chi_g\neq 0$. When $\chi_g=0$, $h_g^{\cN=4}$ is modular and again has a more complicated multiplier system (c.f. \cite{M5}.)
    \end{enumerate}

  \section{Central charge $24$}\label{sec:c=24}
  In this section we discuss three extremal superconformal field theories with central charge 24. Each of these SCFTs can be constructed as a nonlocal $\mathbb Z_2$ orbifold of bosons on a 24-dimensional torus given by $\mathbb R^{24}/\Lambda$ where $\Lambda$ is either the Leech lattice or one of two other Niemeier lattices (c.f. \S \ref{sec:monster}).
  
   \subsection{Extremal theories} \label{sec:morec=24}
  In \cite{DGH} it was discussed how to construct an $\mathcal N=1$ SCFT from a $\mathbb Z_2$ orbifold of the Leech (or a Niemeier) CFT, where the $\mathbb Z_2$  acts on the 24 coordinates $x_i$ as $h: x_i \to -x_i, ~ \forall i$. As discussed in \S \ref{sec:monster}, the original Hilbert space $\mc H$ and the twisted Hilbert space $\mc H^{tw}$ split respectively into subspaces $\mc H_\pm, \mc H^{tw}_\pm$ of invariant and anti-invariant states under the $h$ action (c.f. equations (\ref{eq:Horb}), (\ref{eq:Htwistorb})). 
  
The key observation in \cite{DGH} is that in the twisted sector Hilbert space $\mathcal H^{tw}_-$ there are $2^{12}$ ground states of dimension 3/2; this is precisely the dimension of an $\mathcal N=1$ supercurrent. In fact, the authors show that one can construct a consistent chiral $\mathcal N=1$ SCFT by choosing a linear combination of dimension-3/2 twist fields as a supercurrent. Furthermore, this theory has NS sector Hilbert space given by
  \be\nonumber
  \mathcal H_{NS} = \mathcal H_+ \oplus \mathcal H^{tw}_-,
  \ee
  and Ramond sector Hilbert space given by
    \be\nonumber
  \mathcal H_{R} = \mathcal H_- \oplus \mathcal H^{tw}_+.
  \ee
 The partition function in the NS sector is then given by 
  \bea\nonumber
 \mc Z^{\mc N=1}_{\rm NS}(\Lambda;\t)&=&\Tr_{\rm NS}\,q^{L_0-c/24} ={1\over 2}\frac{\Theta_\Lambda(\t)}{\eta^{24}(\t)}+2^{11} \left ({\eta^{12}(\t)\over \theta_2^{12}(\t)}- {\eta^{12}(\t)\over \theta_3^{12}(\t)}+{\eta^{12}(\t)\over \theta_4^{12}(\t)}\right)\\\nonumber
 &=&K(\t)^2-48K(\t)+12(h+2) \\\label{eq:N=1fns}
 &=& {1\over q} +12h + 4096 q^{1\over 2} + 98580 q +1228800 q^{3\over 2}+ \ldots 
  \eea
  where $\Theta_\Lambda(\t)$ is the lattice theta function, $h$ is the Coxeter number of the root system and the function $K(\t)$ defined in equation \eqref{eqn:Kfct}.
  Again, together with the characters $\Tr_{\rm R}\,q^{L_0-c/24}$ and $\Tr_{\rm NS}(-1)^F q^{L_0-c/24}$, the partition function of equation (\ref{eq:N=1fns}) transforms in a three-dimensional representation of $SL_2(\mathbb Z)$. Furthermore, the function $\Tr_{\rm R}(-1)^F q^{L_0-c/24}= 12(h+2)$ computes the Witten index of the corresponding $\mathcal N=1$ SCFT.
  
  In the case where $\Lambda= \Lambda_L$, the Leech lattice, this is precisely the partition function of an extremal $\mathcal N=1$ SCFT with $k^*=2$ (which we call $\mc E^{\cN=2}_{k^*=2}$) as defined in \cite{Witten}. The orbifold removes all dimension 1 currents in the NS sector; this fact both ensures extremality and precludes the possibility of constructing a superconformal algebra with extended supersymmetry. However, in the case where $\Lambda$ is any of the other Niemeier lattices, a nontrivial current algebra survives the orbifold. The authors of \cite{extN2} show that for $N= A_1^{24}$, one can construct a $\widehat{u(1)_4}$ current algebra, which, together with the supercurrent, satisfies the OPEs of an $\mathcal N=2$ superconformal algebra with central charge 24. Furthermore, they show that the graded partition function in the Ramond sector is precisely the weak Jacobi form which captures the spectrum of an extremal $\mathcal N=2$ SCFT with $m=4$ (which we call $\mc E^{\cN=2}_{m=4}$) according to \cite{GGKMO}:
    \bea\label{eq:c=24N=2Z}
 \mc  Z^{\cN=2}_{m=4}(\Lambda_{A_1^{24}};\t,z) &=& \Tr_{\rm R} (-1)^F y^{J_0} q^{L_0-c/24}\\\nonumber
  &=& {1\over y^4} + 46 + y^4 + \ldots\\\nonumber
  &=&47 {\rm ch}_{{7\over2}; {1},0}(\t,z) + {\rm ch}_{{7\over2}; {1},4}(\t,z)\\\nonumber
&+& (32890 +  2969208 q + \ldots)( {\rm ch}_{{7\over2}; {2},{1}}(\t,z)+{\rm ch}_{{7\over2}; {2},-{1}}(\t,z))\\\nonumber
&+& (14168 + 1659174 q + \ldots)( {\rm ch}_{{7\over2}; {2},2}(\t,z)+{\rm ch}_{{7\over2}; {2},-2}(\t,z))\\\nonumber
&+& (2024 +  485001 q + \ldots)( {\rm ch}_{{7\over2}; {2},{3}}(\t,z)+{\rm ch}_{{7\over2}; {2},-{3}}(\t,z))\\\nonumber
&+& (23 + 61984 q +  \ldots) {\rm ch}_{{7\over2}; {2},4}(\t,z),
  \eea
  where ${\rm ch}_{{\ell}; {h},Q}$ denotes the $\mathcal N=4$ character of central charge $c=3(2\ell+1)$, dimension $h$, and charge $Q$ in the Ramond sector (c.f. appendix \ref{sec:N=2chars}), and we use the fact that the ${\rm ch}_{{\ell}; {h+1},Q}=q{\rm ch}_{{\ell}; {h},Q}$ for the non-BPS characters. 
  
  Similarly, when $N= A_2^{12}$, in \cite{extN4} it is shown that one can construct an $\widehat{su(2)_4}$ current algebra which, along with the supercurrent, generates an $\mathcal N=4$ superconformal algebra with $c=24$. A straightforward computation of the graded partition function illustrates that this theory furnishes an example of an extremal $\mathcal N=4$ SCFT with $c=24$ (which we call $\mc E^{\cN=4}_{m=4}$:
    \bea\label{eq:c=24N=4Z}
  \mc Z^{\cN=4}_{m=4}(\Lambda_{A_2^{12}};\t,z) &=& \Tr_{\rm R} (-1)^F y^{J_3} q^{L_0-c/24}\\\nonumber
  &=& {1\over y^4} +{1\over y^2}+ 56 + y^2+ y^4 + \ldots\\\nonumber
 &=&55 {\rm ch}_{{4}; {1},0}(\t,z) + {\rm ch}_{{4}; {1},2}(\t,z)\\\nonumber
&+& (18876 + 1315512 q + \ldots)( {\rm ch}_{{4}; {2},{1\over 2}}(\t,z)+{\rm ch}_{{4}; {2},-{1\over 2}}(\t,z))\\\nonumber
&+& (12045 + 1152943 q + \ldots)( {\rm ch}_{{4}; {2},1}(\t,z)+{\rm ch}_{{4}; {2},-1}(\t,z))\\\nonumber
&+& (1980 + 391974 q + \ldots)( {\rm ch}_{{4}; {2},{3\over 2}}(\t,z)+{\rm ch}_{{4}; {2},-{3\over 2}}(\t,z))\\\nonumber
&+& (33 + 45990 q +  \ldots) {\rm ch}_{{4}; {2},2}(\t,z),
\eea
where the details of the characters can be found in appendix \ref{sec:N=4chars}.

  \subsection{Symmetry groups and twining functions}\label{sec:symms24}
  Like the extremal theories with central charge 12 discussed in \S \ref{sec:c=12}, the theories in the previous subsection furnish modules for a number of sporadic groups. We first consider $\mc E^{\mc N=1}_{k^*=2}$. The symmetry group of this theory arises from the automorphism group of the Leech lattice and a quantum symmetry coming from the $\mathbb Z_2$ orbifold. As discussed in \cite{DGH}, this is an extension of the group $Co_0$ by a finite abelian group. We do not discuss this theory in more detail here, though it would be interesting to investigate the properties of its twining functions.

Similarly, the discrete symmetry groups of the other two extremal theories we consider in this section arise from the automorphism group of the underlying Niemeier lattice $\Lambda_N$. The Niemeier lattices contain vectors generated by the root systems and additional so-called glue vectors. For the lattices with root systems $A_1^{24}$ and $A_2^{12}$, the glue vectors are specified by elements of the extended binary Golay code and extended ternary Golay code, respectively. See, e.g., \cite{ConwaySloane}, for a detailed description.

  The automorphism group of the $A_1^{24}$ Niemeier lattice is the Mathieu group $M_{24}$. It acts naturally on the 24 copies of the $A_1$ root system in its 24-dimensional (reducible) permutation representation. Furthermore, its action on the glue vectors is inherited from its natural action on the binary Golay code as automorphisms. Note that we must choose a particular $A_1$ root system to construct the affine $\widehat{u(1)_4}$ current algebra which becomes part of the $\mathcal N=2$ SCA with $c=24.$ The choice of this root system breaks the $M_{24}$ symmetry of the theory to an $M_{23}$ subgroup, where the $M_{23}$ fixes the distinguished coordinate direction, say $x_1$, associated with this $A_1$, and acts as a subgroup of $S_{23}$ on the remaining coordinates. 
  In appendix \ref{app:twin}, we discuss the derivation of the twining functions
  \be\label{c=24N=2twin}
  \mc Z^{\cN=2}_{m=4,g}(\t,z)=\Tr_{\rm R}\, g \,(-1)^F y^{J_0} q^{L_0-c/24}
  \ee
  for conjugacy classes $g\in M_{23}$. These functions are weak Jacobi forms of weight zero and index 4 for the group $\Gamma_0(n)$ where $n=o(g)$, and they have the expansion
  \be
    \mc Z^{\cN=2}_{m=4,g}(\t,z)={1\over y^4} + 2 \Tr_{\bf 23}g + y^4+ O(q) \,,
  \ee
  where ${\bf 23}= {\bf 1} + {\bf 22}$ is the 23-dimensional permutation representation of $M_{23}$. 
  
  On the other hand, the automorphism group of the $A_2^{12}$ Niemeier lattice is $2.M_{12}$, an extension of the Mathieu group $M_{12}$, where the $M_{12}$ acts as a subgroup of $S_{12}$ on the 12 root systems, and the extension includes the order two automorphism of the $A_2$ Dynkin diagram. The action of $2.M_{12}$ on the glue vectors of the lattice follows from its action on the ternary Golay code, which specifies the glue vectors. In order to construct an affine $\widehat{su(2)_4}$ current algebra which becomes part of an $\cN=4$ SCA with $c=24$, one chooses a distinguished $A_2$ root system corresponding to two directions, say $x_1, x_2$.  The subgroup of $2.M_{12}$ which preserves the $\mathcal N=4$ SCA is then a copy of $M_{11}$ which fixes $x_1, x_2$ and permutes the other 11 root systems. 
  We discuss the twining functions 
    \be\label{c=24N=4twin}
  \mc Z^{\cN=4}_{m=4,g}(\t,z)=\Tr_{\rm R}g (-1)^F y^{J_0} q^{L_0-c/24}
  \ee
  for certain conjugacy classes $g \in M_{11}$, in appendix \ref{app:twin}. These functions are weak Jacobi forms of weight zero and index 4 for $\Gamma_0(n)$, $n=o(g)$, and they have the expansion
  \be
    \mc Z^{\cN=4}_{m=4,g}(\t,z)={1\over y^4} +{1\over y^2} + (5 \Tr_{\bf 11}g  +1)+ y^2 +y^4+ O(q),
  \ee
  where ${\bf 11} = {\bf 1} + {\bf 10}$ is the 10-dimensional permutation representation of $M_{11}$. 

  Just as discussed in the previous section for central charge 12, the two ECFTs $\mc E^{\mc N=2}_{m=4}$, $\mc E^{\mc N=4}_{m=4}$ with central charge 24 furnish $G$-modules whose graded characters are encoded in the coefficients of certain vector-valued modular forms, where $G$ is the global symmetry group of the theory.   We discuss the properties of these mock modular forms for each case below.
  \begin{enumerate}
  \item $\mc E^{\mc N=2}_{m=4}$: From the discussion in appendix \ref{sec:N=2chars}, it is clear that we can rewrite the graded partition function of equation (\ref{eq:c=24N=2Z}) as
  \be
  \mc Z^{\cN=2}_{m=4}(\t,z) = e\Bigl({3\over 4}\Bigr)(\Psi_{1,-{1\over 2}}(\tau,z))^{-1}  \biggl( 48 \,\tilde\m_{{7\over 2};0}(\t,z) +\sum_{j-{7\over 2}\in \ZZ/7\ZZ} \tilde{h}_j^{\cN=2}(\t) \th_{{7\over 2},j}(\t,z)   \biggr),
  \ee
  where $\tilde{h}^{\cN=2}$ is a weight ${1\over 2}$ vector-valued mock modular form for $SL_2(\ZZ)$ with shadow $48S_{7\over 2}$, defined in \eqref{eq:app-shadow}, and multiplier system inverse to that of $S_{7\over 2}$. 
 Similarly, the $g$-twined functions of equation (\ref{c=24N=2twin}) have an expansion
  \be\label{eq:M23twine}
  \mc Z^{\cN=2}_{m=4,g}(\t,z) = e\Bigl({3\over 4}\Bigr)(\Psi_{1,-{1\over 2}}(\tau,z))^{-1}  \biggl( 2\chi_g \,\tilde\m_{{7\over 2};0}(\t,z) +\sum_{j-{7\over 2}\in \ZZ/7\ZZ} \tilde{h}_{g,j}^{\cN=2}(\t) \th_{{7\over 2},j}(\t,z)   \biggr),
  \ee
  where  $\chi_g = \Tr_{\bf 24}g$ and $\tilde{h}_{g}^{\cN=2}$ is a weight ${1\over 2}$ vector-valued mock modular form for $\Gamma_0(n)$, $n=o(g)$, shadow $2\chi_gS_{7\over 2}$, and multiplier system inverse to that of $S_{7\over 2}$. In appendix \ref{sec:twinetbls}, we present tables of the first several coefficients of the $\tilde h_j^{\cN=2}$ for all $g\in M_{23}$, as well as their decompositions into irreducible $M_{23}$ representations.
    \item $\mc E^{\mc N=4}_{m=4}$: Similarly, the discussion in appendix \ref{sec:N=4chars} indicates that we can write the graded partition function in (\ref{eq:c=24N=4Z}) as
      \be
 \mc Z^{\cN=4}_{m=4}(\t,z) = (\Psi_{1,1}(\tau,z))^{-1}  \biggl( 60 \,\m_{5;0}(\t,z) +\sum_{j\in \ZZ/10\ZZ} \tilde{h}_j^{\cN=4}(\t) \th_{5,j}(\t,z)  \biggr),
  \ee
  where $\tilde{h}^{\cN=4}$ is a weight ${1\over 2}$ vector-valued mock modular form for $SL_2(\ZZ)$ with shadow $60S_5$ and  multiplier system the inverse to that of $S_5$. 
The $g$-twined functions (\ref{c=24N=4twin}) for conjugacy classes $g \in M_{11}$ similarly give rise to vector-valued mock modular forms through the expansion
 \be\label{eq:M11twine}
  \mc Z^{\cN=4}_{m=4,g}(\t,z) = (\Psi_{1,1}(\tau,z))^{-1} \biggl( 5(\Tr_{\bf 12}g) \,\m_{5;0}(\t,z) +\sum_{j\in \ZZ/10\ZZ} \tilde{h}_{g,j}^{\cN=4}(\t) \th_{5,j}(\t,z)   \biggr),
  \ee
where $\tilde{h}_g^{\cN=4}$ is a weight ${1\over 2}$ vector-valued mock modular form for $\Gamma_0(n)$, $n=o(g)$, with shadow $5(\Tr_{\bf 12}g)S_5$ where ${\bf 12}= {\bf 1}+{\bf 11}$ and multiplier system the inverse of that of $S_5$.
\end{enumerate}

\section{Rademacher summability} \label{sec:results} 
Inspired by the relation between the genus zero property of monstrous moonshine and the Rademacher sum construction of the McKay-Thompson series underlined in \cite{Duncan:2009sq, ChengDunc}, we examine the Rademacher expansion at the infinite cusp for the twined functions of the ECFTs introduced in \S \ref{sec:c=12} and \S\ref{sec:c=24}. We begin in section \ref{sec:ConRad} by discussing the $Co_0$-module $\mc V^{s\natural}$, and analyze the other $c=12$ and $c=24$ theories in \S \ref{sec:Rad12} and \S\ref{sec:Rad24}, respectively.  The results presented in \S \ref{sec:Rad12} and \S\ref{sec:Rad24} are obtained by numerically computing the coefficients in equation (\ref{eqn:radvec}) to high accuracy and comparing with the known twining functions described in \S \ref{sec:c=12} and \S\ref{sec:c=24}, respectively.

Finally, in \S \ref{sec:FrickeN=2} we discuss the following curious property of the the twining functions for the theory $\mc E^{\mc N=2}_{m=2}(M_{23}).$ The functions which cannot be expressed as Rademacher sums at the infinite cusp {precisely} correspond to conjugacy classes $g \in M_{23}$ such that $3 | o(g)$. In this case, however, the expansion of these functions about cusps inequivalent to $i\infty$ either has no pole, or the coefficients in such an expansion can be directly related to the coefficients which appear in the expansion of the function at $i\infty$. This property might be interpreted as a generalization of the results of Tuite \cite{tuite1992} reviewed in \S \ref{sec:monster} for McKay-Thompson series for genus zero groups with Atkin-Lehner involutions.

\subsection{The Conway module}\label{sec:ConRad}
The Conway theory $\mathcal{V}^{s\natural}$ \cite{DM-C} furnishes a ${1\over 2}\mathbb Z$-graded $Co_0$-module, i.e., 
\be
\mc V^{s\natural} = \bigoplus_{n=-{1}}^\infty \mc V_{n\over 2}^{s\natural}.
\ee
In \cite{Duncan} it was shown that the action of $Co_0$ is entirely dictated by the eigenvalues of $g$ in its 24-dimensional irreducible representation ($\bf 24$) and therefore only depends on the conjugacy class $[g]$. In the following, we refer to the latter via the Frame shape $\pi_g$ defined in \eqref{eq:Frame}. For each conjugacy class $[g] \in Co_0$, one can define a set of character-valued twined partition functions by taking traces in the NS Hilbert space with and without insertion of $(-1)^F$. These correspond to the previously defined $Z^{s\natural,-}_{{\rm NS},g}(\t),\,Z^{s\natural}_{{\rm NS},g}(\t)$ and take the form 
  \bea\label{eqn:Zsnm}
 \mc Z^{s\natural,-}_{{\rm NS},g}(\t) &=& \sum_{n=-{1}}^\infty (\Tr_{\mc V_{n\over 2}^{s\natural}} \, (-1)^n g)q^{n\over 2}\, .  \\
 \mc Z^{s\natural}_{{\rm NS},g}(\t) &=&  \sum_{n=-{1}}^\infty (\Tr_{\mc V_{n\over 2}^{s\natural}} \, g)q^{n\over 2}\, .\label{eqn:Zsn}
  \eea
As it was previously mentioned, the Conway module $\mc V^{s\natural}$ and the monster module $\mc V^\natural$ have many common features. First of all, in both of these theories the graded traces associated to the particular module are simply related to the Hauptmoduln of the corresponding modular group. However, the fermionic nature of the Conway module is reflected in its half-integer grading, in contrast to the $\mathbb Z$-graded monster module. As a result,  normalized Hauptmoduln arise only after rescaling the twining functions $Z^{s\natural,-}_{{\rm NS},g}(2\t)$ and $Z^{s\natural}_{{\rm NS},g}(2\t)$.
This genus zero property of $\mathcal{V}^{s\natural}$ was shown to hold in \cite{DM-C}; from this the analogue of Theorem \ref{thm:DunFren} for $\mathcal{V}^{\natural}$ directly follows (c.f. Theorem 4.9 in \cite{DM-C}.)

Specifically, all twining functions can be expressed as Rademacher sums at the infinite cusp both (i) as scalar-valued Rademacher sums with respect to appropriate subgroups of genus zero groups appearing in monstrous moonshine, 
and (ii) as vector-valued Rademacher sums where the vector includes \eqref{eqn:Zsnm}, \eqref{eqn:Zsn} and $\mc Z^{s\natural}_{{\rm R},g}(\t)$ and transforms under a modular group containing $\Gamma_0(N)$ and contained in $\mc N(N)$. The explicit modular properties of the twining functions can be found in \cite{DM-C}.

\subsection{Modules with $c=12$}\label{sec:Rad12}
Following \S \ref{sec:symms12} and \cite{M5}, it is clear that the extremal SCFTs $\, \mc E^{\rm Spin(7)}(G)$, $\, \mc E^{\mc N=2}_{m=2}(G)$, $\, \mc E^{\mc N=4}_{m=2}(G)$ with central charge 12 furnish $G$-modules for the global symmetry group $G$ of the theory. We will denote these $G$-modules  by $\mathcal{V}^{\mc A,G}$, where $\mc A$ denotes the extended superconformal algebra and 
    \be
    \mathcal{V}^{\mathcal A, G}= \bigoplus_{r\in \{\alpha\}_{\mathcal A}}\bigoplus_{n=1}^\infty V_{r,n}^{\mathcal A, G}\, ,\quad  \qquad \mc A \in \{{\rm Spin(7)}, \,\mc N=2,\, \mc N=4\}.
    \ee
The corresponding graded characters are the coefficients of certain vector-valued mock modular forms $h^{\mc A}_g$, defined by
    \be
    h^{\mathcal A}_{g,r}(\tau) = a_r q^{-{r^2\over b_{\mathcal A}}}+ \sum_{n=1}^\infty( \Tr_{\mathcal{V}^{\mathcal A,G}_{r,n}}g)q^{n-{ r^2\over b_{\mathcal A}}}\, .
    \ee
    The constants $a_r, b_{\mc A}, \{\alpha\}_{\mathcal A}$ appearing in the expansion are displayed below together with the symmetry groups $G$ on which we  focus in this section. In particular, in the case of the $\cN=4$ theories, we consider two different embeddings of $M_{11}$ into $Co_0$, which we refer to as $M_{11}^{(1)}$ and $M_{11}^{(2)}$. $M_{11}^{(1)}$ can be described as the subgroup of $M_{12}$ which fixes a point in its 12-dimensional permutation representation; on the other hand, $M_{11}^{(2)}$ is the subgroup of $M_{12}$ which fixes a certain length-12 vector in its 12-dimensional permutation representation.
    \begin{center}
    \begin{tabular}{c|c|c|c|c}\toprule
    $\mathcal A$&$\{\alpha\}_{\mathcal A}$&$b_{\mathcal A}$&$\{a_r\}$& $G$\\\midrule
      Spin(7)&$\{1,7\}$&$120$&$\{-1,1\}$&$M_{24}$\\
    $\mathcal N=2$&$\left\{\pm {1\over 2},{3\over 2}\right\}$&$6$&$\{-1,1\}$&$M_{23}, M_{12}$\\
    $\mathcal N=4$&$\{1,2\}$&$12$&$\{-2,-1\}$&$M_{22}, M^{(1)}_{11}, M^{(2)}_{11}$\\\bottomrule
       \end{tabular}
 \end{center}
The functions $h^{\mc A}_{g,r}$ comprise a vector-valued mock modular form of weight $1/2$ and multiplier system $\rho_g$ with respect to $\Gamma_g$ a congruence subgroup of $SL_2(\ZZ)$. Denoting by $n$ the order of $g$, $\Gamma_g$ equals $\Gamma_0(n)$. Moreover, we denote by $\xi$ the smallest cycle in \eqref{eq:Frame}.
 Note that we choose to focus on the Mathieu groups, which are distinguished as all of their twined Jacobi forms have particularly nice behavior at other cusps according to \cite{M5}. However, as reviewed in \S \ref{sec:symms12}, there exists an extremal Spin(7), $\cN=2$, and $\cN=4$ CFTs for each subgroup of $Co_0$ which preserves a one-, two-, or three-plane, respectively. We leave a general analysis of such cases to future work.

In the following we report  the necessary data for the construction of the  functions $h^{\mc A}_g$ via a Rademacher sum 
\be
h^{\mc A}_{g}(\t) = \mathcal{R}^{(\{a_r\})}_{\Gamma_g,\,1/2,\,\rho_g}(\tau)\,,
\ee
and the conjugacy classes of $G$ whose twining function cannot be reproduced by a Rademacher expansion at the infinite cusp. 
    \begin{enumerate}
    \item $\mc E^{\rm Spin(7)}(G)$: The weight 1/2 vector-valued mock modular form $h^{\rm Spin(7)}$ for $SL_2(\mathbb Z)$ is derived from equation \eqref{eq:DecSpin7}. We report here the first few Fourier coefficients
      \bea\nonumber
      h^{\rm Spin(7)}_{1}(\t)&=&q^{-{1\over 120}}(-1 + 1771 q + 35650 q^2 + 374141 q^3 + \ldots)\,,\\\label{eq:Spin7exp}
      h^{\rm Spin(7)}_{7}(\t)&=&q^{-{49\over 120}}(1 + 253 q + 7359 q^2 + 95128 q^3 + \ldots)\, .
      \eea
      These expressions fix the coefficients of the negative $q$-power terms for all the twined versions $h^{\rm Spin(7)}_g$; these polar terms arise from the $G$-invariant NS ground state.
The multiplier system of these mock modular forms is the inverse multiplier of the vector-valued unary theta series $\underline{\tilde S}$, defined in \eqref{eq:spin7shadow}, as long as $\chi_g\neq0$. The latter is completely specified by its representation on the generators of $SL_2(\mathbb Z)$ 
\be
\label{eqn:spin7Mult}
\rho(T)= \begin{pmatrix} e(\frac{1}{120}) &0 \\
0 & e(\frac{49}{120})\\
\end{pmatrix} \, ,\qquad
\rho(S)=\begin{pmatrix} -\sqrt{\frac{2}{5+\sqrt{5}}} & \sqrt{\frac{2}{5-\sqrt{5}}} \\
\sqrt{\frac{2}{5-\sqrt{5}}} & \sqrt{\frac{2}{5+\sqrt{5}}}\\
\end{pmatrix}.
\ee
When the element $g$ has no fixed points the multiplier system is not constrained by that of the shadow; it is given by the inverse of \eqref{eqn:spin7Mult} times  a Frame shape-dependent phase
\be
\nu_g=e\biggl(-\frac{c\, d}{n \,\xi}\biggr)\,.
\ee
The Rademacher series defined in \eqref{eqn:radvec} reproduces $h^{\rm Spin(7)}_g$ for all  conjugacy classes in $M_{24}$ except for the conjugacy classes reported in Table \ref{table:c12poles} and the one with Frame shape $12^2$ for which it has not been found the correct multipler system. 
      \item $\mc E^{\mc N=2}_{m=2}(G)$: From equation \eqref{eq:DecN2} $h^{\cN=2}$ is a weight 1/2 vector-valued mock modular form whose first few coefficients are given by
  \bea\nonumber
  h^{\cN=2}_{1\over 2}(\t)&=&h^{\cN=2}_{-{1\over 2}}(\t)= -q^{-1/24}+770q^{23/24}+13915 q^{47/24}+\ldots \\\label{eq:N=2exp}
  h^{\cN=2}_{{3\over 2}}(\t)&=& q^{-9/24}+231 q^{15/24}+5796 q^{37/24}+\ldots.
  \eea
The multiplier system is given by the inverse of the half-index theta function multiplier, defined in equation \eqref{eqn:halfthetamult}.

In the case $G=M_{23}$, the Rademacher expression \eqref{eqn:radvec} coincides with the vector-valued mock modular form $h_g^{\cN=2}$ for all the conjugacy classes except those for which $3|o(g).$ However, see \S \ref{sec:FrickeN=2} for an analysis of the  structure of these functions at the other poles of $\Gamma_g$.

Similarly, in the case $G=M_{12}$ the functions $h_g^{\cN=2}$ corresponding to the conjugacy classes for which $3|g$ cannot be reproduced by the Rademacher series at the infinite cusp, except for the Frame shape $3^8$. Additionally, the Rademacher expansion of $h_g^{\cN=2}$ at the infinite cusp fails in the case where $g$ has Frame shape $4^28^2$.
The multiplier system of $h_g^{\cN=2}$ for the conjugacy classes in $\pi_g$ with no fixed points and which can be reproduced using the Rademacher expansion is given by the inverse of \eqref{eqn:halfthetamult} times the phase 
\be
\label{eq:phasse}
\nu_g=e\biggl(-\frac{c\, d}{n\,\xi}\biggr)\,.
\ee
    \item $\mc E^{\mc N=4}_{m=2}(G)$:
   Equation \eqref{eq:DecN4} defines $h^{\cN=4}$, a vector-valued mock modular form whose first few coefficients are given by
  \bea\nonumber
  h^{\cN=4}_{1}(\t)&=&-h^{\cN=4}_{-1}(\t)= -2q^{-1/12}+560q^{11/12}+8470 q^{23/12}+\ldots \\\label{eq:N=4exp}
    h^{\cN=4}_{2}(\t)&=&-h^{\cN=4}_{-2}(\t)= -q^{-4/12}-210 q^{8/12}-4444 q^{16/12}+\ldots.
  \eea
The multiplier system of these weight 1/2 mock modular forms with respect to $\G_0(n)$ is the conjugate of the shadow $\chi_g S_{3}$, (c.f. \cite{M5}). 
Due to the symmetry of the theta function and the modular properties of the Jacobi form, it follows that $h^{\cN=4}_{m,r}=-h^{\cN=4}_{m,-r}$. 
Thus, among the 6 components of the mock modular form only three of them are linearly independent. 

In this case we find that the Rademacher sum (\ref{eqn:radvec}) coincides with $h^{\cN=4}_g$ for all conjugacy classes $g\in G$ where $G=M_{22}, M_{11}^{(1)}$. For $G=M_{11}^{(2)}$, the conjugacy class labelled by the Frame shape $2^4 \, 4^4$ has multiplier system given by the inverse of the theta multiplier \eqref{eqn:thetamult} times \eqref{eq:phasse} whereas for $4^2 \, 8^2$ there is no match between the Rademacher sum and the twining function. 

    \end{enumerate}
To summarize, we report below the conjugacy classes corresponding to the mock modular forms that cannot be reconstructed using solely the information at the infinite cusp for these $c=12$ theories. 
  \begin{table}[htb]
  \begin{center}
  \begin{tabular}{c|c}\toprule
  ECFT& Frame shapes with additional poles \\\midrule 
  $\mc E^{\rm Spin(7)}\,(M_{24})$ & $1^6\,3^6 - 1^4\,5^4- 1^2\,2^2\,3^2\,6^2 - 2^2\, 10^2 - 2.4.6.12 - 1.3.5.15$\\
  $\mc E^{\mc N=2}_{m=2}\,(M_{23})$ & $1^6\,3^6 - 1^2\,2^2\,3^2\,6^2 -1.3.5.15$\\
  $\mc E^{\mc N=2}_{m=2}\,(M_{12})$ & $1^6\,3^6-1^2\,2^2\,3^2\,6^2-6^4-4^2\,8^2$\\ 
  $\mc E^{\mc N=4}_{m=2}\,(M_{22})$ & ${\rm None}$\\
  $\mc E^{\mc N=4}_{m=2}\,(M^{(1)}_{11})$ & ${\rm None}$ \\
  $\mc E^{\mc N=4}_{m=2}\,(M^{(2)}_{11})$ & $4^28^2$  \\\bottomrule
  \end{tabular}\caption{\small Pole structure of $h^{\mc A}_{g}(\t)$ for certain extremal theories with central charge 12.}
  \label{table:c12poles}
  \end{center}
  \end{table}

\subsection{Modules with $c=24$}\label{sec:Rad24}
In analogy to the theories with central charge 12, the two extremal CFTs $\mc E^{\mc N=2}_{m=4}$, $\mc E^{\mc N=4}_{m=4}$ with central charge 24 furnish $G$-modules whose graded characters are encoded in the coefficients of certain vector-valued modular forms. We will denote these modules by $\tilde{\mathcal{V}}^{\cA, G}$, 
  \be
    \tilde{\mathcal{V}}^{\mathcal A, G}= \bigoplus_{j\in \{\tilde \alpha\}_{\mathcal A}}\bigoplus_{n=1}^\infty \tilde V_{j,n}^{\mathcal A, G}\, ,\quad  \qquad \mc A \in \{\mc N=2,\, \mc N=4\},
    \ee
    where $G=M_{23}$ and $M_{11}$, respectively. 
  From these considerations and the description given in \S \ref{sec:symms12}, we see that the mock modular forms can be written as
    \be
    \tilde{h}^{\mathcal A}_{g,j}(\tau) = \tilde a_j q^{-{j^2\over \tilde b_{\mathcal A}}}+ \sum_{n=1}^\infty( \Tr_{\tilde{\mathcal{V}}^{\mathcal A,G}_{j,n}}g)q^{n-{ j^2\over \tilde b_{\mathcal A}}}\, ,
    \ee
where the data appearing above can be summarized succinctly in the following table:
      \begin{center}
    \begin{tabular}{c|c|c|c|c}\toprule
    $\mathcal A$&$\{\tilde \alpha\}_{\mathcal A}$&$\tilde b_{\mathcal A}$&$\{a_r\}$&$G$\\\midrule
        $\mathcal N=2$&$\left\{\pm {1\over 2},\pm{3\over 2},\pm {5\over 2}, {7\over 2}\right\}$&$14$&$\{-1,1,-1,1\}$&$M_{23}$\\
    $\mathcal N=4$&$\{1,2,3,4\}$&$20$&$\{-4,-3,-2,-1\}$&$M_{11}$\\\bottomrule
 \end{tabular}
 \end{center}
Furthermore, for all $g\in G$, there is a modular group $\Gamma_g < SL_2(\ZZ)$ with $\Gamma_g=\Gamma_0(n)$, where $n=o(g)$, such that $\tilde{h}^{\mathcal A}_{g}(\tau)$ is a vector-valued mock modular form of weight $1/2$ and multiplier system $\rho_g: \Gamma_g \to \CC^\times$ with respect to $\Gamma_g$.  In appendix \ref{app:twin} we explicitly compute the functions $\tilde h^{\cN=2}_{g}$. Furthermore, we discuss the computation of $\tilde h^{\cN=4}_g$ for three conjugacy classes in $M_{11}$.\footnote{For the other conjugacy classes in $M_{11}$, we  compare $\tilde h^{\cN=4}_g$ with the Rademacher formula by computing the first couple coefficients of the twined function and seeing already that they do not match.}

\begin{enumerate}
  \item $\mc E^{\mc N=2}_{m=4}$: The first few Fourier coefficients of the mock-modular form $\tilde{h}^{\cN=2}$ are
   \bea\nonumber
 \tilde{h}^{\cN=2}_{1\over 2}(\t) = \tilde{h}^{\cN=2}_{-{1\over 2}}(\t) &=& - q^{-1/56} + 32890 q^{55/56} + 2969208 q^{111/56} + \ldots\\\nonumber
 \tilde{h}^{\cN=2}_{3\over 2}(\t) = \tilde{h}^{\cN=2}_{-{3\over 2}}(\t) &=&  q^{-9/56} + 14168 q^{47/56} + 1659174 q^{103/56} + \ldots\\\nonumber
 \tilde{h}^{\cN=2}_{5\over 2}(\t) =  \tilde{h}^{\cN=2}_{-{5\over 2}}(\t) &=& - q^{-25/56} + 2024 q^{31/56} + 485001 q^{87/56} + \ldots\\
 \tilde{h}^{\cN=2}_{7\over 2}(\t)  &=&  q^{-49/56} + 23 q^{7/56} + 61894 q^{63/56} + \ldots. \label{eq:N=2coeffs}
 \eea
As before, the coefficients multiplying the polar $q$-terms are singlets under the action of $g$. The multiplier system is constrained by the multiplier system of the unary theta series $S_{7\over 2}$ and corresponds to the inverse of the half-integral index theta function \eqref{eqn:halfthetamult}.
We find that the functions can be reproduced by a Rademacher sum at the infinite cusp only for $\pi_g \in \{1^{24},\, 1^2\,11^2,\, 1.23\}$. 

    \item $\mc E^{\mc N=4}_{m=4}$: The vector-valued mock modular form $\tilde{h}^{\cN=4}$ has the first few coefficients,
   \bea\nonumber
 \tilde{h}^{\cN=4}_1(\t) = -\tilde{h}^{\cN=4}_{-1}(\t) &=& -4 q^{-1/20} + 18876 q^{19/20} + 1315512 q^{39/20} + \ldots\\\nonumber
 \tilde{h}^{\cN=4}_2(\t) = -\tilde{h}^{\cN=4}_{-2}(\t) &=& -3 q^{-4/20} - 12045 q^{16/20} - 1152943 q^{36/20} + \ldots\\\nonumber
 \tilde{h}^{\cN=4}_3(\t) = -\tilde{h}^{\cN=4}_{-3}(\t) &=& -2 q^{-9/20} + 1980 q^{11/20} + 391974 q^{31/20} + \ldots\\
  \tilde{h}^{\cN=4}_4(\t) = -\tilde{h}^{\cN=4}_{-4}(\t) &=&  -q^{-16/20} - 33 q^{4/20} - 45990 q^{24/20} + \ldots,\label{eq:N=4coeffs}
 \eea
  where $\tilde{h}^{\cN=4}_0(\t)=\tilde{h}^{\cN=4}_5(\t)=0$. The multiplier system is given by the conjugate multiplier system of $5(\Tr_{\bf 12}g)S_5$ and therefore equals the inverse of the theta function multiplier system \eqref{eqn:thetamult}.
We find that the Rademacher sum \eqref{eqn:radvec} correctly reproduces the twining function $\tilde h^{\cN=4}_g$ only for $\pi_g \in \{1^{24},\,1^2\,11^2\}$.
\end{enumerate}

To summarize, we report in Table \ref{table:c24poles} the conjugacy classes which can be reconstructed from solely the information of the infinite cusp and thus do not have poles at any additional cusps. 
  \begin{table}[htb]
  \begin{center}
  \begin{tabular}{c|c}\toprule
 ECFT& Frame shapes with no additional poles \\\midrule 
  $\mc E^{\mc N=2}_{m=4}$ & $1^{24}- 1^2\,11^2 - 1.23$\\
  $\mc E^{\mc N=4}_{m=4}$ & $1^{24}-1^2\,11^2$ \\\bottomrule
  \end{tabular}\caption{\small Pole structure of $\tilde h^{\mc A}_{g}(\t)$ for the two extremal theories with central charge 24. In contrast to Table \ref{table:c12poles}, we report the conjugacy classes where the only pole is at the infinite cusp.}
    \label{table:c24poles}
  \end{center}
  \end{table}

\subsection{Cusp behavior of $h^{\cN=2}_g$}\label{sec:FrickeN=2}
In this section we discuss an intriguing property of the vector-valued mock modular forms $h^{\cN=2}_g$ of $\mc E^{\cN=2}_{m=2}(M_{23})$ for $\pi_g \in \{1^6\,3^6,\, 1^2\,2^2\,3^2\,6^2,\, 1.3.5.15\}$, i.e. $g \in \{3A,\, 6A,\, 15AB\}$ using the standard ATLAS notation \cite{atlas} for these conjugacy classes. These are precisely the functions which are not Rademacher sums at the infinite cusp. They have poles at the cusp at zero, ${1\over 2}$, and ${1\over 5}$, respectively. However, the coefficients in the expansion of these functions around these cusps can be related to the coefficients in the expansion at the infinite cusp, via a relation with functions appearing in the $M_{24}$ ($\ell=2$) case of umbral moonshine. First, let
\be
H_{1A}(\t):={1\over 2}\hat H^{(2)}_{1A}(\t)= q^{-1/8}(-1 + 45 q + 231 q^2 +770 q^3\ldots),
\ee
be the function such that $\hat H^{(2)}_{1A}(\t)$ is the single independent component of a weight ${1\over 2}$ vector-valued mock modular form for $SL_2(\ZZ)$ whose coefficients encode the graded dimensions of an $M_{24}$ module \cite{EOT,UM,UMNL}. Furthermore, let 
\be
H_{g'}(\t):={1\over 2}\hat H^{(2)}_{g'}(\t)
\ee
be the corresponding (weight ${1\over 2}$, vector-valued) mock modular forms for $\G_{g'}$ encoding the graded traces of $g'$ in this module for all conjugacy classes $g' \in M_{24}$, where $\G_{g'}$ is just equal to $\G_0(o(g'))$. We also use below the fact that
\be
H_{3A}(\t):={1\over 2}\hat H^{(2)}_{3A}(\t)= q^{-1/8}(-1 + 0 q + -3 q^2 +5 q^3\ldots)
\ee
for the conjugacy class $g' = 3A$ in $M_{24}$.
Note the following interesting relation between the functions $h^{\cN=2}_{g}$ for all $g\in M_{23}$ and the functions $H_{g'}(\t)$.

We introduce the notation $h^\infty_{g,r}$ to denote the $r$th component of $h^{\cN=2}_g$ expanded about the cusp of $\G_g$ at $\t = i\infty.$ Similarly, we will use the notation $h^{\zeta}_{g,r}$ to denote the $r$th component of $h^{\cN=2}_g$ expanded about the cusp of $\G_g$ at $\t = \zeta.$ Our first observation is that 
\bea\label{eq:Hecke?1}
h^\infty_{1A,{1\over 2}}(\t) &=& {1\over 3} \sum_{\alpha=0}^2 e\left ({\alpha\over 24}\right ) H_{1A}\left ({\t + \alpha \over 3}\right )=q^{-1/24}(-1 + 770 q + 13915 q^2 +132825 q^3 \ldots)\\\nonumber
h^\infty_{1A,{3\over 2}}(\t) &=& {1\over 3} \sum_{\alpha=0}^2 e\left (-{15\alpha\over 24}\right ) H_{1A}\left ({\t + \alpha \over 3}\right )-H_{1A}(3\t)=q^{-9/24}(1 + 231 q + 5796 q^2 +65505q^3 \ldots)
\eea
and
\bea\label{eq:Hecke?2}
h^\infty_{3A,{1\over 2}}(\t) &=& {1\over 3} \sum_{\alpha=0}^2 e\left ({\alpha\over 24}\right ) H_{3A}\left ({\t + \alpha \over 3}\right )=q^{-1/24}(-1 + 5 q + 10 q^2 +21 q^3 \ldots)\\\nonumber
h^\infty_{3A,{3\over 2}}(\t) &=& -{2\over 3} \sum_{\alpha=0}^2 e\left (-{15\alpha\over 24}\right ) H_{3A}\left ({\t + \alpha \over 3}\right )-H_{1A}(3\t)=q^{-9/24}(1 + 6 q + 18 q^2 -15 q^3 \ldots).
\eea
This encodes the relation of the 1A and 3A twining functions of $\mc E^{\cN=2}_{m=2}(M_{23})$ to those of $M_{24}$ umbral moonshine.\footnote{The above equations in (\ref{eq:Hecke?1}) and (\ref{eq:Hecke?2}) look very much like the action of a Hecke operator on $H_{g}$. It would be interesting to explore this connection further.}

Now let's look at the expansion of $h^{\cN=2}_{3A}$ at $\zeta=0$. We find that the components $h^0_{3A,{1\over 2}}(\t),h^0_{3A,{3\over 2}}(\t)$  can be expressed as linear combinations of the functions $h^\infty_{3A,{1\over 2}}(\t),h^\infty_{3A,{3\over 2}}(\t)$ and $H_{1A}(\t)$. Explicitly, the relation is
\bea\nonumber
h^{0}_{3A,{1\over 2}}(\t)&=&H_{1A}\left({\t\over 3}\right ) - 3h^{\infty}_{3A,{1\over 2}}(\t)=2q^{-1/24}+45 q^{7/24}+231 q^{15/24}+755 q^{23/24}+2277q^{31/24}+\ldots\\
h^{0}_{3A,{3\over 2}}(\t)&=&-h^\infty_{3A,{1\over 2}}(\t) -h^\infty_{3A,{3\over 2}}(\t) -H_{1A}(3\t)=q^{-1/24}-6q^{15/24}-5q^{23/24}-18q^{39/24}+\ldots.
\eea
Similarly, consider the following pairs of conjugacy classes: $(g',g)=(2A, 6A)$ and $(g',g)=(5A, 15AB)$ for $g' \in M_{24}$ and $g\in M_{23}$. Then we have a similar relation for the two other functions with additional poles given by
\bea\nonumber
h^{\zeta_g}_{g,{1\over 2}}(\t)&=&H_{g'}\left({\t\over 3}\right ) - 3h^{\infty}_{g,{1\over 2}}(\t)\\\label{eq:FrickeN=2}
h^{\zeta_g}_{g,{3\over 2}}(\t)&=&-h^\infty_{g,{1\over 2}}(\t) -h^\infty_{g,{3\over 2}}(\t) -H_{g'}(3\t),
\eea
where $\zeta_g={1\over 2}$ for $g= 6A$ and $\zeta_g = {1\over 5}$ for $g=15AB$. 

It would be very interesting to understand the origin of these properties, and in particular why they behave similarly to the Hauptmoduln of monstrous moonshine for groups with Atkin-Lehner involutions. For example, consider the McKay-Thompson series for conjugacy class $g=3A$ in the monster group, expanded at the infinite cusp
\be
T_{3A}(\t) = {1\over q} + 783 q + 8672 q^2 + 65367 q^3 + \ldots.
\ee
This is a Hauptmodul for the group $\Gamma_0(3)+3$, which is defined in appendix \ref{app:modgrs} and in particular contains the Fricke involution which takes $\tau \mapsto {-{1\over 3\tau}}$. Such an involution relates the cusp at infinity to the cusp at $\tau=0$, and thus these cusps are equivalent with respect to $\Gamma_0(3)+3$. As a result, the expansion of the Hauptmodul at $\tau=0$, which we will denote $T_{3A}^0(\t)$, is given by
\be
T_{3A}^0(\t)=T_{3A}\left (-{1\over 3\t}\right)=q^{-{1\over 3}} + 783 q^{1\over 3} + 8672 q^{2\over 3} + 65367 q + \ldots = T_{3A}\left ({\tau \over 3}\right ).
\ee
The properties we observe for certain $h^{\cN=2}_g$ in equation (\ref{eq:FrickeN=2}) in this section are strikingly similar to this behavior.

\section{Discussion}\label{sec:disc}
In this work we have investigated the Rademacher summability properties of the twining functions of known extremal CFTs. Inspired by the genus zero property of monstrous moonshine, and its connection to the Rademacher summability of the monstrous McKay-Thompson series at the infinite cusp, we consider a similar expansion for the twined graded characters associated with the other extremal CFTs. Similarly to $\mc V^{\natural}$ and $\mc V^{s\natural}$, we find that $\mc E^{\cN=4}_{m=2}(G)$ for $G=M_{22}$ and $M_{11}^{(1)}$ have the property that all associated twining functions can be written as Rademacher sums at the infinite cusp. However, all of the other cases we consider have at least one conjugacy class whose graded character has pole at additional cusps of the corresponding modular group which are inequivalent to infinity. 

In studying the Rademacher properties of $h^{\mc A}_g$ and $\tilde h^{\mc A}_g$, in \S \ref{sec:results} we examined the Rademacher sum of the corresponding polar term for the group $\G_g=\G_0(n)$. However, in the case of $\mc V^\natural$ and $\mc V^{s\natural}$ it is the case that many of the McKay-Thompson series are Hauptmoduln for  subgroups of $SL_2(\mathbb R)$ with additional Atkin-Lehner involutions (c.f. appendix \ref{app:modgrs}). One obvious question is whether those functions which, according to the results of \S \ref{sec:Rad12} and \S\ref{sec:Rad24}, have poles at cusps inequivalent to $i\infty$ under $\Gamma_g$ are nevertheless Rademacher sums at infinity for a different modular group with more general Atkin-Lehner involutions. However, in the case of half-integral index theta functions it does not seem possible to extend the multiplier system to these groups. 

More generally, in the case of $\mc V^\natural$ and $\mc V^{s\natural}$, the Rademacher summability property applies directly to the twined partition functions. In the cases of the other ECFTs we consider, we study the mock modular forms which arise only after decomposing the partition function into superconformal characters. Though these objects are natural to consider both from a physical and algebraic point of view, one could also consider the Rademacher properties of the twined (flavored) partition function itself. In the Spin(7) case, the (twined) partition function is simply the (twined) partition function of $\mc V^{s\natural}$; from this point of view the functions of the Spin(7) theory are naturally related to Rademacher summable functions (though for different modular groups and with different polar structure.)

On the other hand, in the case of the theories with $\cN=2$ and $\cN=4$ supersymmetry, the flavored partition function is a Jacobi form of index $m$ and has a natural theta expansion into length $2m$ vector-valued modular forms of weight $w=-1/2$. However, these functions in general have at least as many poles as the mock modular forms we studied.
\begin{center}\begin{table}[htb]\begin{small}
  \begin{center}
  \def\arraystretch{1.3}
  \begin{tabular}{c|c|c}\toprule
 ESCFT \,&Twining function & Rademacher summable $[g]$\\\midrule
 $\mc V^\natural$ & $T_g(\t)$ &{\bf All ${\bf [g] \in \mathbb M}$} \cite{Duncan:2009sq}\\
  $\mc V^{s\natural}$ & $\mc Z^{s\natural}_{{\rm NS},g}(\t)$& {\bf All ${\bf [g] \in Co_0}$} \cite{DM-C}\\\hline
 $\mc E^{\rm Spin(7)}(M_{24})$ & $h^{\rm Spin(7)}_g(\t)$  &All but 6 $[g] \in M_{24}$\\
 $\mc E^{\mc N=2}_{m=2}(M_{23})$ &$h^{\cN=2}_g(\t)$  &All but 3 $[g] \in M_{23}$\\
  $\mc E^{\mc N=2}_{m=2}(M_{12})$ & $h^{\cN=2}_g(\t)$  &All but 4 $[g] \in M_{12}$\\
 $\mc E^{\mc N=4}_{m=2}(M_{22})$  &$h^{\cN=4}_g(\t)$ & {\bf All ${\bf [g] \in M_{22}}$}\\
  $\mc E^{\mc N=4}_{m=2}(M_{11}^{(1)})$ &$h^{\cN=4}_g(\t)$  & {\bf All ${\bf [g] \in M_{11}^{(1)}}$}\\
   $\mc E^{\mc N=4}_{m=2}(M_{11}^{(2)})$  &$h^{\cN=4}_g(\t)$& All but 1 $[g] \in M_{11}^{(2)}$\\\hline
    $\mc E^{\mc N=2}_{m=4}$ & $\tilde h^{\cN=2}_g(\t)$ &3 classes $[g]\in M_{23}$\\
 $\mc E^{\mc N=4}_{m=4}$  &$\tilde h^{\cN=4}_g(\t)$ & 2 classes $[g]\in M_{11}$ \\\bottomrule
  \end{tabular}\caption{\small Extremal CFTs whose Rademacher summability properties have been proven already ($\mc V^\natural$ and $\mc V^{s\natural}$) or are considered in \S\ref{sec:results}. In the second column we list the twining functions and in the third column we report our main results. 
  }\label{tbl:ECFTresults}
  \end{center}\end{small}
 \end{table}\end{center}

In the Table \ref{tbl:ECFTresults} we summarize the Rademacher summability of the different ECFTs. In particular, in the third column we report the number of conjugacy classes in the global symmetry group of the theory whose corresponding twining function is a Rademacher sum at the infinite cusp. We have indicated in bold those theories for which all twining functions are Rademacher summable in this sense. 

Our work suggests that the surprising connection between ECFTs and sporadic groups is in fact more general than the connection between ECFTs and Rademacher summability.
We end with the following comments and open questions inspired by our work:
\begin{itemize}
\item It would be interesting to understand more deeply the origin of the curious connection described in \S 6.4, which relates the mock modular forms arising from $\mc E^{\cN=2}_{m=2}(M_{23})$ and the twining functions of Mathieu moonshine. More specifically, we have observed a precise relation between the coefficients in the expansion of $h^{\cN=2}_g$ at  two inequivalent cusps of $\Gamma_0(n)$ where it has poles in the case of $g \in \{3A,6A,15AB\}$. Is this indicative of a larger symmetry of these functions?

\item Furthermore, in the case of $c=12$, as explained in \S \ref{sec:morec=12} there exist ECFTs corresponding to all 1-, 2-, and 3-plane preserving subgroups of $Co_0$. We did not analyze all such ECFTs which arise in this way; it would be interesting to study the Rademacher summability properties of the mock modular forms which arise for all corresponding conjugacy classes of $Co_0$ which preserve a 1-, 2-, or 3-plane in the {\bf 24}. We leave such an analysis to future work. 

\item The form of the partition functions for holomorphic orbifolds of the monster CFT turn out to be highly constrained by the Hauptmodul property and the uniqueness conjecture of the vacuum. In fact, given a generic element $g\in \mathbb{M}$, $\langle g \rangle$-orbifold is either $\mathcal{V}^{\natural}$ or $\mathcal{V}^{L}$, as proved in \cite{Tuite1995}. We expect a similar reasoning to hold for the Conway module, whose uniqueness was proven in \cite{Creutzig:2017fuk}, and its relation to the two other $c=12, \cN=1$ SCFTs, the super-$E_8$ theory and the theory of 24 free fermions.

Most of the other examples of $c=12$ ECFTs analyzed here are different from $\mc V^\natural$ and $\mc V^{s\natural}$ in that not all the mock modular forms appearing in the decomposition of the twining functions are Rademacher summable. However, after a preliminary analysis in the case of $\mc E^{\cN=2}_{m=2}(M_{23})$ we found that the holomorphic $\langle g \rangle$-orbifolds for which $g\in M_{23}$ and $\pi_g \in \{1^8\,2^8,\, 1^6\,3^6,\, 1^4\,5^4,\, 1^3\,7^3\}$ reproduce the original partition function \eqref{EGoftheModel}. It would be interesting to explore this further for the other theories considered in this paper.

\item In this work we did not analyze the case of the $\mc E^{\cN=1}_{k^*=2}$, the ECFT with $Co_0$ symmetry first constructed in \cite{DGH}. One question for the future is to derive the corresponding twining functions of this theory and consider whether they have any special Rademacher summability properties at the infinite cusp of the appropriate modular groups.

\item Another example one could consider is K3 non-linear sigma models (NLSM). These theories are extremal according to the definition of \cite{GGKMO}; however, they are not chiral CFTs. Their symmetry groups and possible twined elliptic genera have been classified; they are related to four-plane-preserving subgroups of the group $Co_0$ \cite{Gaberdiel:2011fg,Cheng:2016org}. It would be interesting to consider in general the Rademacher summability properties of all possible twining functions which can arise for K3 NLSMs. In the case where the symmetry element belongs to the Mathieu group $M_{24}$, it follows from \cite{ChengDunc} that these twining functions\footnote{By twining functions here we mean mock modular forms which arise from a decomposition of the form (\ref{decomposition1}).} are Rademacher sums about the infinite cusp. However a general analysis has yet to be performed.  

One interesting point to note is that in \cite{Cheng:2016org} it was conjectured\footnote{In \cite{Paquette:2017gmb} this conjecture was prove in a physical sense via a derivation from string theory.} that all possible twining functions of K3 CFTs arise from either umbral moonshine or Conway moonshine in a precise way proposed in \cite{Cheng:2014zpa} and \cite{Duncan:2015xoa}, respectively. So in this (roundabout) sense they arise from functions which are Rademacher summable at the infinite cusp due to this property of umbral and Conway moonshine.

It is also the case that the elliptic genus of a  $\langle g \rangle$- orbifold of a K3 CFT either reproduces the K3 elliptic genus or the elliptic genus of $T^4$.\footnote{This is just zero due to fermionic zero modes.} One could investigate in this case whether there is a connection between the Rademacher summability of the $g$-twined functions and whether the $\langle g \rangle$- orbifold yields a K3 or a torus theory.

\item The connection between the Rademacher expansion of a CFT partition function and the path integral of 3d quantum gravity in AdS first suggested in \cite{Farey} primarily served as a source of motivation for our analysis. However, in the case of a $g$-twined partition function of a CFT with discrete symmetry group $G$, a physical interpretation of its Rademacher summability at the infinite cusp is lacking. It would be interesting to find a physical interpretation in instances where such a property holds.

\item The authors of \cite{Paquette:2016xoo} considered a certain compactification of heterotic string theory to two dimensions to provide a physical derivation of the genus zero property of monstrous moonshine. The Hauptmodul property of the monstrous McKay-Thompson series was shown to arise from $T$-duality symmetries which arise upon considering CHL orbifolds of this string compactification. An interesting question is whether the additional ECFTs we consider in this work have any connection with 2d string compactifications, and if this point of view can shed any light on the properties considered in this paper.

\item  We can certainly construct infinite families of 2d CFTs with arbitrarily high central charge and sporadic symmetry groups by considering symmetric products of the theories studied in this paper; however, they will no longer be extremal\cite{GGKMO}. Do these symmetric product theories have any special properties? Are there other theories (extremal or not) with large sporadic group symmetry and $c>24$ which don't arise from this symmetric product construction? Assuming a connection between Rademacher sums and sporadic groups, can one use Rademacher summability techniques to search for such CFTs at higher central charge?

\item Finally, given the previous point, we raise the following question: how ubiquitous are 2d CFTs with sporadic group symmetries? Do such theories play a special role in physics, i.e. in 3d gravity and/or string theory?

\end{itemize}

\section*{Acknowledgements}
We thank Miranda Cheng and Daniel Whalen for collaboration at early stages of this project. We thank Miranda Cheng for comments on a draft.  S.H. is supported by the National Science and Engineering Council of Canada.

 \pagebreak

\appendix
\section{Modular definitions}\label{app:funs}
\subsection{Modular groups}\label{app:modgrs}
In this section we introduce different modular groups encountered in the main text.
We denote by $\Gamma_0(N)$ the Hecke congruence subgroup of level $N$,
\be
\Gamma_0(N) = \left\{ \begin{pmatrix} a & b\\ cN&d\end{pmatrix}\in SL_2(\ZZ) \Big\lvert\, \text{det}(\gamma)=1, ~~c\in \mathbb Z \right\}.
\label{eq:gamma0}
\ee
The Atkin-Lehner involution for $\G_0(N)$ is 
\be
W_e =\left\{ \begin{pmatrix} ae & b\\ cN&de\end{pmatrix}\in GL_2(\ZZ) \Big\lvert\, \text{det}(\gamma)=e, \, e||N \right\},
\ee
where $||$ denotes that $e$ is an exact divisor of $N$, i.e. $e$ divides $N$, $e|N$, and $\bigl(e,\frac{N}{e}\bigr)=1$. Moreover the set of matrices $W_{e_i}$ satisfies 
\begin{align}
\label{eqn:clos1}
&W_e^2=1 \,\text{mod}(\G_0(N)),\\\label{eqn:clos2}
&W_{e_1}W_{e_2}=W_{e_2}W_{e_1}=W_{e_3} \,\text{mod}(\G_0(N)), \qquad e_3= \frac{e_1e_2}{(e_1e_2)^2}.
\end{align}
An important example of Atkin-Lehner involution is the so-called Fricke involution $W_N$, which generates the transformation $\tau\rightarrow -1/N\t$.\\
Next we introduce the modular group $\Gamma_0(n|h)$, defined by
\be
\Gamma_0(n|h) = \left\{ \begin{pmatrix} a & \frac{b}{h}\\ cn&d\end{pmatrix} \Big\lvert\, \text{det}(\gamma)=1 \right\}
\label{eq:gamma0nh}
\ee
where $a,b,c,d\in \ZZ$, $h\in\ZZ$, $h^2|N$ and $N=nh$. For $h$ the largest divisor of $24$, $\G_0(n|h)$ is a subgroup of the normalizer group $\mathcal{N}(N)$ (defined below).
The corresponding Atkin-Lehner involution is  
\be
w_e =\left\{ \begin{pmatrix} ae & \frac{b}{h}\\ cN&de\end{pmatrix} \Big\lvert\, \text{det}(\gamma)=e, \, e||\frac{n}{h} \right\};
\ee
this satisfies a closure condition similar to equation \eqref{eqn:clos2} for $W_e$ with respect to to $\Gamma_0(n|h)$ instead of $\G_0(N)$.
The normalizer group of $\Gamma_0 (N)$ in $SL_2(\RR)$ is 
\be\label{eq:normalizer}
\mathcal{N}(N)= \{\rho\in SL_2(\RR)| \rho \Gamma_0 (N) \rho^{-1} = \Gamma_0 (N)\}\, .
\ee
$\mathcal{N}(N)$ is generated by $\G_0(n|h)$ and its Atkin-Lehner involutions. For an explicit description of the normalizer group the reader is referred to \cite{CN}. 

The groups $\G_g$ with $g\in \mathbb M$ are subgroups of $\mathcal{N}(N)$ of the form $\G_0(n|h)+e_1,e_2,..$ where $n=o(g)$ is the order of $g$, $h|24, h|n$ and $N=nh$. Here $\G_0(n|h)+e_1,e_2,..$ stands for the union of a particular set of Atkin-Lehner involutions ($w_{e_1}, w_{e_2},..$) and $\G_0(n|h)$. From this description it is apparent that $\G_g$ is a subgroup of $\mathcal{N}(N)$ and contains $\Gamma_0(N)$. 

Lastly, we define the group $\Gamma_\theta$, whose Hauptmodul is 
\be\label{eqn:Kfct}
K(\t)= \left (\frac{\eta^2(\t)}{\eta\left ({\t\over 2}\right)\eta(2\t)}\right )^{24} = q^{-1/2}+24 + 276q^{1/2}+\ldots\, ,
\ee
\be
\Gamma_\theta = \left\{ \begin{pmatrix} a & b\\ c&d\end{pmatrix}\in SL_2(\ZZ) \Big\lvert\, c-d \equiv a-b \equiv 1\pmod{2} \right\}.
\label{eq:gammatheta}
\ee

\subsection{Modular and Jacobi forms}\label{app:mods}
We start by defining the Dedekind eta function,
\be
\eta(\tau) = q^{1/24} \prod_{n=1}^\infty (1-q^n),
\ee
and the {Jacobi theta functions} $\th_i(\t,z)$ as follows, 
\begin{align}	\th_1(\t,z)
	&= -i q^{1/8} y^{1/2} \prod_{n=1}^\infty (1-q^n) (1-y q^n) (1-y^{-1} q^{n-1})\,,\\
	\th_2(\t,z)
	&=  q^{1/8} y^{1/2} \prod_{n=1}^\infty (1-q^n) (1+y q^n) (1+y^{-1} q^{n-1})\,,\\
	\th_3(\t,z)
	&=  \prod_{n=1}^\infty (1-q^n) (1+y \,q^{n-1/2}) (1+y^{-1} q^{n-1/2})\,,\\
	\th_4(\t,z)
	&=  \prod_{n=1}^\infty (1-q^n) (1-y \,q^{n-1/2}) (1-y^{-1} q^{n-1/2})\,.
\end{align}
A fundamental object in our discussion is the weight $1/2$ index $m$  theta series, whose components are defined by 
\vspace{-0.3mm}
\begin{equation}
\label{eq:app-vartheta}
\theta_{m,r}(\tau,z) = \sum\limits_{\substack{k\in\mathbb{Z}\\ k\equiv r \, (\text{mod}\, 2m)}}\,q^{k^2/4m}\,y^{k} \, ,
\vspace{0.3mm}
\end{equation}
when $m\in \ZZ_{>0}$ and are otherwise given by 
 \be
 \label{eq:app-varthetahalf}
 \th_{m,r}(\t,z) = \sum_{ k = r \, (\text{mod} \,{2m})} \ex(\tfrac{k}{2})\, q^{k^2/4m} y^k,
 \ee
 for half integer-index $m$, and with $2m \in \ZZ_{>0}$ and $r-m\in \ZZ$.
The modular properties of the theta series are dictated by its transformation under the generators of the modular group $SL(2,\mathbb{Z})$
\begin{equation}
\label{eq:sl2zgen}
T = \begin{pmatrix} 1 & 1 \\ 0 & 1 \end{pmatrix} \, , \qquad    S = \begin{pmatrix} 0 & -1 \\ 1 & 0 \end{pmatrix} \, ,
\end{equation} 
and are thus represented by 
\bea
\label{eq:thetatransf}
\vec{\theta}_{m}(\tau+1,z)&=& \rho(T).\, \vec{\theta}_m(\tau,z) \, ,\\
\vec{\theta}_{m}\Bigl(-\frac{1}{\tau},\frac{z}{\tau} \Bigr)&=& e\Bigl( \frac{mz^2}{\tau}\Bigr)\,\sqrt{-i\tau}\,\rho(S).\,\vec{\theta}_m(\tau,z) \, , 
\eea
where the $2m$-dimensional matrices $\rho(S)$ and $\rho(T)$ define its multiplier system. \\
For $m\in \ZZ$, these take the form 
\be
\label{eqn:thetamult}
\rho(T)_{r,r'}=e\biggl( \frac{r^2}{4m}\biggr)\,\delta_{r,r'}\, ,\qquad   \rho(S)_{r,r'}=\frac{1}{\sqrt{2m}}e\biggl(-\frac{r r'}{2m}\biggr)\, ,
\ee
whereas for $m\in \frac{1}{2}\ZZ$,
\begin{equation}
\label{eqn:halfthetamult}
\rho(T)_{r,r'}=e\biggl( \frac{r^2}{4m}\biggr)\,\delta_{r,r'}\, ,\qquad   \rho(S)_{r,r'}=\frac{1}{\sqrt{2m}}e\biggl(-\frac{r r'}{2m}\biggr)e\biggl(\frac{r- r'}{2}\biggr)\, .
\end{equation}

\subsection{Mock and meromorphic Jacobi forms}\label{app:meroms}
The first instance of mock Jacobi form we consider is the so-called Appell--Lerch sum, defined as
\be\label{eq:AL}
f_u^{(m)}(\tau,z)= \sum_{k\in \mathbb Z} \frac{q^{mk^2} y^{2mk}}{1- yq^k e^{-2\pi iu}}.
\ee
Its completion, following \cite{Zwegers2008}, is
\be\label{def:completion_f}
\hat f_u^{(m)}(\tau,\bar \tau,z)  = f_u^{(m)}(\tau,z)-{1\over2}\sum_{r\in\ZZ/\mathbb Z}R_{m,r}(\tau,u)\theta_{m,r}(\tau,z) 
\ee
with 
\begin{align}\notag
&R_{m,r}(\tau,u)\,=\sum_{k\equiv r\,(\text{mod}\, {2m})}\left({\rm sgn}(k+\tfrac{1}{2}) -E\biggl(\sqrt{{\rm Im}\t\over m} \Bigl(k+2m\frac{{\rm Im}u}{{\rm Im}\t}\Bigr)\biggr) \right)\,
 q^{-{k^2\over 4m}}e^{-2\pi i k u}, \\ \notag
&E(z) \,=\, {\rm sgn}(z) \left(1-\int_{z^2}^\infty dt\,t^{-1/2} \,e^{-\pi t}\right)\,.
\end{align}
Moreover, $\hat f_u^{(m)}(\tau,\bar \tau,z)$ transforms as a (non-holomorphic) Jacobi form of weight 1 and index $m$. 

We denote by $\m_{m;0}(\t,z) =f_0^{(m)}(\t,-z)-f_0^{(m)}(\t,z)$. This specialization of the Appell--Lerch sum has the following relation to the modular group $\SL_2(\ZZ)$: let the (non-holomorphic) completion of $ \m_{m;0} (\t,z)$ be
\be\label{pole_completion}
\hat \m_{m;0} (\t,\bar \t,z) =  \m_{m;0} (\t,z) -\frac{1}{\sqrt{2m}}  \sum_{r \in \ZZ/2m\ZZ} \th_{m,r}(\t,z)  \int^{i\inf}_{-\bar \t}  \bigl(i(\t'+\t)\bigr)^{-1/2} \,\overline{S_{m,r}(-\bar \t')} \, {\rm d}\t'~.
\ee
Then $\hat \m_{m;0}$ transforms like a Jacobi form of weight $1$ and index $m$ for $\SL_2(\ZZ)\ltimes \ZZ^2$ and it has a simple pole at $z=0$.   Here
$S_m = (S_{m,r})$ is the vector-valued cusp form for $\SL_2(\ZZ)$ whose components are given by the unary theta series
\be
\label{eq:app-shadow}
S_{m,r}(\t)   = \frac{1}{2\p i}\frac{\pa}{\pa z} \th_{m,r}(\t,z)\lvert_{z=0}. 
\ee
Note that the explicit form of the theta series $S_{m,r}(\t)$ changes depending on whether $m$ is integer or half-integer because of equations \eqref{eq:app-vartheta}, \eqref{eq:app-varthetahalf}.

For later use, we define two weight one meromorphic Jacobi forms, $\Psi_{1,1}$ of index one, defined as
\be\label{CoverA}
\Psi_{1,1}(\tau,z) = -i \,\frac{\theta_1(\tau,2z)\, \eta(\tau)^3}{(\theta_1(\tau,z))^2}= \frac{y+1}{y-1}- (y^2-y^{-2})q+ \cdots,
\ee
and $\Psi_{1,-\frac{1}{2}}$ of index $-{1\over 2}$, defined as
$$\Psi_{1,-\frac{1}{2}}(\t,z) = -i \,\frac{\eta(\tau)^3}{\theta_1(\tau,z)}= \frac{1}{y^{1/2}-y^{-1/2}} + q\,(y^{1/2}-y^{-1/2}) + O(q^2) .
$$

\section{Superconformal characters and modules}\label{app:B}
In this appendix we review the representation theory and character formulas of the $\cN=4,~ \cN=2, $ and Spin(7) SCAs. 

\subsection{Characters of the Spin(7) algebra}\label{app:spin7char}
Here we briefly review the representation theory of the $\mathcal{SW}(3/2,2)$ superconformal algebra with central charge 12---this is the algebra which arises on the worldsheet of type II string theory compactified on a manifold of Spin(7) holonomy \cite{ShVafa}.\footnote{Throughout this section, we follow the notation used in \cite{exceptional}.} This algebra is an extension of the $c=12$ $\mathcal N=1$ SCA by two additional generators: the stress-energy tensor of a $c=1/2$ Ising model (of dimension 2) and its superpartner (of dimension 5/2).

In \cite{GepnerNoyvert} the unitary representations of the $\mathcal{SW}(3/2,2)$ SCA were classified. There are two algebras---NS and R---which correspond to whether the fermions are 1/2-integer (NS) or integer (R) graded. For our purposes it suffices to work in the NS sector; here the representations are uniquely specified by two quantum numbers and will be labelled $|a,h\rangle$, where $a$ is the dimension of the internal Ising factor, $a \in \{0, 1/16,1/2\}$, and the total dimension is $h$. The result is that there are three massless (BPS) representations with quantum numbers $|0,0\rangle$, $|1/16, 1/2\rangle$, and $|1/2,1\rangle$, and two continuous families of massive (non-BPS) representations with quantum numbers $|0,n\rangle$, and $|1/16, 1/2+n\rangle$, where $n \geq 1/2$.

Conjectural characters for each of these representations were computed in \cite{spin7}, to which we refer for more details and derivations, including a discussion of the characters in the Ramond sector. 
We define the following combination of functions from \eqref{eq:app-vartheta},
\be
\til \th_{m,r} (\t)= \th_{m,r} (\t) + \th_{m,r-m} (\t)  
\ee
which satisfies $\til \th_{m,r}=\til \th_{m,-r}=\til \th_{m,r+m}$, $\til \th_{m,r}(\tau) = \theta_{m/2,r}(\tau/2)$, and
\be
\tilde f_u^{(m)}(\tau,z)=f_u^{(m)}(\tau,z)-f_u^{(m)}(\tau,-z).
\ee
We denote the character of the a non-BPS representation $|a,h\rangle$ by $\chi^{NS}_{a,h}(\t)$, and the characters of the BPS representations by $\tilde{\chi}^{NS}_{a}(\t)$, as they are uniquely specified by their $a$ eigenvalue. The result is that the non-BPS characters are given by
\begin{align}
\chi^{NS}_{0,h}(\t)&=q^{h-{49\over 120}}\, \mc P(\t)\, \Theta^{NS}_{0}(\t)
= q^{h}\, (q^{-1/2}+1 +q^{1/2}+3\,q +\dots)
\end{align}
and
\be
\chi^{NS}_{{1\over 16},h}(\t)=q^{h-{61\over 120}}\, \mc P(\t)\, \Theta^{NS}_{1\over16}(\t)
=  q^{h} \,(q^{-1/2}+2 +3\,q^{1/2}+5\,q +\dots)
\ee
where 
$$\mc P(\t)=\frac{\eta^2(\tau)}{\eta^2 (\tfrac{\t}{2}) \eta^2(2\tau)}\,,$$
and we have defined
\begin{align}
\Theta^{NS}_{0}(\t)&= \Big(\til\th_{30,2}(\t)-\til\th_{30,8}(\t)\Big) \,,\\
\Theta^{NS}_{{1\over 16}}(\t)&=  \Big(\til\th_{30,4}(\t)-\til\th_{30,14}(\t)\Big) \,.
\end{align}
Furthermore, the BPS character of total dimension $h={1\over 2}$ is given by 
\be
\tilde{\chi}^{NS}_{1\over2}(\t)=  \mc P(\t)\,\mu^{NS}(\tau)~,
\ee 
where,
\be\label{def:mu}
\mu^{NS}(\tau)= \big( q^{5\over 8} 
\tilde f^{(5)}_{{\tau\over 2} + {1\over 2}}(6\tau,\tau)+q^{25\over 8}\tilde f^{(5)}_{{\tau\over 2} + {1\over 2}}(6\tau,-2\tau)\big),
\ee
and the other two BPS characters can be found using the BPS relations which relate massless and massive characters:
\be\label{rel_massless_massive1}
\tilde{\chi}^{NS}_0 + \tilde{\chi}^{NS}_{1\over16}= q^{-n}\chi^{NS}_{0,n}\, ,\qquad \qquad
\tilde{\chi}^{NS}_{1\over16} + \tilde{\chi}^{NS}_{1\over2}= q^{-n}\chi^{NS}_{{1\over16},{1\over2}+n}\, .
\ee

\subsubsection*{Spin(7) modules}
The partition function for a module of the Spin(7) superconformal algebra, i.e.
\be
\mc Z^{\rm Spin(7)}_{\rm NS}(\t)=\Tr_{NS} \,q^{L_0-c/24},
\ee
 transforms as a weight zero modular function for the congruence subgroup $\Gamma_\theta$. Furthermore, it follows from the explicit description of the Spin(7) characters above that such a function admits an expansion of the form
 \be
\mc Z^{\rm Spin(7)}_{\rm NS}(\t)=\mc P(\t)\left (A_0 \mu^{NS}(\tau)+F_{1\over 16}(\t)\Theta^{NS}_{1\over 16}(\t)+F_{0}(\t)\Theta^{NS}_{0}(\t)\right )
  \ee
  where we can expand the function $(F_j)$ as
  \bea
    F_{1\over 16}(\t)&=\sum_{n \geq 0} b_j(n) q^{n - 1/120}\\
  F_{0}(\t)&=\sum_{n \geq 0} c_j(n) q^{n - 49/120}.
  \eea
From the properties of the Appell--Lerch sums detailed in appendix \ref{app:meroms}, it follows that $\underline F:=(F_j)$ is a weight 1/2 vector-valued mock modular form for $SL_2(\mathbb Z)$ with shadow given by $A_0 \underline{\tilde S}(\t)$,
where we have defined
\begin{align} \label{eq:spin7shadow}
 \underline{\tilde S}=\bem S_1 \\ S_7\eem
 \end{align}
and
 $\tilde S_\alpha(\tau)= \sum_{k\in\mathbb Z}  k \epsilon^{R}_{\alpha}(k) q^{k^2/120} 
 $ for $\alpha=1,7$ and 
 \begin{align}
 \epsilon^{R}_{1}(k)& = \begin{cases}1&k= 1,29\!\!\! \pmod{60}\\-1&k= -11,-19\!\!\! \pmod{60}\\0&\text{otherwise}\end{cases}\\
\epsilon^{R}_{7}(k)& = \begin{cases}1&k= -7, -23\!\!\! \pmod{60}\\-1&k= 17,13\!\!\! \pmod{60}\\0&\text{otherwise}\end{cases}.
 \end{align}  
 See \cite{exceptional} for more details.

\subsection{$\mathcal N=2$ superconformal characters}\label{sec:N=2chars}
The $\cN=2$ SCA with central charge $c=3(2\ll+1)= 3\hat c$, $\ll \in {1\over 2} \ZZ$, contains an affine $\widehat{u(1)}$ current algebra of level $\ll +{1\over 2}$. In this notation $m=\ell + {1\over 2}.$ The unitary irreducible highest weight representations are labeled by the eigenvalues of $L_0$ and $J_0$, which we call $h$ and $Q$, respectively \cite{Dobrev:1986hq,Kiritsis:1986rv}, and which we denote by $\mathcal{V}^{\cN=2}_{\ll;h,Q}$. There are $2\ll+1$ massless (BPS) representations with eigenvalues $h=\frac{c}{24} = \frac{\hat c}{8}$ and $Q\in\{ -\frac{\hat c}{2}+1,-\frac{\hat c}{2}+2,\dots,\frac{\hat c}{2}-1,\frac{\hat c}{2}\}$, whereas there are $2\ll +1$ continuous families of massive (non-BPS) representations with eigenvalues $h > \frac{\hat c}{8}$ and $Q\in\{ -\frac{\hat c}{2}+1,-\frac{\hat c}{2}+2,\dots,\frac{\hat c}{2}-2,\frac{\hat c}{2}-1,\frac{\hat c}{2}\}$, $Q\neq 0$.

We focus on the graded characters in the Ramond sector, which are defined as
\be
{\rm ch}^{\cN=2}_{\ll;h,Q}(\t,z) = \tr_{\mathcal{V}^{\cN=2}_{\ll;h,Q}} \left( (-1)^{J_0}y^{J_0} q^{L_0-c/24}\right).
\ee
In terms of functions in appendix \ref{app:funs}, the massive characters are
\be
{\rm ch}^{\cN=2}_{\ll;h,Q}(\t,z) = \ex(\tfrac{\ll}{2}) (\Psi_{1,-\frac{1}{2}} (\tau,z))^{-1} q^{h-\frac{c}{24}-\frac{j^2}{4\ell}} \th_{\ll,j}(\t,z) \quad,\quad j = {\rm sgn}(Q)\,(|Q|-1/2)\,,
\ee
and the massless ones (with $Q \neq\frac{\hat c}{2}$) are
\be
{\rm ch}^{\cN=2}_{\ll;c/24,Q}(\t,z) = \ex(\tfrac{\ll+Q+1/2}{2})(\Psi_{1,-\frac{1}{2}} (\t,z) )^{-1}\,y^{Q+\frac{1}{2}} \, f^{(\ll)}_{u}(\t,z+u)\quad,\quad u =\tfrac{1}{2}+\tfrac{(1+2Q)\t}{4\ll}\, .
\ee
Furthermore,  the character ${\rm ch}^{\cN=2}_{\ll;c/24,Q}(\t,z) $ for $Q =\frac{\hat c}{2}$ can be determined by the relation
\begin{align}\notag
{\rm ch}^{{\cal N}=2}_{\ll;c/24,\frac{\hat c}{2}}  & =q^{-n}\Big( {\rm ch}^{{\cal N}=2}_{\ll;n+ c/24,\frac{\hat c}{2}}+ \sum_{k=1}^{\frac{\hat c}{2}-1} (-1)^k \big(
{\rm ch}^{{\cal N}=2}_{\ll,n+c/24,\frac{\hat c}{2}-k} +{\rm ch}^{{\cal N}=2}_{\ll,n+c/24,k-\frac{\hat c}{2}} \big) \Big) \\\label{Nis2_relation_long_short_2}& \;\;+  (-1)^{\frac{\hat c}{2}} {\rm ch}^{{\cal N}=2}_{\ll;c/24,0}\,,
\end{align}
due to the fact that at the unitary bound, several BPS multiplets can combine into a non-BPS multiplet.

\subsubsection*{$\cN=2$ modules}
The graded partition function of a module for the $c= 6m$ $\mathcal N=2$ superconformal algebra in the Ramond sector, i.e.
\be
{\cal Z}^{\cN=2}_{m}(\t,z) = \Tr_{\rm R}\left ( (-1)^{J_0} y^{J_0} q^{L_0-c/24}\right ),
\ee
transforms as a weak Jacobi form of weight zero and index $m$ for $SL_2(\mathbb Z)$ as in the $\mathcal N=4$ case.
Furthermore, from the representation theory discussed above we expect such a partition function to have an expansion
 \be
 {\cal Z}^{\cN=2}_{m}(\t,z) = \ex(\tfrac{\ll}{2}) (\Psi_{1,-\frac{1}{2}})^{-1} \left(C_0\,\til \m_{\ll;0} (\t,z) + \sum_{j -\ll \in \ZZ/2\ll\ZZ} {\til F}^{(\ll)}_{j}(\t) \th_{\ll,j}(\t,z) \right) 
\ee
when the  ${\cal N}=2$ SCA has even central charge, $c=3(2\ll+1)$. (See \cite{M5} for more details.)
In the last equation, we have defined
\[
\til \m_{\ll;0}  = \ex(\tfrac{1}{4})\, y^{1/2} f^{(\ll)}_u(\t,u+z) \quad,\quad u = \frac{1}{2}+  \frac{\t}{4\ll}\,,
\]
and the function ${\til F}^{(\ll)}_{j}(\t)$ satisfies
\be
{\til F}^{(\ll)}_{j}(\t)  ={\til F}^{(\ll)}_{-j}(\t) ={\til F}^{(\ll)}_{j+2\ll}(\t).
\ee

Through its relation to the Appell--Lerch sum,  $\til \m_{\ll;0} $ admits a completion which transforms as  a weight one, half-integral index Jacobi form under the Jacobi group.  Defining  $\widehat{ \til \m_{\ll;0} }$ by replacing $\m_{m;0} $ with $ \til \m_{\ll;0}$ and the integer $m$ with the half-integral $\ll$ in \eq{pole_completion}, we see that $\widehat{ \til \m_{\ll;0} }$ transforms like a Jacobi form of weight $1$ and index $\ll$ under the group $\SL_2(\ZZ)\ltimes \ZZ^2$.
Following the same computation as in the previous section, we hence conclude that ${\til F}^{(\ll)}=({\til F}^{(\ll)}_j)$, where $j-1/2\in \ZZ/2\ll\ZZ$, is a vector-valued mock modular form with a vector-valued shadow $C_0\,  S_\ll = C_0(S_{\ll,j}(\t))$. 

\subsection{$\mathcal N=4$ superconformal characters}\label{sec:N=4chars}
Let $m=\tilde m-1.$The $\cN=4$ SCA with central charge $c=6(\tilde m-1)$, $\tilde m>1$, contains a level $\tilde m-1$ $\widehat{su(2)}$ current algebra (cf. \cite{Eguchi1987}). We will label the unitary irreducible highest weight representations by the eigenvalues of $L_0$ and ${1\over 2} J_0^3$, which we denote by $h$ and $j$, respectively. We discuss representations in the Ramond sector, where a representation with quantum numbers $(h,j)$ will be denoted $\mathcal{V}^{\cN=4}_{m;h,j}$. There are two types of representations: a discrete set of $\tilde m$ massless (BPS) representations, and $\tilde m-1$ continuous families of massive (non-BPS) representations (c.f. \cite{Eguchi1988}.)

The BPS representations have $h={c\over 24}= {\tilde m-1\over 4}$ and $j\in \{0, {1\over 2},\ldots,{\tilde m-1\over 2}\}$, and the non-BPS represenations have $h>{\tilde m-1\over 4}$ and $j\in \{{1\over 2},1, \ldots, {\tilde m-1\over 2}\}$. Their graded characters, defined as
\be
{\rm ch}^{\cN=4}_{m;h,j}(\t,z) = \Tr_{\mathcal{V}^{\cN=4}_{m;h,j}} \left( (-1)^{J_0^3}y^{J_0^3} q^{L_0-c/24}\right),
\ee
were computed in \cite{Eguchi1988a} and can be written in terms of functions defined in appendix \ref{app:funs} as
\be \label{masslesschar}
	{\rm ch}^{\cN=4}_{m;h,j}(\t,z)=
	(\Psi_{1,1}(\tau,z))^{-1}  \m_{\tilde m;j}  (\t,z)
\ee
and
\be \label{massivechar}
{\rm ch}^{\cN=4}_{m;h,j}(\t,z) =
	(\Psi_{1,1}(\tau,z))^{-1} \,q^{h-\frac{c}{24}-\frac{j^2}{\tilde m}} \,  \big(\th_{\tilde m,2j} (\t,z)-\th_{\tilde m,-2j} (\t,z)\big)
\ee
in the massless and massive cases, respectively.

\subsubsection*{$\mathcal N=4$ modules}
 The graded partition function of a module for the $c=6(\tilde m-1)$ ${\cal N}=4$ SCA in the Ramond sector, i.e.
\be
{\cal Z}^{\cN=4}_{m}(\t,z) = \Tr_{\rm R}\left ( (-1)^{J_0^3} y^{J_0^3} q^{L_0-c/24}\right ),
\ee
transforms as a weak Jacobi form of weight zero and index $m$ for $SL_2(\mathbb Z)$. Moreover,
the representation theory of the $\cN=4$ SCA discussed above and the explicit description
of the $\mu$ and $\theta$ functions in appendix \ref{app:funs}
allows one to rewrite the graded partition function as
\begin{align}\label{decomposition1}
{\cal Z}^{\cN=4}_{m}(\t,z) =(\Psi_{1,1}(\tau,z))^{-1} \left( c_0 \, \m_{\tilde m;0}(\t,z) +\sum_{r\in \ZZ/2\tilde m\ZZ}   F^{(\tilde m)}_r(\t)\,\th_{\tilde m,r}(\t,z) \right),
\end{align}
where
the $F^{(\tilde m)}=(F^{(\tilde m)}_r)$, $r\in \ZZ/2\tilde m\ZZ$ obey
\be F^{(\tilde m)}_r(\t) = -F^{(\tilde m)}_{-r}(\t) = F^{(\tilde m)}_{r+2\tilde m}(\t).
\ee
See, for example, \cite{M5}.

The way in which the functions ${\cal Z}^{\cN=4}_m(\t,z) $ and $\hat \m_{\tilde m;0}$ transform under the Jacobi group shows that the non-holomorphic function
\(
\sum_{r\in \ZZ/2\tilde m\ZZ}  \hat F^{(\tilde m)}_r(\t) \, \th_{\tilde m,r}(\t,z)
\)
transforms as a Jacobi form of weight $1$ and index $\tilde m$ under $\SL_2(\ZZ)\ltimes \ZZ^2$, where
\[
 \hat F^{(\tilde m)}_r(\t) =   F^{(\tilde m)}_r(\t)+ c_0 \ex(-\tfrac{1}{8}) \,\frac{ 1}{\sqrt{2\tilde m}}   \int^{i\inf}_{-\bar \t}  (\t'+\t)^{-1/2} \overline{S_{\tilde m,r}(-\bar \t')} \, {\rm d}\t' .
\]
In other words,
$F^{(\tilde m)} = (F^{(\tilde m)}_r)$, $r\in \ZZ/2\tilde m\ZZ$ is a vector-valued mock modular form with a vector-valued shadow $c_0\,  S_{\tilde m}$,
whose $r$-th component is given by $S_{\tilde m,r}(\t)$, with the multiplier for $SL_2(\ZZ)$ given by the inverse of the multiplier system of $ S_{\tilde m}$ (cf. \eq{eqn:thetamult}).

 \section{Twining functions}\label{app:twin}
 In this section we derive the twined partition functions of $\mathcal E^{\cN=2}_{m=4}$ for all conjugacy classes $[g]\in M_{23}$, and we discuss a few such cases for $\mathcal E^{\cN=4}_{m=4}$ and $[g]\in M_{11}$. The approach is similar, so we discuss the two cases in parallel, pointing out distinctions when they occur.
 
 The starting point for each is the Niemeier CFT with target $\mathbb R^{24}/\Lambda_N$, for $N=A_1^{24}$ and $A_2^{12}$ in the $\mathcal N=2$ and $\cN=4$ cases, respectively. 
The partition function of the CFT consists primary states coming from lattice vectors, primary states coming from the 24 currents $i \partial x_i$, and the Virasoro descendents; i.e. it is just given by equation (\ref{eq:NiemZ}). 
For $\Lambda=A_1^{24}$, it will be useful to think of this CFT in the following way. The $i$th copy of the $A_1$ root system furnishes an affine $\widehat{su(2)_1}$ current algebra, generated by the vertex operators
\be
e^{\pm i \sqrt 2 x_i}, ~~i \partial x_i,
\ee
and therefore the partition function $\mc Z^{\Lambda_N}(\t)$ of the theory has a natural decomposition into characters of $\left ( \widehat{su(2)}_1\right )^{24}$. 

There are two irreducible modules of $\widehat{su(2)}_1$---one arising from the vacuum representation, which has a ground state of conformal weight zero, and a second from a highest weight state of conformal weight ${1\over 4}$. We will denote these representations as $[0]$ and $[1]$, respectively.
The characters of these irreducible modules are given by
\bea
\text{ch}_0(\t)&=& \Tr_{[0]}q^{L_0-c/24} = {\theta_3(2\t)\over \eta(\t)},\\
\text{ch}_1(\t)&=& \Tr_{[1]}q^{L_0-c/24} = {\theta_2(2\t)\over \eta(\t)},
\eea
where $c=1$ is the Sugawara central charge of the current algebra.
The full lattice theta function for $\Lambda_N$ with $N=A_1^{24}$ consists of all lattice vectors which are linear combinations of root vectors and glue vectors. The glue vectors can be specified in terms of elements of the extended binary Golay code. This is a length-24 binary code with weight enumerator
\be
x^{24} + 759 x^{16}y^8 + 2576 x^{12}y^{12} + 759 x^8y^{16} + y^{24},
\ee
where the coefficient of the term $x^ny^m$ gives the number of vectors in the code with $n$ zeros and $m$ ones.
Thus we can rewrite the partition function in terms of $\widehat{su(2)}_1$ characters as
\be
\mc Z^{\Lambda_N}(\t)=\text{ch}_0^{24}(\t) + 759 (\text{ch}_0^{16}(\t)\text{ch}_1^8(\t) +  \text{ch}_0^8(\t)\text{ch}_1^{16}(\t) )+ 2576 \text{ch}_0^{12}(\t)\text{ch}_1^{12}(\t) + \text{ch}_1^{24}(\t)
\ee
for $N=A_1^{24}$.

Similarly, in the case of $N=A_2^{12}$, the partition function can naturally be written in terms of characters of $\left ( \widehat{su(3)}_1\right )^{12}$, as each of the 12 $A_2$ root systems furnishes a copy of $\widehat{su(3)}_1$. There are three irreducible $\widehat{su(3)}_1$ modules we will call [i] and with characters we refer to as $\chi_i(\t)$, $i=0,1,2.$ The vacuum module $[0]$ has conformal dimension $h=0$ and the two nontrivial primaries both have conformal dimension $h={1\over 3}$. Furthermore, the glue vectors are now specified in terms of elements of the extended ternary Golay code, which is a length-12 ternary code with weight enumerator
\be
x^{12} + y^{12} + z^{12} + 22(x^6y^6+ y^6z^6+z^6x^6) + 220(x^6y^3z^3+y^6z^3x^3+z^6x^3y^3).
\ee
Therefore, for $N=A_2^{12}$ we can write the partition function in terms of $\widehat{su(3)}_1$ characters by replacing $x,y,z$ in the weight enumerator with $\chi_0,\chi_1,\chi_2$, respectively.
Furthermore, the formulas for these characters are given by
 \be
 \chi_0(\t)=\Tr_{[0]}q^{L_0-c/24}={\theta_3(2\t)\theta_3(6\t) + \theta_2(2\t)\theta_2(6\t)\over \eta^2(\t)}
 \ee
 for the vacuum character, and
 \be
 \chi_i(\t)=\Tr_{[i]}q^{L_0-c/24}={\theta_3(2\t)\theta_3\left({2\t\over 3}\right) + \theta_2(2\t)\theta_2\left({2\t\over 3}\right)\over 2\eta^2(\t)}-{\chi_0(\t)\over 2}
  \ee
for the nontrivial primaries with $i=1,2$, where in this case $c=2$.  Thus the partition function of the theory can be written as
 \be
\mc Z^{\Lambda_N}(\t)=\chi_0^{12}(\t) + 264 \chi_0^{6}(\t)\chi_1^6(\t) +  440\chi_0^3(\t)\chi_{1}^{9}(\t) + 24 \chi_{1}^{12}(\t)
\ee
for $N=A_2^{12}$.

Now we consider a $\mathbb Z_2$ orbifold of the above theories. We will call the $\mathbb Z_2$ symmetry $h$, which acts with a minus sign on the 24 coordinates of the torus:
\be
h: x_i \to -x_i.
\ee
In the case of $N=A_1^{24}$, the orbifold preserves a $\left (\widehat{u(1)}_4\right)^{24}$ current algebra out of the $\left (\widehat{su(2)}_1\right )^{24}$ and, similarly, for $N=A_2^{12}$, the orbifold preserves an $\left (\widehat{su(2)}_4\right)^{12}$ within the $\left (\widehat{su(3)}_1\right )^{12}$. We choose one copy of $\widehat{u(1)}_4$ and $\widehat{su(2)}_4$ to generate the R-symmetry of the $\mathcal N=2$ and $\cN=4$ algebras, respectively.
  In the NS sector the corresponding level-4 current 
is given by the $h$-invariant linear combination
\be\label{eq:u(1)}
J_0=2\left (e^{i \sqrt 2x_1}+e^{-i\sqrt 2 x_1}\right )
\ee
for the $\mathcal N=2$ case and by
\be\label{eq:u(1)b}
J_3=\sqrt 2\left (e^{i \sqrt 2x_1}+e^{-i\sqrt 2 x_1}\right )
\ee
for the $\mathcal N=4$ case \cite{extN2,extN4}.

We consider the Ramond sector partition function graded by these currents and by $(-1)^F$. The Hilbert space is composed of  the anti-invariant states in the untwisted sector and the  invariant states in the twisted sector.  Thus we will compute the trace
\bea
\mc Z_{m=4}^{\mc N=2(4)}(\t,z)&=& \Tr_{\mathcal H_R} (-1)^F q^{L_0-{c \over 24}}y^{J_0(J_3)}\\\nonumber
&=& \Tr_{\mathcal H} \left ( {1-h \over 2}\right ) (-1)^F q^{L_0-{c \over 24}}y^{J_0(J_3)}+\Tr_{\mathcal H^{tw}} \left ( {1+h \over 2}\right ) (-1)^F q^{L_0-{c \over 24}}y^{J_0(J_3)}
\eea
where the first term in the second line implements a projection onto the anti-invariant states in the untwisted sector Hilbert space, $\mathcal H$, and the second term a projection onto the invariant states in the twisted sector Hilbert space, $\mathcal H^{tw}$. Furthermore, we will also consider the twining functions
\be
\mc Z_{m=4,g}^{\mc N=2(4)}(\t,z)= \Tr_{\mathcal H_R} g(-1)^F q^{L_0-{c \over 24}}y^{J_0(J_3)}
\ee
defined for $g\in M_{23}, M_{11}$, respectively. These functions are weak Jacobi forms of weight zero and index four for $\Gamma_0(n)$ where $n=o(g)$. Let
\be\label{eq:untwistZ}
F^{un.}_g(\Lambda;\tau,z):=\Tr_{\mathcal H} \left ( {1-h \over 2}\right ) g(-1)^F q^{L_0-{c \over 24}}y^{J_0(J_3)}=\Tr_{\mathcal H} \left ( {1-h \over 2}\right ) g q^{L_0-{c \over 24}}y^{J_0(J_3)}
\ee
be the $g$-twined trace which is the contribution of the untwisted sector to the partition function $\mc Z_{m=4}^{\mc N=2(4)}(\t,z)$, and 
\be\label{eq:twistZ}
F^{tw}_g(\Lambda;\t,z):=\Tr_{\mathcal H^{tw}} \left ( {1+h \over 2}\right )g (-1)^F q^{L_0-{c \over 24}}y^{J_0(J_3)}
\ee
be the corresponding $g$-twined contribution of the twisted sector. Note that all states in the Hilbert space $\mc H^{un.}$ are bosonic so we can drop the $(-1)^F$ in (\ref{eq:untwistZ}). We will discuss the explicit computation of each of these terms in the next two subsections.

\subsection*{The untwisted sector}
We start with $\Lambda=A_1^{24}$. To implement the trace in the untwisted sector, we need to know the action of $h$ on the $ \widehat{su(2)}_1$ modules, as well as their characters with the $U(1)$ charge included, which we will denote by $\text{ch}_0(\t,z)$ and $\text{ch}_1(\t,z)$. It is straightforward to see that these are given by
\bea
\text{ch}_0(\t,z)=\Tr_{[0]}q^{L_0-c/24}y^{J_0}&=& {\theta_3(2\t,4z)\over \eta(\t)}:=\pfac{z}{},~~~\rm{and}\\
\text{ch}_1(\t,z)=\Tr_{[1]}q^{L_0-c/24}y^{J_0}&=& {\theta_2(2\t,4z)\over \eta(\t)}:=\gfac{z}{},
\eea
where $J_0$ is the zero mode of the $U(1)$ current in equation (\ref{eq:u(1)}). Furthermore, using the explicit description of $h$, it is easy to check that the characters with an $h$-insertion are given by
\bea
\text{ch}^-_0(\t,z)=\Tr_{[0]}hq^{L_0-c/24}y^{J_0}&=& {\theta_4(2\t,4z)\over \eta(\t)}:=\mfac{z}{},~~~\rm{and}\\
\text{ch}^-_1(\t,z)=\Tr_{[1]}hq^{L_0-c/24}y^{J_0}&=&0.
\eea
In order to write the (twined) partition function in terms of these characters, we introduce the following notation,
\bea
\pfac{n}{m}&:=&\text{ch}_0(n\t,0)^m\\\nonumber
\mfac{n}{m}&:=&\text{ch}^-_0(n\t,0)^m\\\nonumber
\gfac{n}{m}&:=&\text{ch}_1(n\t,0)^m.
\eea 
Given this we can evaluate the trace in equation (\ref{eq:untwistZ}) with $g=1$ to compute the contribution of the untwisted states to the Ramond sector partition, which is
 \bea\nonumber
F^{un.}(\Lambda;\tau,z)&=&{ 1\over 2}\left (\pfac{z}{}\pfac{1}{23}-\mfac{z}{}\mfac{1}{23}\right )+{253\over 2}\left (\pfac{z}{}\pfac{1}{7}\gfac{1}{16}+\gfac{z}{}\gfac{1}{7}\pfac{1}{16}\right )\\\nonumber
&+&253\left (\pfac{z}{}\pfac{1}{15}\gfac{1}{8}+\gfac{z}{}\gfac{1}{15}\pfac{1}{8}\right )+644\left (\pfac{z}{}\pfac{1}{11}\gfac{1}{12}+\gfac{z}{}\gfac{1}{11}\pfac{1}{12}\right )\\
&+&{1\over 2}\gfac{z}{}\gfac{1}{23},
 \eea
 where we note that all of the untwisted states are bosonic and thus invariant under $(-1)^F$.
 
 Furthermore, we can compute the $g$-twined trace of equation (\ref{eq:untwistZ}) using an explicit description of the action of $M_{23}$ on the binary Golay code, which we obtain from GAP.\footnote{This open source program lives at https://www.gap-system.org/.} From this we compute the invariant vectors of the Golay code under the 24-dimensional permutation representation of $g$. The results for all conjugacy classes $g$ in $M_{23}$ are given in Table \ref{tbl:A1twin}.

\begin{center}
\begin{table}[htbp]
\begin{center}
\small
{\renewcommand{\arraystretch}{1.5}
\begin{tabular}{c|c|c}\toprule
$M_{23}$ $[g]$ & Frame shape & $F^{un.}_g(\Lambda;\tau,z)$\\\midrule
2A & $1^{8}2^8$ &    \begin{tabular}{@{}c@{}}${1\over 2}\left (\pfac{z}{}\pfac{1}{7}- \mfac{z}{}\mfac{1}{7}+\gfac{z}{}\gfac{1}{7} \right )\left(\pfac{2}{8}+ \gfac{2}{8}\right )$\\
$+7\left (\pfac{z}{}\pfac{1}{7}- \mfac{z}{}\mfac{1}{7}+\gfac{z}{}\gfac{1}{7} \right )\pfac{2}{4} \gfac{2}{4}$\\
$+14 \left (\pfac{z}{}\pfac{1}{3}\gfac{1}{4}+\gfac{z}{}\gfac{1}{3}\pfac{1}{4} \right )\pfac{2}{2}\gfac{2}{2}\left(\pfac{2}{2}+ \gfac{2}{2}\right )^2$\end{tabular}\\\hline
3A & $1^{6}3^6$ &    \begin{tabular}{@{}c@{}}${1\over 2}\left (\pfac{z}{}\pfac{1}{5}\pfac{3}{6}-\mfac{z}{}\mfac{1}{5}\mfac{3}{6}\right)+{1\over 2}\gfac{z}{}\gfac{1}{5}\gfac{3}{6}$\\
$+ {1\over 2}\left(\pfac{1}{5}\pfac{3}{}\gfac{z}{}\gfac{3}{5}+\gfac{1}{5}\gfac{3}{}\pfac{z}{}\pfac{3}{5}\right)$\\
$+ 5\left(\pfac{z}{}\pfac{1}{3}\pfac{3}{4}\gfac{1}{2}\gfac{3}{2}+\gfac{z}{}\gfac{1}{3}\gfac{3}{4}\pfac{1}{2}\pfac{3}{2}\right)$\\
$+ {5\over 2}\left(\pfac{z}{}\pfac{1}{}\pfac{3}{2}\gfac{1}{4}\gfac{3}{4}+\gfac{z}{}\gfac{1}{}\gfac{3}{2}\pfac{1}{4}\pfac{3}{4}\right)$\\
$+ {5\over 2}\left(\pfac{z}{}\pfac{1}{4}\pfac{3}{}\gfac{1}{}\gfac{3}{5}+\gfac{z}{}\gfac{1}{4}\gfac{3}{}\pfac{1}{}\pfac{3}{5}\right)$\\
$+5\left( \pfac{z}{}\pfac{1}{2}\gfac{1}{3}+ \gfac{z}{}\gfac{1}{2}\pfac{1}{3}\right)\pfac{3}{3}\gfac{3}{3} $\end{tabular}\\\hline
4A & $1^{4}2^24^4$ &    \begin{tabular}{@{}c@{}}${1\over 2}\left (\left (\pfac{z}{}\pfac{1}{3}- \mfac{z}{}\mfac{1}{3}\right)\pfac{2}{2}+\gfac{z}{}\gfac{1}{3}\gfac{2}{2} \right )\left(\pfac{4}{2}+ \gfac{4}{2}\right )^2$\\
$+2\left (\left (\pfac{z}{}\pfac{1}{3}- \mfac{z}{}\mfac{1}{3}\right)\gfac{2}{2}+\gfac{z}{}\gfac{1}{3}\pfac{2}{2} \right )\pfac{4}{2}\gfac{4}{2}$
\\$+2\left (\pfac{z}{}\pfac{1}{}\gfac{1}{2}+\gfac{z}{}\gfac{1}{}\pfac{1}{2}\right)\pfac{2}{}\gfac{2}{}\left(\pfac{4}{}\gfac{4}{3}+\pfac{4}{3}\gfac{4}{}\right )$\end{tabular}\\\hline
5A & $1^{4}5^4$ &    \begin{tabular}{@{}c@{}}${1\over 2}\left (\pfac{z}{}\pfac{1}{3}\pfac{5}{4}-\mfac{z}{}\mfac{1}{3}\mfac{5}{4}\right)+{1\over 2}\gfac{z}{}\gfac{1}{3}\gfac{5}{4}$\\
$+ {1\over 2}\left(\pfac{1}{3}\pfac{5}{}\gfac{z}{}\gfac{5}{3}+\gfac{1}{3}\gfac{5}{}\pfac{z}{}\pfac{5}{3}\right)$\\
$+ {3\over 2}\left(\pfac{z}{}\pfac{1}{2}\pfac{5}{}\gfac{1}{}\gfac{5}{3}+\gfac{z}{}\gfac{1}{2}\gfac{5}{}\pfac{1}{}\pfac{5}{3}\right)$\\
$+{3\over 2}\left( \pfac{z}{}\pfac{1}{}\gfac{1}{2}+ \gfac{z}{}\gfac{1}{}\pfac{1}{2}\right)\pfac{5}{2}\gfac{5}{2} $\end{tabular}\\\hline
6A & $1^22^23^26^2$ &    \begin{tabular}{@{}c@{}}${1\over 2}\left (\pfac{z}{}\pfac{1}{}\pfac{3}{2}- \mfac{z}{}\mfac{1}{}\mfac{3}{2}+\gfac{z}{}\gfac{1}{}\gfac{3}{2} \right )\left(\pfac{2}{}\pfac{6}{}+ \gfac{2}{}\gfac{6}{}\right )^2$\\
$+{1\over 2}\left (\pfac{z}{}\gfac{1}{}+\gfac{z}{}\pfac{1}{} \right )\pfac{3}{}\gfac{3}{}\left(\pfac{2}{}\gfac{6}{}+ \gfac{2}{}\pfac{6}{}\right )^2$\end{tabular}\\\hline
7A & $1^{3}7^3$ &    \begin{tabular}{@{}c@{}} $ {1\over 2}\left (\pfac{z}{}\pfac{1}{2}\pfac{7}{3}-\mfac{z}{}\mfac{1}{2}\mfac{7}{3}\right)+{1\over 2}\gfac{z}{}\gfac{1}{2}\gfac{7}{3}$\\
$+ {1\over 2}\left (\pfac{z}{}\gfac{1}{2}\gfac{7}+\gfac{z}{}\pfac{1}{2}\pfac{7}{}\right)\pfac{7}{}\gfac{7}{}$\\
$+\left(\pfac{z}{}\pfac{7}{}+\gfac{z}{}\gfac{7}{}\right)\pfac{1}{}\pfac{7}{}\gfac{1}{}\gfac{7}{}$\end{tabular} \\\hline

8A & $1^{2}2.4.8^2$ &    \begin{tabular}{@{}c@{}} ${1\over 2}\left(\pfac{z}{}\pfac{1}{}-\mfac{z}{}\mfac{1}{}\right)\pfac{2}{}\pfac{4}{}\pfac{8}{2}+{1\over 2}\gfac{z}{}\gfac{1}{}\gfac{2}{}\gfac{4}{}\gfac{8}{2}$\\
$+{1\over 2}\left(\pfac{z}{}\pfac{1}{}-\mfac{z}{}\mfac{1}{}\right)\pfac{2}{}\pfac{4}{}\gfac{8}{2}+{1\over 2}\gfac{z}{}\gfac{1}{}\gfac{2}{}\gfac{4}{}\pfac{8}{2}$\\
$+{1\over 2}\left(\pfac{z}{}\pfac{1}{}-\mfac{z}{}\mfac{1}{}\right)\pfac{2}{}\gfac{4}{}\pfac{8}{}\gfac{8}{}+{1\over 2}\gfac{z}{}\gfac{1}{}\gfac{2}{}\pfac{4}{}\pfac{8}{}\gfac{8}{}$
\end{tabular} \\\hline
11AB & $1^{2}11^2$ &   \begin{tabular}{@{}c@{}} $  {1\over 2}\left(({\bf z})_+({\bf 1})_+({\bf 11})^2_+-({\bf z})_-({\bf 1})_-({\bf 11})^2_-\right )+ {1\over 2} (\tilde{\bf z})(\tilde{\bf 1})(\widetilde{\bf 11})^2$\\
$+ {1\over 2}\left(({\bf z})_+(\tilde{\bf 1})({\bf 11})_+(\widetilde{\bf 11})+(\tilde{\bf z})_+({\bf 1})_+({\bf 11})_+(\widetilde{\bf 11})\right )$\end{tabular}  \\\hline
14AB & $1.2.7.14$ &  \begin{tabular}{@{}c@{}} $
{1\over 2}\left(({\bf z})_+({\bf 2})_+({\bf 7})_+({\bf 14})_+-({\bf z})_-({\bf 2})_+({\bf 7})_-({\bf 14})_+\right)+{1\over 2}(\tilde{\bf z})(\tilde{\bf 2})(\tilde{\bf 7})(\widetilde{\bf 14}) 
$\\
$+{1\over 2}\left(({\bf z})_+(\tilde{\bf 2})({\bf 7})_+(\widetilde{\bf 14})-({\bf z})_-(\tilde{\bf 2})({\bf 7})_-(\widetilde{\bf 14})+(\tilde{\bf z})({\bf 2})_+(\tilde{\bf 7})({\bf 14})_+\right)$\end{tabular}  \\\hline
15AB & $1.3.5.15$ &  \begin{tabular}{@{}c@{}} $
{1\over 2}\left(({\bf z})_+({\bf 3})_+({\bf 5})_+({\bf 15})_+-({\bf z})_-({\bf 3})_-({\bf 5})_-({\bf 15})_-\right)+{1\over 2}(\tilde{\bf z})(\tilde{\bf 3})(\tilde{\bf 5})(\widetilde{\bf 15}) 
$\\
$+{1\over 2}\left(({\bf z})_+(\tilde{\bf 3})(\tilde{\bf 5})({\bf 15})_++(\tilde{\bf z})({\bf 3})_+({\bf 5})_+(\widetilde{\bf 15})\right)$\end{tabular}  \\\hline
23AB & $1.23$ &    ${1\over 2}\left(({\bf z})_+({\bf 23})_+-({\bf z})_-({\bf 23})_-\right)$\\ \bottomrule
\end{tabular}}
\end{center}
\caption{The twining functions $F^{un.}_g(A_1^{24};\tau,z)$ of the untwisted sector of $\mc E^{\mc N=2}_{m=4}$ under for all conjugacy classes $[g]\in M_{23}$. We label them by their Frame shapes corresponding to their embedding into the {\bf 24} of $Co_0$.}\label{tbl:A1twin}\end{table}
\end{center}

Similarly, we now consider the functions $F^{un.}_g(\Lambda;\tau,z)$ for $\Lambda=A_2^{12}$. First we need the $\widehat{su(3)}$ characters including a chemical potential for the Cartan $J_3$ of the invariant $\widehat{su(2)}_4$. These are given by
 \be
 \chi_0(\t,z)=\Tr_{[0]}q^{L_0-c/24}y^{J_3}= {\theta_3(2\t,4z)\theta_3(6\t) + \theta_2(2\t,4z)\theta_2(6\t)\over 2\eta^2(\t)} :=\Pfac{z}{}
 \ee
 for the vacuum character and
 \be
  \chi_i(\t,z)=\Tr_{[i]}q^{L_0-c/24}y^{J_3}={\theta_3(2\t,4z)\theta_3\left({2\t\over 3}\right) + \theta_2(2\t,4z)\theta_2\left({2\t\over 3}\right)\over 2\eta^2(\t)}-{\chi_0(\t,z)\over 2}:=\Gfac{z}{}
 \ee
 for the nontrivial primaries with $i=1,2$.
 Finally, we also need the trace of $h$ in these modules, which we compute to be 
 \be
 \chi_0^-(\t,z)=\Tr_{[0]} h q^{L_0-c/24}y^{J_3}=\frac{\theta_4(2\t,4z)\theta_4(2\t)}{\eta^2(\t)}:=\Mfac{z}{}
 \ee
 and
 \be
   \chi^-_i(\t,z)=\Tr_{[i]}h q^{L_0-c/24}y^{J_3}=0,~~i=1,2.
 \ee
 
 Putting all of these components together, we compute the partition function of the orbifold theory in the untwisted sector with a projection onto anti-invariant states under $h$ to be
 \bea\nonumber
F^{un.}(\Lambda;\tau,z)&=&{ 1\over 2}\left (\Pfac{z}{}\Pfac{1}{11}-\Mfac{z}{}\Mfac{1}{11}\right )+66\left (\Pfac{z}{}\Pfac{1}{5}\Gfac{1}{6}+\Gfac{z}{}\Gfac{1}{5}\Pfac{1}{6}\right )+55\Pfac{z}{}\Pfac{1}{2}\Gfac{1}{9}\\
&+&165\Gfac{z}{}\Pfac{1}{3}\Gfac{1}{8}+12\Gfac{z}{}\Gfac{1}{11}.
 \eea
 As an example, we consider elements in conjugacy classes $[g] \in \{3A,5A,11AB\}$ of $M_{11}$. Again we use GAP to obtain an action of $M_{11}$ in its 11-dimensional permutation representation on the ternary Golay code, which we then use to compute the invariant vectors of the theory under this action. The results are reported in Table \ref{tbl:A2twin}.
\begin{center}
\begin{table}[htbp]
\begin{center}
\small
{\renewcommand{\arraystretch}{1.5}
\begin{tabular}{c|c|c}\toprule
$L_{2}(11)$ $[g]$ & Frame shape&  $F^{un.}_g(\Lambda;\tau,z)$\\\midrule
3A &$1^63^6$ &    \begin{tabular}{@{}c@{}}${1\over 2}\left(\Pfac{z}{}\Pfac{1}{2}\Pfac{3}{3}-\Mfac{z}{}\Mfac{1}{2}\Mfac{3}{3}\right)+3\Gfac{z}{}\Gfac{1}{2}\Gfac{3}{3}$\\
$+ 3\Pfac{z}{}\Pfac{1}{2}\Pfac{3}{}\Gfac{3}{2}+3\Gfac{z}{}\Gfac{1}{2}\Gfac{3}{}\Pfac{3}{2}+3\Gfac{z}{}\Gfac{1}{2}\Gfac{3}{2}\Pfac{3}{}+\Pfac{z}{}\Pfac{1}{2}\Gfac{3}{3}$\end{tabular}\\\hline
5A & $1^45^4$ &   \begin{tabular}{@{}c@{}}${1\over 2}\left(\Pfac{z}{}\Pfac{1}{}\Pfac{5}{2}-\Mfac{z}{}\Mfac{1}{}\Mfac{5}{2}\right)+2\Gfac{z}{}\Gfac{1}{}\Gfac{5}{2}$\\
$+ \Pfac{z}{}\Pfac{5}{}\Gfac{1}{}\Gfac{5}{}+ \Pfac{1}{}\Pfac{5}{}\Gfac{z}{}\Gfac{5}{}$\end{tabular}\\\hline
11AB & $1^211^2$ &    ${1\over2}\left (\Pfac{z}{}\Pfac{11}{}-\Mfac{z}{}\Mfac{11}{}\right )+ \Gfac{z}{}\Gfac{11}{}$ \\\bottomrule
\end{tabular}}
\end{center}
\caption{The twining functions $F^{un.}_g(A_2^{12};\tau,z)$ of the untwisted sector of $\mc E^{\mc N=4}_{m=4}$ for certain conjugacy classes $[g]\in M_{11}$. We label them by their Frame shapes corresponding to their embedding into the {\bf 24} of $Co_0$.}\label{tbl:A2twin}\end{table}
\end{center}

\subsection*{The twisted sector}
Finally, we need a description of the twisted sector Hilbert space, and the action of $h$ on the twisted states.
After we include the $U(1)$ grading, the contribution of the twisted sector in (\ref{eq:twistZ}) to the full partition function is 
\be\label{eq:Ftw}
F^{tw}(\Lambda;\t,z)= 2^{11}\frac{\theta_2(\t,2z)}{\theta_2(\t,0)} \left ( \frac{\theta_3(\t,2z)}{\theta_3(\t,0)}\frac{\eta^{24}(\t)}{\eta^{24}(\t/2)}-\frac{\theta_4(\t,2z)}{\theta_4(\t,0)}\frac{\eta^{24}(2\t)\eta^{24}(\t/2)}{\eta^{48}(\t)}\right )
\ee
for both $\Lambda=A_1^{24}$ and $\Lambda=A_2^{12}$.

  The twisted sector Hilbert spaces of all $\mathbb Z_2$ orbifolds of a Niemeier CFT are isomorphic and have an action of the group $Co_0$. Once we grade by the additional $U(1)$ charge as in equation (\ref{eq:Ftw}), the $Co_0$ symmetry is broken to subgroups which preserve a two-plane in the 24-dimensional representation. In particular, since both $M_{23}$ and $L_{2}(11)$ satisfy this constraint, we can define a consistent action of elements of these groups on the twisted sector Hilbert spaces. The action for a given conjugacy class $g$ of these groups follows from the 24-dimensional permutation representation of $g$ as follows. 
  Define
  \be
  \eta_g(\t):=q \prod_{n>0}\prod_{i=1}^{12}(1-\lambda_i^{-1}q^n)(1-\lambda_iq^n)
  \ee
  and
    \be
  \eta_{-g}(\t):=q \prod_{n>0}\prod_{i=1}^{12}(1+\lambda_i^{-1}q^n)(1+\lambda_iq^n)
  \ee
  where $\{\lambda_i\}$ are the 24 eigenvalues of $g$ in its 24-dimensional permutation representation, specified by the Frame shape $\pi_g$ as in equation (\ref{eq:Frame}). Then the trace of $g$ in the twisted sector is given by
  \be\label{eq:Ftwg}
F^{tw}_g(\Lambda;\t,z)= c_g\frac{\theta_2(\t,2z)}{\theta_2(\t,0)} \left ( \frac{\theta_3(\t,2z)}{\theta_3(\t,0)}\frac{\eta_g(\t)}{\eta_g(\t/2)}-\frac{\theta_4(\t,2z)}{\theta_4(\t,0)}\frac{\eta_{-g}(\t)}{\eta_{-g}(\t/2)}\right )
\ee
where the constant $c_g$ is defined by
\be
c_g:= 2^{{1\over 2}(\text{\# of cycles of }\pi_g) -1}.
\ee

From this and the results in the previous section we can reconstruct all the twining functions of $\mc E^{\mc N=2}_{m=4}$  under elements of $M_{23}$. We present the first several coefficients of these functions and their decompositions into irreducible $M_{23}$ representations in the tables in the next section.
\pagebreak

 \section{Tables}\label{sec:tables}
 In this section we present certain useful tables. In \S \ref{chartables} we present character tables of certain groups mentioned in the text. In \S \ref{sec:twinetbls} we present the first several coefficients and decompositions of the vector-valued mock modular forms arising from $\mc E^{\cN=2}_{m=4}$ for conjugacy classes $[g] \in M_{23}$.
\subsection{Irreducible characters}\label{chartables}
Below, we make use of the following standard notation: $b_n=(-1+i\sqrt{n})/2, \, \overline{b_n}=(-1-i\sqrt{n})/2$, $\b_n=(-1+\sqrt{n})/2,\, \overline{\b_n}=(-1-\sqrt{n})/2$ and $a_n=i\sqrt{n}, \, \overline{a_n}=-i\sqrt{n}$.

\begin{table}[htbp]
\vspace{0.5cm}

\centering
\caption{Character Table of $M_{23}$.}\smallskip
\begin{small}
\begin{tabular}{c@{ }|@{\;}r@{ }r@{ }r@{ }r@{ }r@{ }r@{ }r@{ }r@{ }r@{ }r@{ }r@{ }r@{ }r@{ }r@{ }r@{ }r@{ }r@{ }r}\toprule
$[g]$&&1A	&2A	&3A	&4A	&5A	&6A  &6B  &7AB  &8A  &11A  &11B  &14A  &14B  &15A  &15B  &23A  &23B  \\
\midrule
$[g^{2}]$ &   & 1A & 1A & 3A & 2A & 5A & 3A & 7A & 7B & 4A & 11B & 11A & 7A & 7B & 15A & 15B & 23A & 23B \\

$[g^{3}]$ &   & 1A & 2A & 1A & 4A & 5A & 2A & 7B & 7A & 8A & 11A & 11B & 14B & 14A & 5A & 5A & 23A & 23B \\

$[g^{5}]$ &   & 1A & 2A & 3A & 4A &A & 6A & 7B & 7A & 8A & 11A & 11B & 14B & 14A & 3A & 3A & 23B & 23A \\

$[g^{7}]$ &   & 1A & 2A & 3A & 4A & 5A & 6A & 1A & 1A & 8A & 11B & 11A & 2A & 2A & 15B & 15A & 23B & 23A \\

$[g^{11}]$ &   &1A & 2A & 3A & 4A & 5A & 6A & 7A & 7B & 8A & 1A & 1A & 14A & 14B & 15B & 15A & 23B & 23A \\

$[g^{23}]$ &   & 1A & 2A & 3A & 4A & 5A & 6A & 7A & 7B & 8A & 11A & 11B & 14A & 14B & 15A & 15B & 1A & 1A \\

	\midrule
$\chi_{1}$ &  & 1 & 1 & 1 & 1 & 1 & 1 & 1 & 1 & 1 & 1 & 1 & 1 & 1 & 1 & 1 & 1 & 1 \\

$\chi_{2}$ &  & 22 & 6 & 4 & 2 & 2 & 0 & 1 & 1 & 0 & 0 & 0 & -1 & -1 & -1 & -1 & -1 & -1 \\

$\chi_{3}$ &  & 45 & -3 & 0 & 1 & 0 & 0 & $b_7$ & $\overline{b_7}$& -1 & 1 & 1 & -$b_7$ & -$\overline{b_7}$ & 0 & 0 & -1 & -1 \\

$\chi_{4}$ &  & 45 & -3 & 0 & 1 & 0 & 0 & $\overline{b_7}$ & $b_7$& -1 & 1 & 1 & -$\overline{b_7}$ & -${b_7}$ & 0 & 0 & -1 & -1 \\

$\chi_{5}$ &  & 230 & 22 & 5 & 2 & 0 & 1 & -1 & -1 & 0 & -1 & -1 & 1 & 1 & 0 & 0 & 0 & 0 \\

$\chi_{6}$ &  & 231 & 7 & 6 & -1 & 1 & -2 & 0 & 0 & -1 & 0 & 0 & 0 & 0 & 1 & 1 & 1 & 1 \\

$\chi_{7}$ &  & 231 & 7 & -3 & -1 & 1 & 1 & 0 & 0 & -1 & 0 & 0 & 0 & 0 & ${b_{15}}$ & $\overline{b_{15}}$ & 1 & 1 \\

$\chi_{8}$ &  & 231 & 7 & -3 & -1 & 1 & 1 & 0 & 0 & -1 & 0 & 0 & 0 & 0 & $\overline{b_{15}}$ & $b_{15}$& 1 & 1 \\

$\chi_{9}$ &  & 253 & 13 & 1 & 1 & -2 & 1 & 1 & 1 & -1 & 0 & 0 & -1 & -1 & 1 & 1 & 0 & 0 \\

$\chi_{10}$ &  & 770 & -14 & 5 & -2 & 0 & 1 & 0 & 0 & 0 & 0 & 0 & 0 & 0 & 0 & 0 & ${b_{23}}$ & $\overline{b_{23}}$ \\

$\chi_{11}$ &  & 770 & -14 & 5 & -2 & 0 & 1 & 0 & 0 & 0 & 0 & 0 & 0 & 0 & 0 & 0 & $\overline{b_{23}}$ & ${b_{23}}$ \\

$\chi_{12}$ &  & 896 & 0 & -4 & 0 & 1 & 0 & 0 & 0 & 0 & ${b_{11}}$ & $\overline{b_{11}}$ & 0 & 0 & 1 & 1 & -1 & -1 \\

$\chi_{13}$ &  & 896 & 0 & -4 & 0 & 1 & 0 & 0 & 0 & 0 & $\overline{b_{11}}$ & ${b_{11}}$ & 0 & 0 & 1 & 1 & -1 & -1 \\

$\chi_{14}$ &  & 990 & -18 & 0 & 2 & 0 & 0 & $b_7$ & $\overline{b_7}$ & 0 & 0 & 0 & $b_7$ & $\overline{b_7}$ & 0 & 0 & 1 & 1 \\

$\chi_{15}$ &  & 990 & -18 & 0 & 2 & 0 & 0 & $\overline{b_7}$ & ${b_7}$ & 0 & 0 & 0 & $\overline{b_7}$ & ${b_7}$ & 0 & 0 & 1 & 1 \\

$\chi_{16}$ &  & 1035 & 27 & 0 & -1 & 0 & 0 & -1 & -1 & 1 & 1 & 1 & -1 & -1 & 0 & 0 & 0 & 0 \\

$\chi_{17}$ &  & 2024 & 8 & -1 & 0 & -1 & -1 & 1 & 1 & 0 & 0 & 0 & 1 & 1 & -1 & -1 & 0 & 0 \\
\bottomrule
\end{tabular}
\end{small}
\end{table}

	\begin{table}[htbp]
\begin{center}
\caption{Character table of $M_{11}$.}
\begin{small}
\begin{tabular}{c|rrrrrrrrrrr}\toprule
$[g]$&1A&2A&3A&4A&5A&6A&8A&8B&11A&11B\\\midrule
$[g^2]$&1A&1A&3A&2A&5A&3A&4A&4A&11B&11A\\
$[g^3]$&1A&2A&1A&4A&5A&2A&8A&8B&11A&11B\\
$[g^5]$&1A&2A&3A&4A&1A&6A&8B&8A&11A&11B\\
\midrule
${\chi}_{1}$&1&1&1&1&1&1&1&1&1&1\\
${\chi}_{2}$&10&2&1&2&0&-1&0&0&-1&-1\\
${\chi}_{3}$&10&-2&1&0&0&1&$a_2$&$\overline{a_2}$&-1&-1\\
${\chi}_{4}$&10&-2&1&0&0&1&$\overline{a_2}$&$a_2$&-1&-1\\
${\chi}_{5}$&11&3&2&-1&1&0&-1&-1&0&0\\
${\chi}_{6}$&16&0&-2&0&1&0&0&0&$\beta_{11}$&$\overline{\beta_{11}}$\\
${\chi}_{7}$&16&0&-2&0&1&0&0&0&$\overline{\beta_{11}}$&$\beta_{11}$\\
${\chi}_{8}$&44&4&-1&0&-1&1&0&0&0&0\\
${\chi}_{9}$&45&-3&0&1&0&0&-1&-1&1&1\\
${\chi}_{10}$&55&-1&1&-1&0&-1&1&1&0&0\\
\bottomrule
\end{tabular}
\end{small}
\label{tbl:M11}
\end{center}
\end{table}
\pagebreak
\subsection{Coefficients and decompositions}\label{sec:twinetbls}
\begin{center}
\begin{table}[htb]\begin{center}
\footnotesize


\caption{\footnotesize The twined series for $M_{23}$. The table displays the Fourier coefficients multiplying $q^{-D/56}$ in the $q$-expansion of the function $\tilde{h}^{\cN=2}_{g,1}(\t)$.}
\begin{tabular}{c@{ }|@{\;}r@{ }r@{ }r@{ }r@{ }r@{ }r@{ }r@{ }r@{ }r@{ }r@{ }r@{ }r@{ }r}\toprule
$[g]$&1A	&2A	&3A	&4A	&5A	&6AB  &7AB  &8A  &11AB  &14AB  &15AB  &23AB  \\
	\midrule
-1 & -1 & -1 & -1 & -1 & -1 & -1 & -1 & -1 & -1 & -1 & -1 & -1 \\

79 & 32890 & 490 & 76 & 22 & 10 & 4 & 4 & 0 & 0 & 0 & 1 & 0 \\

159 & 2969208 & 10136 & 585 & 80 & 18 & 17 & 4 & 0 & 0 & 0 & 0 & 0 \\

239 & 101822334 & 88670 & 3192 & 374 & 54 & -16 & 5 & 2 & -2 & 1 & -3 & 0 \\

319 & 2065775107 & 636803 & 12550 & 947 & 132 & 74 & 11 & -1 & 0 & -1 & 0 & 0 \\

399 & 29747513059 & 3408531 & 42757 & 2399 & 269 & -15 & 12 & 5 & 0 & 0 & 2 & 0 \\

479 & 334821538370 & 16448690 & 136784 & 5582 & 530 & 80 & 11 & -4 & 0 & -1 & -1 & 0 \\

559 & 3122115821404 & 68126268 & 386305 & 12996 & 824 & 33 & 20 & -4 & -3 & 0 & 5 & 0 \\

639 & 25061866943436 & 262901388 & 1026324 & 26780 & 1586 & 360 & 29 & 4 & 4 & 1 & -1 & 1 \\

719 & 177895424302751 & 922681999 & 2615528 & 53771 & 2666 & -320 & 29 & -7 & 1 & 1 & -7 & 0 \\

799 & 1138785187015234 & 3070987058 & 6274135 & 104846 & 4574 & 1079 & 31 & 4 & 0 & -1 & 5 & -1 \\

879 & 6672991048411185 & 9574047505 & 14472639 & 201593 & 7415 & -401 & 57 & 9 & 3 & 1 & -1 & 0 \\

959 & 36211921311763437 & 28624358621 & 32442711 & 369065 & 11122 & 1079 & 58 & -5 & -2 & -2 & 1 & 0 \\

1039 & 183681040795024267 & 81543759179 & 70065910 & 662651 & 17967 & -70 & 61 & 19 & 0 & 1 & 0 & -1 \\

1119 & 877475502920966100 & 224506987348 & 147298461 & 1169604 & 27740 & 3409 & 88 & -16 & -3 & 0 & -4 & 0 \\

%
%
%
%
%
%
%
%
%
%
	\bottomrule
\end{tabular}
\end{center}
\end{table}

\begin{table}[htb]\begin{center}
\footnotesize
\centering
\caption{\footnotesize The twined series for $M_{23}$. The table displays the Fourier coefficients multiplying $q^{-D/56}$ in the $q$-expansion of the function $\tilde{h}^{\cN=2}_{g,2}(\t)$.}
\begin{tabular}{c@{ }|@{\;}r@{ }r@{ }r@{ }r@{ }r@{ }r@{ }r@{ }r@{ }r@{ }r@{ }r@{ }r@{ }r}\toprule
$[g]$&1A	&2A	&3A	&4A	&5A	&6AB  &7AB  &8A  &11AB  &14AB  &15AB  &23AB  \\
	\midrule
-9 & 1 & 1 & 1 & 1 & 1 & 1 & 1 & 1 & 1 & 1 & 1 & 1 \\

71 & 14168 & 392 & 74 & 20 & 8 & 2 & 0 & 2 & 0 & 0 & -1 & 0 \\

151 & 1659174 & 6278 & 465 & 94 & 24 & 5 & -1 & 2 & 0 & -1 & 0 & 0 \\

231 & 63544239 & 70287 & 2367 & 279 & 59 & 27 & -4 & -1 & 0 & 0 & 2 & 0 \\

311 & 1373777350 & 471990 & 10699 & 786 & 125 & -21 & 1 & 0 & 0 & 1 & -1 & 0 \\

391 & 20649050170 & 2768410 & 36727 & 2114 & 200 & 115 & -7 & 2 & 1 & 1 & 2 & 0 \\

471 & 239838441957 & 13053893 & 113958 & 5229 & 457 & -34 & -5 & -3 & 4 & -1 & -2 & -1 \\

551 & 2291638384937 & 56517657 & 337376 & 11397 & 842 & 120 & -8 & -1 & 0 & 0 & -4 & 0 \\

631 & 18760451739204 & 216334868 & 899886 & 23784 & 1479 & 50 & -7 & -6 & -1 & 1 & 6 & 0 \\

711 & 135352127137850 & 778525770 & 2278664 & 48830 & 2600 & 528 & -7 & 12 & -4 & 1 & -1 & 0 \\

791 & 878471971333176 & 2585630360 & 5566971 & 97056 & 3901 & -469 & -18 & 8 & 0 & -2 & 1 & 1 \\

871 & 5209082274923427 & 8188169219 & 12900135 & 183083 & 6807 & 1475 & -9 & -5 & -1 & -1 & 0 & 0 \\

951 & 28562269988425239 & 24491271063 & 28872441 & 336679 & 10799 & -567 & -32 & 3 & 0 & 0 & -4 & 1 \\

1031 & 146211017617763307 & 70510224443 & 62961633 & 610623 & 17107 & 1481 & -15 & -3 & 0 & 1 & -2 & 0 \\

1111 & 704198296122633807 & 194427334975 & 132796005 & 1086555 & 26522 & -191 & -37 & -15 & 0 & -1 & 5 & 0 \\

%
%
%
%
%
%
%
%
%
%
	\bottomrule
\end{tabular}\end{center}

\end{table}

\begin{table}[htb]\begin{center}
\footnotesize
\centering
\caption{\footnotesize The twined series for $M_{23}$. The table displays the Fourier coefficients multiplying $q^{-D/56}$ in the $q$-expansion of the function $\tilde{h}^{\cN=2}_{g,3}(\t)$.}
\begin{tabular}{c@{ }|@{\;}r@{ }r@{ }r@{ }r@{ }r@{ }r@{ }r@{ }r@{ }r@{ }r@{ }r@{ }r@{ }r}\toprule
$[g]$&1A	&2A	&3A	&4A	&5A	&6AB  &7AB  &8A  &11AB  &14AB  &15AB  &23AB  \\
	\midrule
-25 & -1 & -1 & -1 & -1 & -1 & -1 & -1 & -1 & -1 & -1 & -1 & -1 \\

55 & 2024 & 120 & 26 & 12 & 4 & 6 & 1 & 2 & 0 & 1 & 1 & 0 \\

135 & 485001 & 2953 & 234 & 41 & 11 & -2 & -1 & 1 & 0 & -1 & -1 & 0 \\

215 & 23912778 & 37850 & 1704 & 206 & 38 & 8 & 1 & -4 & -1 & 1 & -1 & 0 \\

295 & 594404250 & 276954 & 7008 & 634 & 105 & 12 & -1 & -2 & 0 & -1 & 3 & 1 \\

375 & 9795220335 & 1719215 & 25389 & 1679 & 215 & 77 & 5 & 3 & -1 & 1 & -1 & 0 \\

455 & 121610515928 & 8440360 & 85280 & 3852 & 333 & -56 & 6 & 2 & 0 & -2 & 0 & 0 \\

535 & 1223045193953 & 37766625 & 248780 & 9089 & 693 & 264 & 9 & 1 & 0 & 1 & 0 & -1 \\

615 & 10431487439956 & 148238340 & 677512 & 19744 & 1221 & -96 & 2 & 2 & 0 & -2 & -3 & 0 \\

695 & 77848480769761 & 545254705 & 1771723 & 40485 & 2236 & 259 & 2 & 7 & -2 & 2 & -2 & 0 \\

775 & 519869748402405 & 1843176725 & 4327128 & 78673 & 3720 & 56 & 2 & 3 & 5 & -2 & 3 & 0 \\

855 & 3159048430391220 & 5930043604 & 10148229 & 152428 & 5665 & 1033 & 10 & -20 & 4 & 2 & 4 & 0 \\

935 & 17694698437501954 & 17975169890 & 23094682 & 285770 & 9394 & -910 & 15 & -14 & 0 & -1 & -8 & 0 \\

1015 & 92296742373818321 & 52381498417 & 50515790 & 519049 & 14871 & 2710 & 31 & 13 & -2 & 3 & 0 & 0 \\

1095 & 452022567897804867 & 145967611235 & 107402373 & 917355 & 23372 & -1123 & 26 & -1 & -4 & -2 & 8 & -1 \\

%
%
%
%
%
%
%
%
%
%
	\bottomrule
\end{tabular}\end{center}

\end{table}

\begin{table}[htb]\begin{center}
\footnotesize
\centering
\caption{\footnotesize The twined series for $M_{23}$. The table displays the Fourier coefficients multiplying $q^{-D/56}$ in the $q$-expansion of the function $\tilde{h}^{\cN=2}_{g,4}(\t)$.}
\begin{tabular}{c@{ }|@{\;}r@{ }r@{ }r@{ }r@{ }r@{ }r@{ }r@{ }r@{ }r@{ }r@{ }r@{ }r@{ }r}\toprule
$[g]$&1A	&2A	&3A	&4A	&5A	&6AB  &7AB  &8A  &11AB  &14AB  &15AB  &23AB  \\
	\midrule
-49 & 1 & 1 & 1 & 1 & 1 & 1 & 1 & 1 & 1 & 1 & 1 & 1 \\

31 & 23 & 7 & 5 & 3 & 3 & 1 & 2 & 1 & 1 & 0 & 0 & 0 \\

111 & 61984 & 1008 & 109 & 36 & 14 & 9 & 6 & -2 & -1 & 0 & -1 & -1 \\

191 & 4994473 & 12841 & 814 & 105 & 23 & -2 & 8 & 1 & 0 & -4 & -1 & 0 \\

271 & 159121844 & 126116 & 3851 & 384 & 79 & 47 & 14 & 6 & 2 & 4 & 1 & 1 \\

351 & 3066459912 & 791976 & 14742 & 1104 & 107 & -18 & 12 & 4 & -1 & -4 & 2 & 0 \\

431 & 42526230351 & 4396655 & 52188 & 2871 & 301 & 44 & 24 & -5 & 0 & 4 & -2 & 0 \\

511 & 465019661864 & 19995832 & 157790 & 6236 & 549 & 34 & -19 & -2 & -1 & 3 & 0 & 0 \\

591 & 4237704983457 & 83898753 & 443403 & 14105 & 1002 & 267 & 36 & -3 & 0 & 8 & 3 & 0 \\

671 & 33383739990645 & 313694485 & 1187433 & 29821 & 1780 & -215 & 44 & 1 & 0 & -8 & -2 & 0 \\

751 & 233270628632745 & 1105509481 & 2962041 & 60217 & 2700 & 769 & 60 & -7 & 0 & -8 & 6 & -1 \\

831 & 1473401102910159 & 3610317407 & 7067235 & 114659 & 4804 & -289 & 66 & -3 & -6 & 4 & -5 & 0 \\

911 & 8534324476198088 & 11260342856 & 16341254 & 218344 & 7803 & 758 & 92 & 24 & 7 & 0 & -6 & 0 \\

991 & 45845203718962384 & 33250870352 & 36222793 & 402864 & 12389 & -31 & 100 & 16 & 0 & 0 & 8 & 0 \\

1071 & 230465514424059585 & 94612982465 & 77950314 & 723905 & 19565 & 2570 & -99 & -15 & 0 & 1 & -1 & 1 \\

%
%
%
%
%
%
%
%
%
%
	\bottomrule
\end{tabular}\end{center}

\end{table}



\begin{table}[htbp]\begin{center}
\footnotesize
\vspace{0.5cm}
\caption{\footnotesize The table shows the decomposition of the Fourier coefficients multiplying $q^{-D/56}$ in the function $\tilde{h}^{\cN=2}_{g,1}(\t)$ into irreducible representations $\chi_n$ of $M_{23}$.}\smallskip
\begin{tabular}{c@{ }|@{\;}r@{ }r@{ }r@{ }r@{ }r@{ }r}\toprule
&$\chi_1$	&$\chi_2$	&$\chi_3$	&$\chi_4$	&$\chi_5$	&$\chi_6$   \\
	\midrule
-1 & -1 & 0 & 0 & 0 & 0 & 0 \\

79 & 3 & 6 & 0 & 0 & 8 & 4 \\

159 & 13 & 50 & 4 & 4 & 172 & 109 \\

239 & 75 & 520 & 361 & 361 & 3132 & 2637 \\

319 & 555 & 6234 & 8431 & 8431 & 52201 & 48823 \\

399 & 4516 & 72901 & 127496 & 127496 & 699946 & 683877 \\

479 & 39919 & 762273 & 1458831 & 1458831 & 7687965 & 7629252 \\

559 & 334018 & 6894918 & 13697115 & 13697115 & 70963150 & 70890020 \\

639 & 2561300 & 54661450 & 110264073 & 110264073 & 567249277 & 568242219 \\

719 & 17798711 & 385781306 & 783730795 & 783730795 & 4018617602 & 4030917545 \\

799 & 112816113 & 2462973668 & 5020155321 & 5020155321 & 25701388426 & 25795911388 \\

879 & 657802189 & 14413077161 & 29426216833 & 29426216833 & 150534019095 & 151134807838 \\

959 & 3560695812 & 78161432884 & 159711506399 & 159711506399 & 816701680872 & 820091947106 \\

1039 & 18036997856 & 396321128501 & 810190283194 & 810190283194 & 4142106992628 & 4159658528152 \\

1119 & 86103293155 & 1892920594014 & 3870600609373 & 3870600609373 & 19786191912051 & 19870958706758 \\

%
%
%
%
%
%
%
%
%
%

\end{tabular}

\begin{tabular}{c@{ }|@{\;}r@{ }r@{ }r@{ }r@{ }r@{ }r}\toprule
&$\chi_7$	&$\chi_8$	&$\chi_9$	&$\chi_{10}$&$\chi_{11}$&$\chi_{12}$	 \\
	\midrule
-1 & 0 & 0 & 0 & 0 & 0 & 0 \\

79 & 1 & 1 & 4 & 1 & 1 & 2 \\

159 & 84 & 84 & 128 & 184 & 184 & 249 \\

239 & 2474 & 2474 & 2975 & 7288 & 7288 & 8876 \\

319 & 48214 & 48214 & 54404 & 152910 & 152910 & 181177 \\

399 & 681735 & 681735 & 754547 & 2228718 & 2228718 & 2611937 \\

479 & 7622433 & 7622433 & 8384529 & 25191152 & 25191152 & 29406002 \\

559 & 70870712 & 70870712 & 77765357 & 235322052 & 235322052 & 274222110 \\

639 & 568190993 & 568190993 & 622851914 & 1890404790 & 1890404790 & 2201283134 \\

719 & 4030786690 & 4030786690 & 4416567105 & 13423360272 & 13423360272 & 15625363600 \\

799 & 25795597950 & 25795597950 & 28258570095 & 85943199892 & 85943199892 & 100024910245 \\

879 & 151134084106 & 151134084106 & 165547158795 & 503648520358 & 503648520358 & 586120983088 \\

959 & 820090325240 & 820090325240 & 898251754417 & 2733239573184 & 2733239573184 & 3180668696724 \\

1039 & 4159655024839 & 4159655024839 & 4555976147351 & 13864390059718 & 13864390059718 & 16133598865348 \\

1119 & 19870951342688 & 19870951342688 & 21763871927454 & 66233398639795 & 66233398639795 & 77072943845016 \\

%
%
%
%
%
%
%
%
%
%
%
\bottomrule
\end{tabular}

\vspace{1mm}

\begin{tabular}{c@{ }|@{\;}r@{ }r@{ }r@{ }r@{ }r} 
&$\chi_{13}$	&$\chi_{14}$	&$\chi_{15}$	&$\chi_{16}$	&$\chi_{17}$ \\
	\midrule

-1 & 0 & 0 & 0 & 0 & 0 \\

79 & 2 & 1 & 1 & 7 & 7 \\

159 & 249 & 225 & 225 & 400 & 614 \\

239 & 8876 & 9311 & 9311 & 11209 & 20448 \\

319 & 181177 & 196277 & 196277 & 215961 & 411688 \\

399 & 2611937 & 2864311 & 2864311 & 3052375 & 5912176 \\

479 & 29406002 & 32384527 & 32384527 & 34136386 & 66480999 \\

559 & 274222110 & 302544979 & 302544979 & 317457009 & 619667980 \\

639 & 2201283134 & 2430487575 & 2430487575 & 2545442969 & 4973369259 \\

719 & 15625363600 & 17258520366 & 17258520366 & 18058720902 & 35299442572 \\

799 & 100024910245 & 110498190645 & 110498190645 & 115573171916 & 225958546463 \\

879 & 586120983088 & 647547609561 & 647547609561 & 677144766331 & 1324034573732 \\

959 & 3180668696724 & 3514164059458 & 3514164059458 & 3674386668984 & 7184990032658 \\

1039 & 16133598865348 & 17825641954808 & 17825641954808 & 18637288293732 & 36444893250715 \\

%
%
%
%
%
%
%
%
%
%
%

\bottomrule
\end{tabular}\end{center}
\end{table}


\begin{table}[htbp]\begin{center}
\footnotesize
\vspace{0.5cm}
\caption{\footnotesize The table shows the decomposition of the Fourier coefficients multiplying $q^{-D/56}$ in the function $\tilde{h}^{\cN=2}_{g,2}(\t)$ into irreducible representations $\chi_n$ of $M_{23}$.}\smallskip
\begin{tabular}{c@{ }|@{\;}r@{ }r@{ }r@{ }r@{ }r@{ }r}\toprule
&$\chi_1$	&$\chi_2$	&$\chi_3$	&$\chi_4$	&$\chi_5$	&$\chi_6$   \\
	\midrule
-9 & 1 & 0 & 0 & 0 & 0 & 0 \\

71 & 2 & 5 & 0 & 0 & 7 & 3 \\

151 & 10 & 37 & 3 & 3 & 108 & 67 \\

231 & 60 & 371 & 211 & 211 & 2094 & 1692 \\

311 & 401 & 4320 & 5558 & 5558 & 35182 & 32682 \\

391 & 3347 & 51686 & 88067 & 88067 & 489393 & 475958 \\

471 & 29191 & 549308 & 1043607 & 1043607 & 5517942 & 5468790 \\

551 & 247970 & 5076761 & 10046497 & 10046497 & 52142001 & 52052092 \\

631 & 1925413 & 40964481 & 82518209 & 82518209 & 424787070 & 425421787 \\

711 & 13572600 & 293700302 & 596218175 & 596218175 & 3058208800 & 3067141035 \\

791 & 87112694 & 1900466946 & 3872364194 & 3872364194 & 19828140356 & 19899850416 \\

871 & 513770437 & 11252794339 & 22969949734 & 22969949734 & 117516030536 & 117981038495 \\

951 & 2809241950 & 61654430595 & 125970832520 & 125970832520 & 644191847539 & 646855266217 \\

1031 & 14359666110 & 315486247432 & 644909227030 & 644909227030 & 3297183714290 & 3311123599862 \\

%
%
%
%
%
%
%
%
%
%
%
%

\end{tabular}

\begin{tabular}{c@{ }|@{\;}r@{ }r@{ }r@{ }r@{ }r@{ }r}\toprule
&$\chi_7$	&$\chi_8$	&$\chi_9$	&$\chi_{10}$&$\chi_{11}$&$\chi_{12}$	 \\
	\midrule

-9 & 0 & 0 & 0 & 0 & 0 & 0 \\

71 & 0 & 0 & 2 & 0 & 0 & 0 \\

151 & 45 & 45 & 74 & 100 & 100 & 137 \\

231 & 1580 & 1580 & 1932 & 4481 & 4481 & 5533 \\

311 & 32142 & 32142 & 36420 & 101485 & 101485 & 120436 \\

391 & 474150 & 474150 & 525770 & 1545133 & 1545133 & 1812904 \\

471 & 5463084 & 5463084 & 6012239 & 18038594 & 18038594 & 21063676 \\

551 & 52035254 & 52035254 & 57111733 & 172694257 & 172694257 & 201278319 \\

631 & 425376804 & 425376804 & 466340794 & 1414993671 & 1414993671 & 1647801950 \\

711 & 3067027234 & 3067027234 & 3360726669 & 10212802503 & 10212802503 & 11888586088 \\

791 & 19899571950 & 19899571950 & 21800037596 & 66296462153 & 66296462153 & 77160348340 \\

871 & 117980393857 & 117980393857 & 129233185927 & 393155335789 & 393155335789 & 457538780532 \\

951 & 646853822454 & 646853822454 & 708508249448 & 2155841669413 & 2155841669413 & 2508762643117 \\

1031 & 3311120452151 & 3311120452151 & 3626606693880 & 11036094133318 & 11036094133318 & 12842424391871 \\

1111 & 15947023462552 & 15947023462552 & 17466176597842 & 53154055601266 & 53154055601266 & 61853163157258 \\

%
%
%
%
%
%
%
%
%
%
%
\bottomrule
\end{tabular}
\vspace{1mm}

\begin{tabular}{c@{ }|@{\;}r@{ }r@{ }r@{ }r@{ }r}
&$\chi_{13}$	&$\chi_{14}$	&$\chi_{15}$	&$\chi_{16}$	&$\chi_{17}$ \\
	\midrule

-9 & 0 & 0 & 0 & 0 & 0 \\

71 & 0 & 0 & 0 & 5 & 3 \\

151 & 137 & 125 & 125 & 229 & 343 \\

231 & 5533 & 5714 & 5714 & 7145 & 12797 \\

311 & 120436 & 130213 & 130213 & 144101 & 273914 \\

391 & 1812904 & 1985578 & 1985578 & 2122817 & 4105045 \\

471 & 21063676 & 23189159 & 23189159 & 24465217 & 47625182 \\

551 & 201278319 & 222025040 & 222025040 & 233079356 & 454856422 \\

631 & 1647801950 & 1819248831 & 1819248831 & 1905627188 & 3722950603 \\

711 & 11888586088 & 13130671614 & 13130671614 & 13740785934 & 26857884945 \\

791 & 77160348340 & 85238123579 & 85238123579 & 89156647614 & 174307658489 \\

871 & 457538780532 & 505484997007 & 505484997007 & 528601141162 & 1033572367727 \\

951 & 2508762643117 & 2771795449104 & 2771795449104 & 2898203583666 & 5667189790804 \\

1031 & 12842424391871 & 14189261724139 & 14189261724139 & 14835429985991 & 29010332044013 \\

1111 & 61853163157258 & 68340924042730 & 68340924042730 & 71450643678386 & 139722462064585 \\

%
%
%
%
%
%
%
%
%
%

\bottomrule
\end{tabular}\end{center}
\end{table}


\begin{table}[htbp]\begin{center}
\footnotesize
\vspace{0.5cm}
\caption{\footnotesize The table shows the decomposition of the Fourier coefficients multiplying $q^{-D/56}$ in the function $\tilde{h}^{\cN=2}_{g,3}(\t)$ into irreducible representations $\chi_n$ of $M_{23}$.}\smallskip
\begin{tabular}{c@{ }|@{\;}r@{ }r@{ }r@{ }r@{ }r@{ }r}\toprule
&$\chi_1$	&$\chi_2$	&$\chi_3$	&$\chi_4$	&$\chi_5$	&$\chi_6$   \\
	\midrule
-25 & -1 & 0 & 0 & 0 & 0 & 0 \\

55 & 2 & 2 & 0 & 0 & 3 & 0 \\

135 & 4 & 17 & 0 & 0 & 44 & 26 \\

215 & 35 & 192 & 70 & 70 & 910 & 692 \\

295 & 228 & 2109 & 2333 & 2333 & 15904 & 14401 \\

375 & 1815 & 25661 & 41343 & 41343 & 235739 & 227084 \\

455 & 15674 & 283294 & 527166 & 527166 & 2813624 & 2778595 \\

535 & 135681 & 2728183 & 5353413 & 5353413 & 27892476 & 27802131 \\

615 & 1082201 & 22844512 & 45852109 & 45852109 & 236430984 & 236628378 \\

695 & 7845613 & 169151973 & 342809579 & 342809579 & 1759756137 & 1764351037 \\

775 & 51675284 & 1125397975 & 2291272507 & 2291272507 & 11736660597 & 11777354738 \\

855 & 311949208 & 6826465150 & 13929053191 & 13929053191 & 71275567395 & 71552199331 \\

935 & 1741436074 & 38202099659 & 78037446817 & 78037446817 & 399108330301 & 400742735902 \\

1015 & 9067633922 & 199170750739 & 407094744895 & 407094744895 & 2081435347715 & 2090191121215 \\

1095 & 44366697752 & 975187152952 & 1993866656330 & 1993866656330 & 10192904243304 & 10236401971933 \\

%
%
%
%
%
%
%
%
%
%
%

\end{tabular}

\begin{tabular}{c@{ }|@{\;}r@{ }r@{ }r@{ }r@{ }r@{ }r}\toprule
&$\chi_7$	&$\chi_8$	&$\chi_9$	&$\chi_{10}$&$\chi_{11}$&$\chi_{12}$	 \\
	\midrule
-25 & 0 & 0 & 0 & 0 & 0 & 0 \\

55 & 0 & 0 & 1 & 0 & 0 & 0 \\

135 & 14 & 14 & 27 & 25 & 25 & 38 \\

215 & 609 & 609 & 788 & 1643 & 1643 & 2065 \\

295 & 14053 & 14053 & 16128 & 43581 & 43581 & 52061 \\

375 & 225834 & 225834 & 251423 & 731026 & 731026 & 859812 \\

455 & 2774317 & 2774317 & 3057500 & 9137701 & 9137701 & 10679771 \\

535 & 27789758 & 27789758 & 30517710 & 92128896 & 92128896 & 107420534 \\

615 & 236594479 & 236594479 & 259438583 & 786646427 & 786646427 & 916233373 \\

695 & 1764262516 & 1764262516 & 1933413743 & 5873450857 & 5873450857 & 6837772000 \\

775 & 11777138395 & 11777138395 & 12902535131 & 39231891343 & 39231891343 & 45662596088 \\

855 & 71551692177 & 71551692177 & 78378155440 & 238424129443 & 238424129443 & 277474384470 \\

935 & 400741580942 & 400741580942 & 438943677467 & 1335557549215 & 1335557549215 & 1554211032206 \\

1015 & 2090188596103 & 2090188596103 & 2289359341885 & 6966572026982 & 6966572026982 & 8106871287186 \\

1095 & 10236396601532 & 10236396601532 & 11211583746696 & 34119303660620 & 34119303660620 & 39703341303712 \\

%
%
%
%
%
%
%
%
%
%
%
\bottomrule
\end{tabular}

\vspace{1mm}

\begin{tabular}{c@{ }|@{\;}r@{ }r@{ }r@{ }r@{ }r}
&$\chi_{13}$	&$\chi_{14}$	&$\chi_{15}$	&$\chi_{16}$	&$\chi_{17}$ \\
	\midrule

-25 & 0 & 0 & 0 & 0 & 0 \\

55 & 0 & 0 & 0 & 1 & 0 \\

135 & 38 & 30 & 30 & 78 & 103 \\

215 & 2065 & 2080 & 2080 & 2799 & 4845 \\

295 & 52061 & 55872 & 55872 & 63071 & 118714 \\

375 & 859812 & 939215 & 939215 & 1011049 & 1948452 \\

455 & 10679771 & 11745983 & 11745983 & 12423389 & 24153700 \\

535 & 107420534 & 118443822 & 118443822 & 124470503 & 242778649 \\

615 & 916233373 & 1011381183 & 1011381183 & 1059877938 & 2070176920 \\

695 & 6837772000 & 7551522153 & 7551522153 & 7904063364 & 15447739906 \\

775 & 45662596088 & 50440859848 & 50440859848 & 52765036392 & 103154220956 \\

855 & 277474384470 & 306544968511 & 306544968511 & 320579896279 & 626812915588 \\

935 & 1554211032206 & 1717144636530 & 1717144636530 & 1795503032799 & 3510906233262 \\

1015 & 8106871287186 & 8957019447270 & 8957019447270 & 9365049511187 & 18313001324552 \\

1095 & 39703341303712 & 43867672430453 & 43867672430453 & 45864145536600 & 89687451269313 \\

%
%
%
%
%
%
%
%
%
%
%

\bottomrule
\end{tabular}\end{center}
\end{table}


\begin{table}[htbp]\begin{center}
\footnotesize
\vspace{0.5cm}
\caption{\footnotesize The table shows the decomposition of the Fourier coefficients multiplying $q^{-D/56}$ in the function $\tilde{h}^{\cN=2}_{g,4}(\t)$ into irreducible representations $\chi_n$ of $M_{23}$.}\smallskip
\begin{tabular}{c@{ }|@{\;}r@{ }r@{ }r@{ }r@{ }r@{ }r}\toprule
&$\chi_1$	&$\chi_2$	&$\chi_3$	&$\chi_4$	&$\chi_5$	&$\chi_6$   \\
	\midrule
-49 & 1 & 0 & 0 & 0 & 0 & 0 \\

31 & 1 & 1 & 0 & 0 & 0 & 0 \\

111 & 4 & 10 & 0 & 0 & 15 & 6 \\

191 & 15 & 69 & 10 & 10 & 245 & 172 \\

271 & 109 & 746 & 572 & 572 & 4753 & 4045 \\

351 & 719 & 8794 & 12676 & 12676 & 76096 & 71969 \\

431 & 6211 & 102911 & 182780 & 182780 & 996449 & 976116 \\

511 & 54133 & 1051489 & 2029244 & 2029244 & 10653186 & 10587505 \\

591 & 449634 & 9337434 & 18600800 & 18600800 & 96246988 & 96195371 \\

671 & 3396944 & 72726076 & 146918168 & 146918168 & 755302028 & 756828089 \\

751 & 23297382 & 505622990 & 1027806326 & 1027806326 & 5268663130 & 5285372319 \\

831 & 145823768 & 3185848487 & 6495661454 & 6495661454 & 33250375888 & 33374695435 \\

911 & 840907061 & 18431146756 & 37635327650 & 37635327650 & 192515167760 & 193289021419 \\

991 & 4506789615 & 98947558783 & 202202119656 & 202202119656 & 1033940285159 & 1038249102575 \\

1071 & 22628187385 & 497248672021 & 1016558364226 & 1016558364226 & 5197058914779 & 5219123838308 \\

%
%
%
%
%
%
%
%
%
%

\end{tabular}

\begin{tabular}{c@{ }|@{\;}r@{ }r@{ }r@{ }r@{ }r@{ }r}\toprule
&$\chi_7$	&$\chi_8$	&$\chi_9$	&$\chi_{10}$&$\chi_{11}$&$\chi_{12}$	 \\
	\midrule
-49 & 0 & 0 & 0 & 0 & 0 & 0 \\

31 & 0 & 0 & 0 & 0 & 0 & 0 \\

111 & 3 & 3 & 8 & 1 & 1 & 4 \\

191 & 131 & 131 & 192 & 326 & 326 & 422 \\

271 & 3864 & 3864 & 4584 & 11441 & 11441 & 13896 \\

351 & 71227 & 71227 & 79986 & 227680 & 227680 & 269022 \\

431 & 973518 & 973518 & 1076328 & 3188386 & 3188386 & 3734146 \\

511 & 10579624 & 10579624 & 11630930 & 35000973 & 35000973 & 40841472 \\

591 & 96173267 & 96173267 & 105510368 & 319449559 & 319449559 & 372208482 \\

671 & 756768664 & 756768664 & 829494146 & 2518305187 & 2518305187 & 2932230208 \\

751 & 5285224408 & 5285224408 & 5790846500 & 17602308990 & 17602308990 & 20489229802 \\

831 & 33374342002 & 33374342002 & 36560188886 & 111198256076 & 111198256076 & 129415838256 \\

911 & 193288204547 & 193288204547 & 211719348700 & 644138990502 & 644138990502 & 749610922070 \\

991 & 1038247291426 & 1038247291426 & 1137194846082 & 3460365519723 & 3460365519723 & 4026806724990 \\

1071 & 5219119941435 & 5219119941435 & 5716368606934 & 17395758923417 & 17395758923417 & 20242906870222 \\

%
%
%
%
%
%
%
%
%
%
%

\bottomrule
\end{tabular}

\vspace{1mm}

\begin{tabular}{c@{ }|@{\;}r@{ }r@{ }r@{ }r@{ }r}
&$\chi_{13}$	&$\chi_{14}$	&$\chi_{15}$	&$\chi_{16}$	&$\chi_{17}$ \\
	\midrule

-49 & 0 & 0 & 0 & 0 & 0 \\

31 & 0 & 0 & 0 & 0 & 0 \\

111 & 4 & 1 & 1 & 14 & 14 \\

191 & 422 & 405 & 405 & 632 & 1024 \\

271 & 13896 & 14621 & 14621 & 17398 & 31919 \\

351 & 269022 & 292364 & 292364 & 319046 & 610695 \\

431 & 3734146 & 4097893 & 4097893 & 4358824 & 8450520 \\

511 & 40841472 & 44996504 & 44996504 & 47382036 & 92324399 \\

591 & 372208482 & 410707009 & 410707009 & 430804236 & 841061633 \\

671 & 2932230208 & 3237782827 & 3237782827 & 3390298956 & 6624684857 \\

751 & 20489229802 & 22631442859 & 22631442859 & 23678982898 & 46287128401 \\

831 & 129415838256 & 142968950388 & 142968950388 & 149529067462 & 292352194117 \\

911 & 749610922070 & 828178149562 & 828178149562 & 866014527476 & 1693351770023 \\

991 & 4026806724990 & 4449040146378 & 4449040146378 & 4651835981222 & 9096369338035 \\

1071 & 20242906870222 & 22365973077886 & 22365973077886 & 23384220869164 & 45727565759616 \\

%
%
%
%
%
%
%
%
%
%

\bottomrule
\end{tabular}
\end{center}
\end{table}
\end{center}


\clearpage

\bibliographystyle{JHEP}
\bibliography{ECFTrefs}

    \end{document}